

\input amssym.tex 

\def\unredoffs{}
\tolerance=1000\hfuzz=2pt
\catcode`\@=11 
\ifx\hyperdef\UNd@FiNeD\def\hyperdef#1#2#3#4{#4}\def\hyperref#1#2#3#4{#4}\def\href#1#2{#2}\fi
\magnification=1200\unredoffs\baselineskip=16pt plus 2pt minus 1pt
\def\Date#1{\vfill\leftline{#1}\tenpoint\supereject%
\footline={\hss\tenrm\hyperdef\hypernoname{page}\folio\folio\hss}}%

{\count255=\time\divide\count255 by 60 \xdef\hourmin{\number\count255}
 \multiply\count255 by-60\advance\count255 by\time
 \xdef\hourmin{\hourmin:\ifnum\count255<10 0\fi\the\count255}
}
\def\date{\number\day.\number\month.\number\year\ at \hourmin}


\def\nolabels{\def\wrlabeL##1{}\def\eqlabeL##1{}\def\reflabeL##1{}}
\def\writelabels{\def\wrlabeL##1{\leavevmode\vadjust{\rlap{\smash%
{\line{{\escapechar=` \hfill\rlap{\sevenrm\hskip.03in\string##1}}}}}}}%
\def\eqlabeL##1{{\escapechar-1\rlap{\sevenrm\hskip.05in\string##1}}}%
\def\reflabeL##1{\noexpand\llap{\noexpand\sevenrm\string\string\string##1}}}
\nolabels

\global\newcount\secno \global\secno=0
\global\newcount\meqno \global\meqno=1
\def\s@csym{}

\def\newsec#1\par{\global\advance\secno by1%
{\toks0{#1}\message{(\the\secno. \the\toks0)}}%
\global\subsecno=0\eqnres@t\let\s@csym\secsym\xdef\secn@m{\the\secno}\noindent
{\bf\hyperdef\hypernoname{section}{\the\secno}{\the\secno.} #1}%
\writetoca{{\string\hyperref{}{section}{\the\secno}{\bf \the\secno\quad}} {\bf #1}}\par%
\nobreak\medskip\nobreak\noindent\ignorespaces}
\def\eqnres@t{\xdef\secsym{\the\secno.}\global\meqno=1\bigbreak\bigskip}
\def\sequentialequations{\def\eqnres@t{\bigbreak}}\xdef\secsym{}

\global\newcount\subsecno \global\subsecno=0
\def\subsec#1\par{\global\advance\subsecno by1%
{\toks0{#1}\message{(\s@csym\the\subsecno. \the\toks0)}}%
\global\subsubsecno=0%
\ifnum\lastpenalty>9000\else\bigbreak\fi
\noindent{\it\hyperdef\hypernoname{subsection}{\secn@m.\the\subsecno}%
{\secn@m.\the\subsecno.} #1}\writetoca{\string\hskip1.45cm
{\string\hyperref{}{subsection}{\secn@m.\the\subsecno}{\secn@m.\the\subsecno.}}
{#1}}\par\nobreak\medskip\nobreak\noindent\ignorespaces}

\def\appendix#1#2{\global\meqno=1\global\subsecno=0\xdef\secsym{\hbox{#1.}}%
\bigbreak\bigskip\noindent{\bf Appendix \hyperdef\hypernoname{appendix}{#1}%
{#1.} #2}{\toks0{(#1. #2)}\message{\the\toks0}}%
\xdef\s@csym{#1.}\xdef\secn@m{#1}%
\writetoca{{\string\hyperref{}{appendix}{#1}{\bf {#1}\quad}} {\bf #2}}%
\par\nobreak\medskip\nobreak}

%
\def\checkm@de#1#2{\ifmmode{\def\f@rst##1{##1}\hyperdef\hypernoname{equation}%
{#1}{#2}}\else\hyperref{}{equation}{#1}{#2}\fi}
\def\eqnn#1{\DefWarn#1\xdef #1{(\noexpand\relax\noexpand\checkm@de%
{\s@csym\the\meqno}{\secsym\the\meqno})}%
\wrlabeL#1\writedef{#1\leftbracket#1}\global\advance\meqno by1}
\def\f@rst#1{\c@t#1a\em@ark}\def\c@t#1#2\em@ark{#1}
\def\eqna#1{\DefWarn#1\wrlabeL{#1$\{\}$}%
\xdef #1##1{(\noexpand\relax\noexpand\checkm@de%
{\s@csym\the\meqno\noexpand\f@rst{##1}1}{\hbox{$\secsym\the\meqno##1$}})}
\writedef{#1\numbersign1\leftbracket#1{\numbersign1}}\global\advance\meqno by1}
\def\eqn#1#2{\DefWarn#1%
\xdef #1{(\noexpand\hyperref{}{equation}{\s@csym\the\meqno}%
{\secsym\the\meqno})}$$#2\eqno(\hyperdef\hypernoname{equation}%
{\s@csym\the\meqno}{\secsym\the\meqno})\eqlabeL#1$$%
\writedef{#1\leftbracket#1}\global\advance\meqno by1}
\def\xeqn{\expandafter\xe@n}\def\xe@n(#1){#1}
\def\xeqna#1{\expandafter\xe@n#1}
\def\eqns#1{(\e@ns #1{\hbox{}})}
\def\e@ns#1{\ifx\UNd@FiNeD#1\message{eqnlabel \string#1 is undefined.}%
\xdef#1{(?.?)}\fi{\let\hyperref=\relax\xdef\next{#1}}%
\ifx\next\em@rk\def\next{}\else%
\ifx\next#1\xeqn#1\else\def\n@xt{#1}\ifx\n@xt\next#1\else\xeqna#1\fi
\fi\let\next=\e@ns\fi\next}

\def\DefWarn#1{\ifx\UNd@FiNeD#1\else
\immediate\write16{*** WARNING: the label \string#1 is already defined ***}\fi}
%
\newskip\footskip\footskip14pt plus 1pt minus 1pt 
\def\footnotefont{\ninepoint}\def\f@t#1{\footnotefont #1\@foot}
\def\f@@t{\baselineskip\footskip\bgroup\footnotefont\aftergroup\@foot\let\next}
\setbox\strutbox=\hbox{\vrule height9.5pt depth4.5pt width0pt}
\global\newcount\ftno \global\ftno=0
\def\foot{\global\advance\ftno by1\def\foot@rg{\hyperref{}{footnote}%
{\the\ftno}{\the\ftno}\xdef\foot@rg{\noexpand\hyperdef\noexpand\hypernoname%
{footnote}{\the\ftno}{\the\ftno}}}\footnote{$^{\foot@rg}$}}
%
%
%
\global\newcount\refno \global\refno=1
\newwrite\rfile
\def\ref{[\hyperref{}{reference}{\the\refno}{\the\refno}]\nref}
\def\nref#1{\DefWarn#1%
\xdef#1{[\noexpand\hyperref{}{reference}{\the\refno}{\the\refno}]}%
\writedef{#1\leftbracket#1}%
\ifnum\refno=1\immediate\openout\rfile=\jobname.refs\fi
\chardef\wfile=\rfile\immediate\write\rfile{\noexpand\item{[\noexpand\hyperdef%
\noexpand\hypernoname{reference}{\the\refno}{\the\refno}]\ }%
\reflabeL{#1\hskip.31in}\pctsign}\global\advance\refno by1\findarg}
\def\findarg#1#{\begingroup\obeylines\newlinechar=`\^^M\pass@rg}
{\obeylines\gdef\pass@rg#1{\writ@line\relax #1^^M\hbox{}^^M}%
\gdef\writ@line#1^^M{\expandafter\toks0\expandafter{\striprel@x #1}%
\edef\next{\the\toks0}\ifx\next\em@rk\let\next=\endgroup\else\ifx\next\empty%
\else\immediate\write\wfile{\the\toks0}\fi\let\next=\writ@line\fi\next\relax}}
\def\striprel@x#1{} \def\em@rk{\hbox{}}
\def\lref{\begingroup\obeylines\lr@f}
\def\lr@f#1#2{\DefWarn#1\gdef#1{\let#1=\UNd@FiNeD\ref#1{#2}}\endgroup\unskip}
\def\semi{;\hfil\break}
\def\addref#1{\immediate\write\rfile{\noexpand\item{}#1}} 
\def\listrefs{\vfill\supereject\immediate\closeout\rfile\writestoppt
\baselineskip=\footskip\centerline{{\bf References}}\bigskip{\parindent=20pt%
\frenchspacing\escapechar=` \input \jobname.refs\vfill\eject}\nonfrenchspacing}
\def\startrefs#1{\immediate\openout\rfile=\jobname.refs\refno=#1}
\def\xref{\expandafter\xr@f}\def\xr@f[#1]{#1}
\def\refs#1{\count255=1[\r@fs #1{\hbox{}}]}
\def\r@fs#1{\ifx\UNd@FiNeD#1\message{reflabel \string#1 is undefined.}%
\nref#1{need to supply reference \string#1.}\fi%
\vphantom{\hphantom{#1}}{\let\hyperref=\relax\xdef\next{#1}}%
\ifx\next\em@rk\def\next{}%
\else\ifx\next#1\ifodd\count255\relax\xref#1\count255=0\fi%
\else#1\count255=1\fi\let\next=\r@fs\fi\next}
%

%
\newwrite\ffile\global\newcount\figno \global\figno=1
\def\fig{fig.~\hyperref{}{figure}{\the\figno}{\the\figno}\nfig}
\def\nfig#1{\DefWarn#1%
\xdef#1{fig.~\noexpand\hyperref{}{figure}{\the\figno}{\the\figno}}%
\writedef{#1\leftbracket fig.\noexpand~\xfig#1}%
\ifnum\figno=1\immediate\openout\ffile=\jobname.figs\fi\chardef\wfile=\ffile%
{\let\hyperref=\relax
\immediate\write\ffile{\noexpand\medskip\noexpand\item{Fig.\ %
\noexpand\hyperdef\noexpand\hypernoname{figure}{\the\figno}{\the\figno}. }
\reflabeL{#1\hskip.55in}\pctsign}}\global\advance\figno by1\findarg}
\def\xfig{\expandafter\xf@g}\def\xf@g fig.\penalty\@M\ {}
\def\figs#1{figs.~\f@gs #1{\hbox{}}}
\def\f@gs#1{{\let\hyperref=\relax\xdef\next{#1}}\ifx\next\em@rk\def\next{}\else
\ifx\next#1\xfig #1\else#1\fi\let\next=\f@gs\fi\next}
%
\def\figin{\epsfcheck\figin}\def\figins{\epsfcheck\figins}
\def\epsfcheck{\ifx\epsfbox\UnDeFiNeD
\message{(NO epsf.tex, FIGURES WILL BE IGNORED)}
\gdef\figin##1{\vskip2in}\gdef\figins##1{\hskip.5in}
\else\message{(FIGURES WILL BE INCLUDED)}%
\gdef\figin##1{##1}\gdef\figins##1{##1}\fi}
\def\DefWarn#1{}
\def\figinsert{\goodbreak\topinsert}
\def\ifig#1#2#3{\DefWarn#1\xdef#1{fig.~\the\figno}
\writedef{#1\leftbracket fig.\noexpand~\the\figno}%
\figinsert\figin{\centerline{#3}}
\smallskip
\leftskip=0pt \rightskip=0pt
\baselineskip12pt\noindent
{{\bf Fig.~\the\figno}\ \ninepoint #2}
\medskip
\global\advance\figno by1\par\endinsert}
\newwrite\lfile
{\escapechar-1\xdef\pctsign{\string\%}\xdef\leftbracket{\string\{}
\xdef\rightbracket{\string\}}\xdef\numbersign{\string\#}}
\def\writedefs{\immediate\openout\lfile=label.defs \def\writedef##1{%
{\let\hyperref=\relax\let\hyperdef=\relax\let\hypernoname=\relax
 \immediate\write\lfile{\string\def\string##1\rightbracket}}}}%
\def\writestop{\def\writestoppt{\immediate\write\lfile{\string\pageno
 \the\pageno\string\startrefs\leftbracket\the\refno\rightbracket
 \string\def\string\secsym\leftbracket\secsym\rightbracket
 \string\secno\the\secno\string\meqno\the\meqno}\immediate\closeout\lfile}}
\def\writestoppt{}\def\writedef#1{}

\def\seclab#1{\DefWarn#1%
\xdef #1{\noexpand\hyperref{}{section}{\the\secno}{\the\secno}}%
\writedef{#1\leftbracket#1}\wrlabeL{#1=#1}\par%
\nobreak\medskip\nobreak\noindent\ignorespaces}
\def\subseclab#1\par{\DefWarn#1%
\xdef #1{\noexpand\hyperref{}{subsection}{\the\secno.\the\subsecno}%
{\the\secno.\the\subsecno}}\writedef{#1\leftbracket#1}\wrlabeL{#1=#1}\par%
\nobreak\medskip\nobreak\noindent\ignorespaces}
\def\applab#1{\DefWarn#1%
\xdef #1{\noexpand\hyperref{}{appendix}{\secn@m}{\secn@m}}%
\writedef{#1\leftbracket#1}\wrlabeL{#1=#1}}
\def\appsublab#1{\DefWarn#1%
\xdef #1{\noexpand\hyperref{}{appendix}{\secn@m.\the\subsecno}{\secn@m.\the\subsecno}}%
\writedef{#1\leftbracket#1}\wrlabeL{#1=#1}}
\newwrite\tfile \def\writetoca#1{}
\def\leaderfill{\leaders\hbox to 1em{\hss.\hss}\hfill}
\def\writetoc{\immediate\openout\tfile=\jobname.toc
   \def\writetoca##1{{\edef\next{\write\tfile{\noindent ##1
   \string\leaderfill{
   \string\hyperref{}{page}{\noexpand\number\pageno}%
   {\noexpand\number\pageno}} \par}}\next}}
}
\newread\ch@ckfile
\def\listtoc{\immediate\closeout\tfile\immediate\openin\ch@ckfile=\jobname.toc
\ifeof\ch@ckfile\message{no file \jobname.toc, no table of contents this pass}%
\else\closein\ch@ckfile\centerline{\bf Contents}\nobreak\medskip%
{\baselineskip=16pt\footnotefont\parskip=0pt\catcode`\@=11\input\jobname.toc
\catcode`\@=12\bigbreak\bigskip}\fi}
\catcode`\@=12 
\def\tenpoint{\def\rm{\fam0\tenrm}
\textfont0=\tenrm \scriptfont0=\sevenrm \scriptscriptfont0=\fiverm
\textfont1=\teni  \scriptfont1=\seveni  \scriptscriptfont1=\fivei
\textfont2=\tensy \scriptfont2=\sevensy \scriptscriptfont2=\fivesy
\textfont\itfam=\tenit \def\it{\fam\itfam\tenit}\def\footnotefont{\ninepoint}%
\textfont\bffam=\tenbf \def\bf{\fam\bffam\tenbf}\def\sl{\fam\slfam\tensl}\rm}
\font\ninerm=cmr9 \font\sixrm=cmr6 \font\ninei=cmmi9 \font\sixi=cmmi6
\font\ninesy=cmsy9 \font\sixsy=cmsy6 \font\ninebf=cmbx9
\font\nineit=cmti9 \font\ninesl=cmsl9 \skewchar\ninei='177
\skewchar\sixi='177 \skewchar\ninesy='60 \skewchar\sixsy='60
\def\ninepoint{\def\rm{\fam0\ninerm}
\textfont0=\ninerm \scriptfont0=\sixrm \scriptscriptfont0=\fiverm
\textfont1=\ninei \scriptfont1=\sixi \scriptscriptfont1=\fivei
\textfont2=\ninesy \scriptfont2=\sixsy \scriptscriptfont2=\fivesy
\textfont\itfam=\ninei \def\it{\fam\itfam\nineit}\def\sl{\fam\slfam\ninesl}%
\textfont\bffam=\ninebf \def\bf{\fam\bffam\ninebf}\rm}
%
\hyphenation{anom-aly anom-alies coun-ter-term coun-ter-terms}

\global\newcount\subsubsecno \global\subsubsecno=0
\def\subsubsec#1\par{\global\advance\subsubsecno by1%
{\toks0{#1}\message{(\the\secno\the\subsecno\the\subsubsecno. \the\toks0)}}%
\ifnum\lastpenalty>9000\else\bigbreak\fi
\noindent{\it\hyperdef\hypernoname{subsubsection}{\the\secno.\the\subsecno\the\subsubsecno}%
{\the\secno.\the\subsecno.\the\subsubsecno.} #1}
\par\nobreak\medskip\nobreak\noindent\ignorespaces}

\def\DefWarn#1{}
\def\tikzcaption#1#2{\DefWarn#1\xdef#1{Fig.~\the\figno}
\writedef{#1\leftbracket Fig.\noexpand~\the\figno}%
{
\smallskip
\leftskip=20pt \rightskip=20pt \baselineskip12pt\noindent
{{\bf Fig.~\the\figno}\ \ninepoint #2}
\bigskip
\global\advance\figno by1 \par}}

\def\ntoalpha#1{%
\ifcase#1%
@%
\or A\or B\or C\or D\or E\or F\or G\or H\or I
\fi
}

\global\newcount\appno \global\appno=1
\def\applab#1{\xdef #1{\ntoalpha\appno}\writedef{#1\leftbracket#1}\wrlabeL{#1=#1}
\global\advance\appno by1}

\def\preprint#1 #2\par{\rightline{\vbox{\baselineskip12pt\hbox{#1}\hbox{#2}}}\vskip2cm}
%
\def\title#1\par{\centerline{\bf #1}\nopagenumbers\pageno=0}
\def\author#1\par{\bigskip\bigskip\centerline{#1}}

\newcount\addressno

\def\email#1#2{
\footnote{\null}{\kern-\parindent \llap{$^#1$\hskip1pt}email: #2}}

\def\startcenter{%
  \par
  \begingroup
  \leftskip=0pt plus 1fil
  \rightskip=\leftskip
  \parindent=0pt
  \parfillskip=0pt
}
\def\stopcenter{\endgroup}

\def\address{\bigskip%
  \ifnum\the\addressno=0\else\stopcenter\endgroup\fi
  \advance\addressno by 1%
  \begingroup
  \startcenter
  \it
  \obeylines
  \addressAux
}
\def\addressAux#1{#1}

\def\abstract{\stopcenter\endgroup\bigskip\bigskip\noindent}

\def\Dsl{\,\raise.15ex\hbox{/}\mkern-13.5mu D} 
\def\dsl{\raise.15ex\hbox{/}\kern-.57em\partial}
 
\def\boxeqn#1{\vcenter{\vbox{\hrule\hbox{\vrule\kern3pt\vbox{\kern3pt
	\hbox{${\displaystyle #1}$}\kern3pt}\kern3pt\vrule}\hrule}}}


\def\ap{{\alpha^{\prime}}}

\def\a{\alpha}
\def\b{{\beta}}

\def\d{{\delta}}

\def\t{{\theta}}

\def\half{{1\over 2}}
\def\p{{\partial}}

\def\bar{\overline}
\def\({\left(}
\def\){\right)}
\def\dz{{\rm d}z}

\def\cZ{{\cal Z}}


\def\Box{\square}


\def\len#1{{%
\def\Dlen{\left|\mkern-1mu #1\mkern -0.5mu\right|}%
\def\Sslen{\left|\mkern-1.3mu #1\mkern -1.3mu\right|}%
\def\SSlen{\left|\mkern-2.8mu #1\mkern-1.3mu\right|}%
\mathchoice{\Dlen}{\Dlen}{\Sslen}{\SSlen}}}

\def\perm#1{{\rm perm}#1}

\def\sfrac#1/#2{\kern.1em\raise.5ex\hbox{\the\scriptfont0 #1}%
\kern-.1em/\kern-.15em\lower.25ex\hbox{\the\scriptfont0 #2}}

\font\tenshuffle=shuffle10 \font\sevenshuffle=shuffle7 \font\fiveshuffle=shuffle7 at 5pt
\def\shuffle{{%
\def\Dshuffle{\mathbin{\hbox{\tenshuffle\char'001}}}%
\def\Sshuffle{\mathbin{\hbox{\sevenshuffle\char'001}}}%
\def\SSshuffle{\mathbin{\hbox{\fiveshuffle\char'001}}}%
\mathchoice{\Dshuffle}{\Dshuffle}{\Sshuffle}{\SSshuffle}}}


\def\qed{\hbox{\hskip 3pt
\vbox{\hrule\hbox to 7pt{\vrule height 7pt\hfill\vrule}
\hrule}}\hskip3pt}

\overfullrule=0pt\relax

\frenchspacing

\newread\instream \openin\instream= label.defs
\ifeof\instream \message{No labels in advance yet. Wait till next pass.}
\else \closein\instream \input label.defs
\fi
\writedefs

\def\arXiv:#1].{\hepthStrip#1 \nil}
\def\hepthStrip#1 #2\nil{\href{http://arxiv.org/abs/#1}{arXiv:#1 #2\unskip}].}

\def\proof{\medskip\noindent{\bf Proof.\enspace}}
\def\dd{{\rm d}}
\def\frac#1#2{{#1\over #2}}
\def\reg{{\rm reg}}
\def\sign{{\rm sign}}
\def\eom{{\rm eom}}
\def\ord{{\mathop{\rm ord}}}

\lref\BGapwww{
{\tt http://repo.or.cz/BGap.git}
}

\preprint \phantom{M}

\title Non-abelian $Z$-theory:

\title Berends--Giele recursion for the $\ap$-expansion of disk integrals

\author
Carlos R. Mafra$^{a,b}$\email{a}{c.r.mafra@soton.ac.uk} and
Oliver Schlotterer$^c$\email{c}{olivers@aei.mpg.de}

\address
$^a$ STAG Research Centre and Mathematical Sciences,
University of Southampton, UK
\medskip
$^b$Institute for Advanced Study, School of Natural Sciences,
Einstein Drive, Princeton, NJ 08540, USA
\medskip
$^c$Max--Planck--Institut f\"ur Gravitationsphysik,
Albert--Einstein--Institut,
Am M\"uhlenberg 1, 14476 Potsdam, Germany

\abstract
We present a recursive method to calculate the $\ap$-expansion
of disk integrals arising in tree-level scattering of open strings
which resembles the approach of Berends and Giele to gluon amplitudes.
Following an earlier interpretation of disk integrals as doubly partial
amplitudes of an effective theory of scalars dubbed as $Z$-theory, we pinpoint
the equation of motion of $Z$-theory from the Berends--Giele recursion for
its tree amplitudes. A computer implementation of this method
including explicit results for the recursion up to order $\ap^7$
is made available on the website {\tt http://repo.or.cz/BGap.git}.

\Date {September 2016}


\lref\GreenGA{
  M.~B.~Green and M.~Gutperle,
  ``Symmetry breaking at enhanced symmetry points,''
Nucl.\ Phys.\ B {\bf 460}, 77 (1996).
[hep-th/9509171].
}

\lref\inprog{
J.~J.~M.~Carrasco, C.~R.~Mafra and O.~Schlotterer,
  to appear
}

\lref\abZtheory{
J.~J.~M.~Carrasco, C.~R.~Mafra and O.~Schlotterer,
  ``Abelian Z-theory: NLSM amplitudes and alpha'-corrections from the open string,''
[arXiv:1608.02569 [hep-th]].
}

\lref\howeSYM{
S.~J.~Gates, Jr. and S.~Vashakidze,
  ``On $D=10$, $N=1$ Supersymmetry, Superspace Geometry and Superstring Effects,''
Nucl.\ Phys.\ B {\bf 291}, 172 (1987).
\semi
M.~Cederwall, B.~E.~W.~Nilsson and D.~Tsimpis,
  ``The Structure of maximally supersymmetric Yang-Mills theory: Constraining higher order corrections,''
JHEP {\bf 0106}, 034 (2001).
[hep-th/0102009].
\semi
 M.~Cederwall, B.~E.~W.~Nilsson and D.~Tsimpis,
  ``D = 10 superYang-Mills at ${\cal O}(\ap^2)$,''
JHEP {\bf 0107}, 042 (2001).
[hep-th/0104236].
\semi
M.~Cederwall, B.~E.~W.~Nilsson and D.~Tsimpis,
  ``Spinorial cohomology of Abelian D=10 superYang-Mills at ${\cal O}(\ap^3)$,''
JHEP {\bf 0211}, 023 (2002).
[hep-th/0205165].
\semi
N.~Berkovits and P.~S.~Howe,
  ``The Cohomology of superspace, pure spinors and invariant integrals,''
JHEP {\bf 0806}, 046 (2008).
[arXiv:0803.3024 [hep-th]].
\semi
  P.~S.~Howe, U.~Lindstrom and L.~Wulff,
  ``D=10 supersymmetric Yang-Mills theory at $\ap^4$,''
JHEP {\bf 1007}, 028 (2010).
[arXiv:1004.3466 [hep-th]].
}

\lref\chapoton{
F.~Chapoton, ``The anticyclic operad of moulds'',
International Mathematics Research Notices 2007 (2007).
[math/0609436].
}

\lref\white{
  C.D.~White,
  ``Exact solutions for the biadjoint scalar field,''
[arXiv:1606.04724 [hep-th]].
}

\lref\harmonicP{
  E.~Remiddi and J.~A.~M.~Vermaseren,
  ``Harmonic polylogarithms,''
Int.\ J.\ Mod.\ Phys.\ A {\bf 15}, 725 (2000).
[hep-ph/9905237].
}

\lref\datamine{
  J.~Blumlein, D.~J.~Broadhurst and J.~A.~M.~Vermaseren,
  ``The Multiple Zeta Value Data Mine,''
Comput.\ Phys.\ Commun.\  {\bf 181}, 582 (2010).
[arXiv:0907.2557 [math-ph]].
}

\lref\FORM{
	J.A.M.~Vermaseren,
  	``New features of FORM,''
	[math-ph/0010025].
\semi
	J.~Kuipers, T.~Ueda, J.A.M.~Vermaseren and J.~Vollinga,
	``FORM version 4.0,''
	Comput.\ Phys.\ Commun.\  {\bf 184}, 1453 (2013).
	[arXiv:1203.6543 [cs.SC]].
}

\lref\thibon{
J.-Y. Thibon,
``Lie idempotents in descent algebras''
(lecture notes), Workshop on Hopf Algebras and Props, Boston, March 5 - 9, 2007 Clay Mathematics Institute.
}

\lref\BrownUM{
  F.~Brown,
  ``The Massless higher-loop two-point function,''
Commun.\ Math.\ Phys.\  {\bf 287}, 925 (2009).
[arXiv:0804.1660 [math.AG]]. \semi
C.~Bogner and F.~Brown,
  ``Symbolic integration and multiple polylogarithms,''
PoS LL {\bf 2012}, 053 (2012).
[arXiv:1209.6524 [hep-ph]]. \semi
C.~Anastasiou, C.~Duhr, F.~Dulat and B.~Mistlberger,
  ``Soft triple-real radiation for Higgs production at N3LO,''
JHEP {\bf 1307}, 003 (2013).
[arXiv:1302.4379 [hep-ph]].
}

\lref\BognerNDA{
  C.~Bogner,
  ``MPL -- A program for computations with iterated integrals on moduli spaces of curves of genus zero,''
Comput.\ Phys.\ Commun.\  {\bf 203}, 339 (2016).
[arXiv:1510.04562 [physics.comp-ph]].
}

\lref\BognerMHA{
  C.~Bogner and F.~Brown,
  ``Feynman integrals and iterated integrals on moduli spaces of curves of genus zero,''
Commun.\ Num.\ Theor.\ Phys.\  {\bf 09}, 189 (2015).
[arXiv:1408.1862 [hep-th]].
}

\lref\StiebergerWEA{
  S.~Stieberger,
  ``Closed superstring amplitudes, single-valued multiple zeta values and the Deligne associator,''
J.\ Phys.\ A {\bf 47}, 155401 (2014).
[arXiv:1310.3259 [hep-th]].
}

\lref\BarreiroDPA{
  L.~A.~Barreiro and R.~Medina,
  ``RNS derivation of N-point disk amplitudes from the revisited S-matrix approach,''
Nucl.\ Phys.\ B {\bf 886}, 870 (2014).
[arXiv:1310.5942 [hep-th]].
}

\lref\GoncharovIEA{
  A.~B.~Goncharov,
  ``Multiple polylogarithms and mixed Tate motives,''
[math/0103059 [math.AG]].
}

\lref\DuhrZQ{
  C.~Duhr, H.~Gangl and J.~R.~Rhodes,
  ``From polygons and symbols to polylogarithmic functions,''
JHEP {\bf 1210}, 075 (2012).
[arXiv:1110.0458 [math-ph]].
}

\lref\AblingerCF{
  J.~Ablinger, J.~Bl\"umlein and C.~Schneider,
  ``Analytic and Algorithmic Aspects of Generalized Harmonic Sums and Polylogarithms,''
J.\ Math.\ Phys.\  {\bf 54}, 082301 (2013).
[arXiv:1302.0378 [math-ph]]. \semi
J.~Ablinger and J.~Bl\"umlein,
  ``Harmonic Sums, Polylogarithms, Special Numbers, and their Generalizations,''
[arXiv:1304.7071 [math-ph]].
}

\lref\TourkineBAK{
  P.~Tourkine and P.~Vanhove,
  ``Higher-loop amplitude monodromy relations in string and gauge theory,''
[arXiv:1608.01665 [hep-th]].
}

\lref\PanzerIDA{
  E.~Panzer,
  ``Feynman integrals and hyperlogarithms,''
[arXiv:1506.07243 [math-ph]].
}
\lref\hyperint{
  E.~Panzer,
  ``Algorithms for the symbolic integration of hyperlogarithms with applications to Feynman integrals,''
Comput.\ Phys.\ Commun.\  {\bf 188}, 148 (2015).
[arXiv:1403.3385 [hep-th]].
}

\lref\fourptids{
	C.~R.~Mafra,
  	``Pure Spinor Superspace Identities for Massless Four-point Kinematic Factors,''
	JHEP {\bf 0804}, 093 (2008).
	[arXiv:0801.0580 [hep-th]].
}

\lref\BerendsME{
	F.~A.~Berends and W.~T.~Giele,
  	``Recursive Calculations for Processes with n Gluons,''
	Nucl.\ Phys.\ B {\bf 306}, 759 (1988).
}
\lref\PSBCJ{
	C.~R.~Mafra, O.~Schlotterer and S.~Stieberger,
	``Explicit BCJ Numerators from Pure Spinors,''
	JHEP {\bf 1107}, 092 (2011).
	[arXiv:1104.5224 [hep-th]].
}
\lref\BGPS{
	C.~R.~Mafra and O.~Schlotterer,
  	``Berends-Giele recursions and the BCJ duality in superspace and components,''
	JHEP {\bf 1603}, 097 (2016).
	[arXiv:1510.08846 [hep-th]].
}
\lref\BGSym{
	F.~A.~Berends and W.~T.~Giele,
	``Multiple Soft Gluon Radiation in Parton Processes,''
	Nucl.\ Phys.\ B {\bf 313}, 595 (1989).
}
\lref\Gauge{
	S.~Lee, C.R.~Mafra and O.~Schlotterer,
  	``Non-linear gauge transformations in $D=10$ SYM theory and the BCJ duality,''
	JHEP {\bf 1603}, 090 (2016).
	[arXiv:1510.08843 [hep-th]].
}
\lref\DPellis{
	F.~Cachazo, S.~He and E.Y.~Yuan,
	``Scattering of Massless Particles: Scalars, Gluons and Gravitons,''
	JHEP {\bf 1407}, 033 (2014).
	[arXiv:1309.0885 [hep-th]].
}
\lref\KKsym{
	R.~Kleiss and H.~Kuijf,
	``Multi - Gluon Cross-sections and Five Jet Production at Hadron Colliders,''
	Nucl.\ Phys.\ B {\bf 312}, 616 (1989).
}
\lref\oldMomKer{
	Z.~Bern, L.~J.~Dixon, M.~Perelstein and J.~S.~Rozowsky,
	``Multileg one loop gravity amplitudes from gauge theory,''
	Nucl.\ Phys.\ B {\bf 546}, 423 (1999).
	[hep-th/9811140].
}

\lref\BjerrumBohrRD{
  N.~E.~J.~Bjerrum-Bohr, P.~H.~Damgaard and P.~Vanhove,
  ``Minimal Basis for Gauge Theory Amplitudes,''
Phys.\ Rev.\ Lett.\  {\bf 103}, 161602 (2009).
[arXiv:0907.1425 [hep-th]].
}

\lref\StiebergerHQ{
  S.~Stieberger,
  ``Open \& Closed vs. Pure Open String Disk Amplitudes,''
[arXiv:0907.2211 [hep-th]].
}

\lref\MomKer{
	N.~E.~J.~Bjerrum-Bohr, P.~H.~Damgaard, T.~Sondergaard and P.~Vanhove,
	``The Momentum Kernel of Gauge and Gravity Theories,''
	JHEP {\bf 1101}, 001 (2011).
	[arXiv:1010.3933 [hep-th]].
}
\lref\Polylogs{
	J.~Broedel, O.~Schlotterer and S.~Stieberger,
	``Polylogarithms, Multiple Zeta Values and Superstring Amplitudes,''
	Fortsch.\ Phys.\  {\bf 61}, 812 (2013).
	[arXiv:1304.7267 [hep-th]].
}

\lref\StiebergerHZA{
  S.~Stieberger and T.~R.~Taylor,
  ``Superstring Amplitudes as a Mellin Transform of Supergravity,''
Nucl.\ Phys.\ B {\bf 873}, 65 (2013).
[arXiv:1303.1532 [hep-th]].
}

\lref\BroedelAZA{
  J.~Broedel, O.~Schlotterer, S.~Stieberger and T.~Terasoma,
  ``All order $\alpha^{\prime}$-expansion of superstring trees from the Drinfeld associator,''
Phys.\ Rev.\ D {\bf 89}, no. 6, 066014 (2014).
[arXiv:1304.7304 [hep-th]].
}

\lref\nptTreeI{
	C.~R.~Mafra, O.~Schlotterer and S.~Stieberger,
	``Complete N-Point Superstring Disk Amplitude I. Pure Spinor Computation,''
	Nucl.\ Phys.\ B {\bf 873}, 419 (2013).
	[arXiv:1106.2645 [hep-th]].
}
\lref\nptTreeII{
	C.~R.~Mafra, O.~Schlotterer and S.~Stieberger,
	``Complete N-Point Superstring Disk Amplitude II. Amplitude
	and Hypergeometric Function Structure,''
	Nucl.\ Phys.\ B {\bf 873}, 461 (2013).
	[arXiv:1106.2646 [hep-th]].
}

\lref\StiebergerRR{
  S.~Stieberger,
  ``Constraints on Tree-Level Higher Order Gravitational Couplings in Superstring Theory,''
Phys.\ Rev.\ Lett.\  {\bf 106}, 111601 (2011).
[arXiv:0910.0180 [hep-th]].
}

\lref\nptMethod{
	C.~R.~Mafra, O.~Schlotterer, S.~Stieberger and D.~Tsimpis,
	``A recursive method for SYM n-point tree amplitudes,''
	Phys.\ Rev.\ D {\bf 83}, 126012 (2011).
	[arXiv:1012.3981 [hep-th]].
}

\lref\EOMBBs{
	C.~R.~Mafra and O.~Schlotterer,
  	``Multiparticle SYM equations of motion and pure spinor BRST blocks,''
	JHEP {\bf 1407}, 153 (2014).
	[arXiv:1404.4986 [hep-th]].
}
\lref\reutenauer{
	C.~Reutenauer,
	``Free Lie Algebras'', London Mathematical Society Monographs, 1993.
}
\lref\Ree{
	R.~Ree, ``Lie elements and an algebra associated with shuffles'',
	Ann. Math. {\bf 62}, No. 2 (1958), 210--220.
}

\lref\Griffing{
G.~Griffing, ``Dual Lie Elements and a Derivation for the Cofree Coassociative Coalgebra'', 
Proceedings of the American Mathematical Society {\bf 123}, No. 11 (1995), 3269--3277.
}

\lref\BGschocker{
	M.~Schocker,
	``Lie elements and Knuth relations,'' Canad. J. Math. {\bf 56} (2004), 871-882.
	[math/0209327].
}
\lref\Selivanov{
	K.~G.~Selivanov,
	``Postclassicism in tree amplitudes,''
	[hep-th/9905128].
}

\lref\moreSelivanov{
A.~A.~Rosly and K.~G.~Selivanov,
  ``On amplitudes in selfdual sector of Yang-Mills theory,''
Phys.\ Lett.\ B {\bf 399}, 135 (1997).
[hep-th/9611101].
\semi
	A.~A.~Rosly and K.~G.~Selivanov,
  	``Gravitational SD perturbiner,''
	[hep-th/9710196].
}
\lref\Bardeen{
  W.~A.~Bardeen,
  ``Selfdual Yang-Mills theory, integrability and multiparton amplitudes,''
Prog.\ Theor.\ Phys.\ Suppl.\  {\bf 123}, 1 (1996).
}

\lref\KLT{
	H.~Kawai, D.~C.~Lewellen and S.~H.~H.~Tye,
	``A Relation Between Tree Amplitudes of Closed and Open Strings,''
	Nucl.\ Phys.\ B {\bf 269}, 1 (1986).
}

\lref\mathMZV{
K.~Aomoto, ``Special values of hyperlogarithms and linear difference schemes,''
 Illinois J. Math. {\bf 34} (2), 191 (1990).
T.~Terasoma, ``Selberg integrals and multiple zeta values,'' Compositio Mathematica {\bf 133}, 1 (2002).}
\lref\BrownmathMZV{
F.~Brown, ``Multiple zeta values and periods of moduli spaces ${\cal M}_{0,n}$,''
Ann. Sci. Ec. Norm. Super. {\bf 42} (4), 371 (2009),
[math/0606419].
}

\lref\OprisaWU{
	R.~Medina, F.~T.~Brandt and F.~R.~Machado,
  	``The Open superstring five point amplitude revisited,''
	JHEP {\bf 0207}, 071 (2002).
	[hep-th/0208121].
\semi
	L.~A.~Barreiro and R.~Medina,
  	``5-field terms in the open superstring effective action,''
	JHEP {\bf 0503}, 055 (2005).
	[hep-th/0503182].
\semi
  D.~Oprisa and S.~Stieberger,
  ``Six gluon open superstring disk amplitude, multiple hypergeometric series and Euler-Zagier sums,''
[hep-th/0509042]. \semi
S.~Stieberger and T.~R.~Taylor,
  ``Multi-Gluon Scattering in Open Superstring Theory,''
Phys.\ Rev.\ D {\bf 74}, 126007 (2006).
[hep-th/0609175].
}

\lref\psf{
 	N.~Berkovits,
	``Super-Poincare covariant quantization of the superstring,''
	JHEP {\bf 0004}, 018 (2000)
	[arXiv:hep-th/0001035].
}
\lref\Duhr{
	C.~Duhr, S.~Hoeche and F.~Maltoni,
	``Color-dressed recursive relations for multi-parton amplitudes,''
	JHEP {\bf 0608}, 062 (2006).
	[hep-ph/0607057].
}

\lref\SchnetzHQA{
  O.~Schnetz,
  ``Graphical functions and single-valued multiple polylogarithms,''
Commun.\ Num.\ Theor.\ Phys.\  {\bf 08}, 589 (2014).
[arXiv:1302.6445 [math.NT]]. \semi
F.~Brown,
  ``Single-valued Motivic Periods and Multiple Zeta Values,''
SIGMA {\bf 2}, e25 (2014).
[arXiv:1309.5309 [math.NT]].
}

\lref\BCJ{
	Z.~Bern, J.~J.~M.~Carrasco and H.~Johansson,
  	``New Relations for Gauge-Theory Amplitudes,''
	Phys.\ Rev.\ D {\bf 78}, 085011 (2008).
	[arXiv:0805.3993 [hep-ph]].
}

\lref\FTlimit{
	C.R.~Mafra,
	``Berends-Giele recursion for double-color-ordered amplitudes,''
	JHEP {\bf 1607}, 080 (2016).
	[arXiv:1603.09731 [hep-th]].
}
\lref\Bardeen{
  W.~A.~Bardeen,
  ``Selfdual Yang-Mills theory, integrability and multiparton amplitudes,''
Prog.\ Theor.\ Phys.\ Suppl.\  {\bf 123}, 1 (1996).
}

\lref\HuangTAG{
  Y.~t.~Huang, O.~Schlotterer and C.~Wen,
  ``Universality in string interactions,''
[arXiv:1602.01674 [hep-th]].
}

\lref\SchlottererNY{
  O.~Schlotterer and S.~Stieberger,
  ``Motivic Multiple Zeta Values and Superstring Amplitudes,''
J.\ Phys.\ A {\bf 46}, 475401 (2013).
[arXiv:1205.1516 [hep-th]].
}

\lref\DrummondVZ{
  J.~M.~Drummond and E.~Ragoucy,
  ``Superstring amplitudes and the associator,''
JHEP {\bf 1308}, 135 (2013).
[arXiv:1301.0794 [hep-th]].
}

\lref\WWW{ J.~Broedel, O.~Schlotterer and S.~Stieberger,
{\tt http://mzv.mpp.mpg.de}
}

\lref\StiebergerHBA{
  S.~Stieberger and T.~R.~Taylor,
  ``Closed String Amplitudes as Single-Valued Open String Amplitudes,''
Nucl.\ Phys.\ B {\bf 881}, 269 (2014).
[arXiv:1401.1218 [hep-th]].
}

\lref\BohrMomKer{
	N.~E.~J.~Bjerrum-Bohr, P.~H.~Damgaard, B.~Feng and T.~Sondergaard,
	``Gravity and Yang-Mills Amplitude Relations,''
	Phys.\ Rev.\ D {\bf 82}, 107702 (2010).
	[arXiv:1005.4367 [hep-th]].
}
\lref\grosswitten{
  D.~J.~Gross and E.~Witten,
  ``Superstring Modifications of Einstein's Equations,''
Nucl.\ Phys.\ B {\bf 277}, 1 (1986).
}

\lref\GreenSW{
  M.~B.~Green, J.~H.~Schwarz and L.~Brink,
  ``N=4 Yang-Mills and N=8 Supergravity as Limits of String Theories,''
Nucl.\ Phys.\ B {\bf 198}, 474 (1982).
}

\lref\PuhlfuerstGTA{
  G.~Puhlf\"urst and S.~Stieberger,
  ``Differential Equations, Associators, and Recurrences for Amplitudes,''
Nucl.\ Phys.\ B {\bf 902}, 186 (2016).
[arXiv:1507.01582 [hep-th]].
}

\lref\CachazoXEA{
  F.~Cachazo, S.~He and E.~Y.~Yuan,
  ``Scattering Equations and Matrices: From Einstein To Yang-Mills, DBI and NLSM,''
JHEP {\bf 1507}, 149 (2015).
[arXiv:1412.3479 [hep-th]].
}

\lref\BoelsJUA{
  R.~H.~Boels,
  ``On the field theory expansion of superstring five point amplitudes,''
Nucl.\ Phys.\ B {\bf 876}, 215 (2013).
[arXiv:1304.7918 [hep-th]].
}

\lref\BernUE{
  Z.~Bern, J.~J.~M.~Carrasco and H.~Johansson,
  ``Perturbative Quantum Gravity as a Double Copy of Gauge Theory,''
Phys.\ Rev.\ Lett.\  {\bf 105}, 061602 (2010).
[arXiv:1004.0476 [hep-th]].
}

\lref\BornINF{
  M.~Born and L.~Infeld,
  ``Foundations of the new field theory,''
Proc.\ Roy.\ Soc.\ Lond.\ A {\bf 144}, 425 (1934).
}

\lref\KampfVHA{
L.~Susskind and G.~Frye,
  ``Algebraic aspects of pionic duality diagrams,''
Phys.\ Rev.\ D {\bf 1}, 1682 (1970). \semi
H.~Osborn,
  ``Implications of adler zeros for multipion processes,''
Lett.\ Nuovo Cim.\  {\bf 2S1}, 717 (1969), [Lett.\ Nuovo Cim.\  {\bf 2}, 717 (1969)].
\semi
J.~R.~Ellis and B.~Renner,
  ``On the relationship between chiral and dual models,''
Nucl.\ Phys.\ B {\bf 21}, 205 (1970). \semi
  K.~Kampf, J.~Novotny and J.~Trnka,
  ``Tree-level Amplitudes in the Nonlinear Sigma Model,''
JHEP {\bf 1305}, 032 (2013).
[arXiv:1304.3048 [hep-th]].
}

\lref\ChangZZA{
J.~A.~Cronin,
  ``Phenomenological model of strong and weak interactions in chiral U(3) x U(3),''
Phys.\ Rev.\  {\bf 161}, 1483 (1967).
\semi
S.~Weinberg,
  ``Dynamical approach to current algebra,''
Phys.\ Rev.\ Lett.\  {\bf 18}, 188 (1967).
\semi
S.~Weinberg,
  ``Nonlinear realizations of chiral symmetry,''
Phys.\ Rev.\  {\bf 166}, 1568 (1968).
\semi
L.~S.~Brown,
  ``Field Theory Of Chiral Symmetry,''
Phys.\ Rev.\  {\bf 163}, 1802 (1967).
\semi
  P.~Chang and F.~Gursey,
  ``Unified Formulation of Effective Nonlinear Pion-Nucleon Lagrangians,''
Phys.\ Rev.\  {\bf 164}, 1752 (1967).
}

\lref\BroedelVLA{
B.~Enriquez, 
``Analogues elliptiques des nombres multizetas,''
 to appear in: Bull.\ Soc.\ Math.\ France.
 [arXiv:1301.3042 [math.NT]].\semi
  J.~Broedel, C.~R.~Mafra, N.~Matthes and O.~Schlotterer,
  ``Elliptic multiple zeta values and one-loop superstring amplitudes,''
JHEP {\bf 1507}, 112 (2015).
[arXiv:1412.5535 [hep-th]]. \semi
 J.~Broedel, N.~Matthes and O.~Schlotterer,
  ``Relations between elliptic multiple zeta values and a special derivation algebra,''
J.\ Phys.\ A {\bf 49}, no. 15, 155203 (2016).
[arXiv:1507.02254 [hep-th]].
}

\lref\DHokerLBK{
E.~D'Hoker, M.B.~Green and P.~Vanhove,
  ``On the modular structure of the genus-one Type II superstring low energy expansion,''
JHEP {\bf 1508}, 041 (2015).
[arXiv:1502.06698 [hep-th]]\semi
E.~D'Hoker, M.~B.~Green and P.~Vanhove,
  ``Proof of a modular relation between 1-, 2- and 3-loop Feynman diagrams on a torus,''
[arXiv:1509.00363 [hep-th]]. \semi
  A.~Basu,
  ``Poisson equation for the Mercedes diagram in string theory at genus one,''
Class.\ Quant.\ Grav.\  {\bf 33}, no. 5, 055005 (2016).
[arXiv:1511.07455 [hep-th]]. \semi
E.~D'Hoker, M.~B.~Green, O.~Gurdogan and P.~Vanhove,
  ``Modular Graph Functions,''
[arXiv:1512.06779 [hep-th]]. \semi
E.~D'Hoker and M.~B.~Green,
  ``Identities between Modular Graph Forms,''
[arXiv:1603.00839 [hep-th]]. \semi
A.~Basu,
  ``Poisson equation for the three loop ladder diagram in string theory at genus one,''
[arXiv:1606.02203 [hep-th]]\semi
A.~Basu,
  ``Proving relations between modular graph functions,''
[arXiv:1606.07084 [hep-th]]\semi
A.~Basu,
  ``Simplifying the one loop five graviton amplitude in type IIB string theory,''
[arXiv:1608.02056 [hep-th]]. \semi
  E.~D'Hoker and J.~Kaidi,
  ``Hierarchy of Modular Graph Identities,''
[arXiv:1608.04393 [hep-th]].
}

\lref\BoelsBV{
R.~Boels, K.~J.~Larsen, N.~A.~Obers and M.~Vonk,
  ``MHV, CSW and BCFW: Field theory structures in string theory amplitudes,''
JHEP {\bf 0811}, 015 (2008).
[arXiv:0808.2598 [hep-th]].$\!$;
  C.~Cheung, D.~O'Connell and B.~Wecht,
  ``BCFW Recursion Relations and String Theory,''
JHEP {\bf 1009}, 052 (2010).
[arXiv:1002.4674 [hep-th]]. \semi
  R.~H.~Boels, D.~Marmiroli and N.~A.~Obers,
  ``On-shell Recursion in String Theory,''
JHEP {\bf 1010}, 034 (2010).
[arXiv:1002.5029 [hep-th]].
}

\lref\BrittoFQ{
  R.~Britto, F.~Cachazo, B.~Feng and E.~Witten,
  ``Direct proof of tree-level recursion relation in Yang-Mills theory,''
Phys.\ Rev.\ Lett.\  {\bf 94}, 181602 (2005).
[hep-th/0501052].
}

\lref\BIref{
  R.~R.~Metsaev, M.~Rakhmanov and A.~A.~Tseytlin,
  ``The {Born-Infeld} Action as the Effective Action in the Open Superstring Theory,''
Phys.\ Lett.\ B {\bf 193}, 207 (1987).
}
\lref\cresson{
J. Cresson,
``Calcul Moulien'', [math/0509548].
}

\listtoc
\writetoc
\filbreak

\newsec Introduction

It is well known that string theory reduces to supersymmetric field theories
involving non-abelian gauge bosons and gravitons when the size of the strings
approaches zero. Hence, one might obtain
a glimpse into
the inner workings of the full string theory
by studying the corrections that are induced by strings of finite size,
set by the length scale $\sqrt{\ap}$.  One approach to study such $\ap$-corrections
to field theory is through the calculation of string scattering amplitudes, see e.g. \refs{\GreenSW,
\grosswitten}. Within this framework, higher-derivative corrections
are encoded in the $\ap$-expansion of
certain integrals defined on the Riemann surface that encodes
the string interactions.

In this work, we will mostly study tree-level scattering of open strings,
where the Riemann surface has the topology of a disk.
As will be reviewed in section~\sectwo, the $\ap$-corrections to super-Yang--Mills (SYM) field theory
arise from iterated integrals over the disk boundary. These integrals 
can be characterized by two words $P$ and $Q$ formed from the $n$ external legs which 
refer to the integration domain $P=(p_1,p_2, \ldots,p_n)$ and integrand
$Q=(q_1,q_2, \ldots,q_n)$ in
\eqn\ZintdefIntro{
Z(P|q_1,q_2,\ldots,q_n) \equiv \ap^{n-3}\!\!\!\!
\int \limits_{D(P)} {\dz_1 \ \dz_2 \ \cdots  \ \dz_n \over {\rm vol}(SL(2,\Bbb R))}
 { \prod_{i<j}^n |z_{ij}|^{\ap s_{ij}}  \over z_{q_1 q_2} z_{q_2 q_3} \ldots z_{q_{n-1} q_n} z_{q_n q_1}}\,.
}
This paper concerns the calculation of the $\ap$-expansion of these disk integrals
in a {\it recursive\/} manner for any given domain $P$ and integrand $Q$. This
technical accomplishment is accompanied by conceptual advances concerning the
interpretation of disk integrals \ZintdefIntro\ in the light of double-copy
structures among field and string theories.

As the technical novelty of this paper, we set up a Berends--Giele (BG) recursion
\BerendsME\ that allows to compute the $\ap$-expansion of the integrals $Z(P|Q)$
and generalizes a recent BG recursion \FTlimit\ for their field-theory limit to
{\it all orders\/} of $\ap$. As a result of this setup, once a {\it finite} number
of terms in the BG recursion at the $w^{\rm th}$ order in $\ap$ is known, the
expansion of disk integrals at {\it any} multiplicity is obtained up to the same
order $\ap^w$. The recursion is driven by simple deconcatenation operations acting
on the words $P$ and $Q$, which are trivially automated on a computer. The
resulting ease to probe $\ap$-corrections at large multiplicities is
unprecedented in modern all-multiplicity approaches \refs{\Polylogs, \BroedelAZA}
to the $\ap$-expansion of disk integrals.

The conceptual novelty of this article is related to the interpretation of string disk integrals \ZintdefIntro\
as {\it tree-level amplitudes} in an effective\foot{The word ``effective'' deserves particular
emphasis since the high-energy properties of $Z$-theory (and its
quantum corrections)
are left for future investigations.} theory of bi-colored scalar fields $\Phi$ dubbed as $Z$-theory
\abZtheory. These scalars will be seen to satisfy an equation of motion of schematic structure,
\eqn\appetizer{
\Box \Phi = \Phi^2 + \ap^2 \zeta_2 (\p^2 \Phi^3 +\Phi^4)
+ \ap^3 \zeta_3 (\p^4 \Phi^3 +\p^2 \Phi^4 + \Phi^5) + {\cal O}(\ap^4)\,.
}
The above equation of motion is at the heart of the recursive method
proposed in this paper; solving it using a perturbiner \Selivanov\ expansion in terms of
recursively defined
coefficients $\phi_{A|B}$ is equivalent to a Berends--Giele recursion\foot{For
a recent derivation of Berends--Giele recursions for tree amplitudes
from a perturbiner solution of the field-theory equations of
motion, see \refs{\Gauge,\FTlimit}.
An older account can be found in \refs{\Selivanov,\Bardeen}.}
that computes the $\ap$-expansion of the disk
integrals \ZintdefIntro\ as if they were
tree amplitudes of an effective field theory,
\eqn\ZfromPhi{
Z(A,n|B,n) = s_A \phi_{A|B}\,.
}
Therefore this paper gives a precise meaning to the perspective on disk integrals as $Z$-theory 
amplitudes \abZtheory\ by pinpointing its underlying equation of motion.
After this fundamental conceptual shift to extract the $\ap$-expansion of
disk integrals from the equation of motion of $\Phi$, its form to all orders in $\ap$ is proposed to be
\eqnn\fullBGintro
$$\displaylines{
\frac{1}{2} \Box \Phi =
\sum_{p=2}^{\infty}(-\ap)^{p-2}  \int^\eom \prod_{i<j}^{p} |z_{ij}|^{\ap
\partial_{ij}}  \hfil\fullBGintro\hfilneg\cr
\times\Bigl(\,
\sum_{l=1}^{p-1}
\frac{ [\Phi_{12\ldots l} , \Phi_{p,p-1\ldots l+1}] }{(z_{12}z_{23}\ldots z_{l-1,l})
 (z_{p,p-1} z_{p-1,p-2} \ldots z_{l+2,l+1})}
+ \perm(2,3,\ldots,p{-}1)\, \Bigr)\,.
}$$
The detailed description of the
above result will be explained in section~\regpoly, but here we note its
remarkable structural similarity with a certain representation of the
superstring disk amplitude for massless external states \nptTreeI. The
$(n{-}2)!$-term representation which led to the all-order proposal \fullBGintro\
has played a fundamental role in the all-multiplicity derivation of
local tree-level numerators \refs{\PSBCJ,\FTlimit} which obey the
duality between color and kinematics~\BCJ.


\subsec $Z$-theory and double copies

The relevance of the disk integrals \ZintdefIntro\  
is much broader than what the higher-derivative completion of field theory might lead one to
suspect. They have triggered deep insights
into the anatomy of numerous field theories through the fact that closed-string
tree-level integrals (encoding $\ap$-corrections to supergravity theories) boil
down to squares of disk integrals through the KLT relations \KLT.  In a
field-theory context, this double-copy connection between open and closed strings 
became a crucial hint in understanding quantum-gravity interactions as a square of 
suitably-arranged gauge-theory building blocks \refs{\BCJ, \BernUE}. 

Double-copy structures have recently been identified in the tree-level
amplitudes of additional field theories \CachazoXEA. For instance, classical
Born--Infeld theory \BornINF\ which governs the low-energy effective action of
open superstrings \BIref\ turned out to be a double copy of gauge theories and an
effective theory of pions known as the non-linear sigma model (NLSM) \ChangZZA, see
\KampfVHA\ for its tree-level amplitudes. As a string-theory incarnation of the
Born--Infeld double copy, tree-level amplitudes of the NLSM have been identified as
the low-energy limit of the disk integrals in the scattering of {\it abelian} gauge
bosons \abZtheory. This unexpected emergence of pion amplitudes exemplifies that 
disk integrals also capture the interactions of particles that cannot be found in the naive 
string spectrum\foot{See \GreenGA\ for a string-theory realization of the NLSM through 
toroidal compactifications in presence of worldsheet boundary condensates.}.

Moreover, the entire tree-level S-matrix of massless open-superstring states can be
presented as a double copy of SYM with $\ap$-dependent disk integrals \Polylogs. 
Their $Z$-theory interpretation in \abZtheory\ was driven by the quest to identify the
second double-copy ingredient of the open superstring besides SYM. In view of the 
biadjoint-scalar and NLSM interactions in the low-energy
limit of $Z$-theory, its full-fledged $\ap$-dependence 
describes effective higher-derivative deformations of these two scalar
field theories \abZtheory. As a double-copy component to complete SYM
to the massless open-superstring S-matrix, the collection of effective
interactions encompassed by $Z$-theory deserve further investigations.

In this work, we identify the equation of motion \fullBGintro\ of the full non-abelian $Z$-theory, where the
integration domain of the underlying disk integrals endows the putative scalars
$\Phi$ with a second color degree of freedom. By the results of \Polylogs, disk
integrals in their interpretation as $Z$-theory amplitudes obey the duality between
color and kinematics due to Bern, Carrasco and Johansson (BCJ) \BCJ\ in one of
their color orderings. Hence, the effective theories gathered in $Z$-theory are
of particular interest to advance our understanding of the BCJ duality.
The abelian limit of $Z$-theory arises from disk
integrals without any notion of color ordering in the integration domain and has
been studied in \abZtheory\ as a factory for BCJ-satisfying $\ap$-corrections to the NLSM.
The present article extends this endeavor such as to efficiently compute
the doubly-partial amplitudes of effective bi-colored theories with BCJ duality
in one of the gauge groups and explicitly known field equations \fullBGintro.

\subsec Outline

This paper is organized as follows: Following a review of disk integrals and the
Berends--Giele description of their field-theory limit in section \sectwo, the
Berends--Giele recursion for their $\ap$-corrections and the resulting field
equations of non-abelian $Z$-theory are presented in section \secthree. The
mathematical tools to control the equations of motion to all orders in the fields
and derivatives by means of suitably regularized polylogarithms are elaborated in
section \regpoly. In section \secfive, the Berends--Giele recursion is extended to
closed-string integrals over surfaces with the topology of a sphere before we conclude
in section~\secsix. Numerous appendices and ancillary files complement the
discussions in the main text.

The BG recursion that generates all terms up to the $\ap^{7}$-order in the
$\ap$-expansion of disk integrals at arbitrary multiplicity as well as
the auxiliary computer programs used in their derivations can be
downloaded from \BGapwww.

\newsec Review and preliminaries
\par\seclab\sectwo

\noindent
In this section, we review the definitions and symmetries of the disk integrals
under investigations as well as their appearances in tree amplitudes of massless
open-string states. We also review the recent Berends--Giele approach to their
field-theory limit in order to set the stage for the generalization to
$\alpha'$-corrections.

\subsec String disk integrals

We define a cyclic chain $C(Q)$ of worldsheet propagators $z_{ij}^{-1}$ with $z_{ij} \equiv z_i - z_j$ on
words $Q\equiv q_1q_2 \ldots q_n$ of length $n$ as
\eqn\Zposdef{
C(Q) \equiv {1\over z_{q_1q_2}z_{q_2q_3} \cdots z_{q_n q_1}}\, \ .
}
Then, the iterated disk integrals on the real line
that appear in the computation of open-superstring tree-level amplitudes are completely
specified by two words $P$ and $Q$,
\eqn\Zintdef{
Z(P|Q) \equiv 
\ap^{n-3}
\int \limits_{D(P)}\! {\dz_1\dz_2\cdots \dz_n \over {\rm vol}(SL(2,\Bbb R))}
\prod_{i<j}^n |z_{ij}|^{\ap s_{ij}} C(Q) \,,
}
where $P\equiv p_1 p_2 \ldots p_n$ encodes the domain of the iterated integrals,
\eqn\domain{
D(P) \equiv \{ (z_1,z_2,\ldots,z_n) \in \Bbb R^n, \ -\infty < z_{p_1} < z_{p_2} < \ldots < z_{p_n} < \infty \} \ .
}
Mandelstam variables $s_{ij\ldots p}$ involving legs $i,j,\ldots,p$ are defined via region momenta $k_{ij\ldots p}$,
\eqn\mandmom{
k_{ij\ldots p} \equiv k_i + k_j +\ldots +k_p \ , \ \ \ \ \ \ s_{ij\ldots p}  \equiv {1\over 2} \, k_{ij\ldots p} ^2 \ ,
}
and the more standard open-string conventions for the normalization of $\alpha'$ (which would cause proliferation 
of factors of two) can be recovered by
globally setting $\alpha' \rightarrow 2\alpha'$ everywhere in this work.
In the sequel, we refer to the word $P$ as the {\it integration region} or {\it domain} and 
to $Q$ as the {\it integrand} of \Zintdef, where $P$ is understood to be a permutation of~$Q$. 
The inverse volume $ {\rm vol}(SL(2,\Bbb R))$ of the conformal Killing group of 
the disk instructs to mod out by the redundancy of M\"obius transformation
$z \rightarrow {az+b\over cz+d}$ (with $ad-bc=1$). This amounts to
fixing three positions such as $(z_{1},z_{{n-1}},z_{n})=(0,1,\infty)$ and to inserting a
compensating Jacobian:
\eqn\volsl{
\int \limits_{D(12\ldots n)}\! {\dz_1\dz_2\cdots \dz_n \over {\rm vol}(SL(2,\Bbb R))}
= z_{1,n-1} z_{1,n} z_{n-1,n}
\! \! \! \! \! \! \! \! \! \!\! \! \! \! \! \!    \int\limits_{z_1\leq z_{2} \leq z_{3} \leq \ldots \leq z_{n-2} \leq z_{n-1}} 
\! \! \! \! \! \! \! \! \! \! \! \! \! \! \! \! 
 \dz_{2}\, \dz_{3}\, \ldots \, \dz_{n-2}\,.
}
Given that the words $P$ and $Q$ in the disk integrals \Zintdef\ 
encode the integration region $D(P)$ in \domain\ and the integrand $C(Q)$ in \Zposdef,
respectively, there
is in general no relation between $Z(P|Q)$ and $Z(Q|P)$. This can already be seen
from the different symmetries w.r.t.\ variable $P$ at fixed $Q$ on the one hand and 
variable $Q$ at fixed $P$ on the other hand.

\subsubsec Symmetries of disk integrals in the integrand

The manifest cyclic symmetry and reflection (anti-)symmetry of the integrand $C(Q)$ in \Zposdef\ 
directly propagates to the disk integrals
\eqn\Qcyc{
Z(P|q_2q_3 \ldots q_n q_1) = Z(P|q_1q_2 \ldots q_n)\,,\qquad
Z(P|\tilde Q) = (-1)^\len{Q} Z(P|Q)\,,
}
where $\len{Q}=n$ denotes the length of the word $Q=q_1q_2\ldots q_n$, and the
tilde in $\tilde Q= q_n \ldots q_2 q_1$ is a shorthand for its reversal. Moreover, the disk
integrals satisfy \Polylogs\ the Kleiss--Kuijf relations \KKsym,
\eqn\KKZint{
Z(P|A,1,B,n) = (-1)^\len{A} Z(P|1,\tilde A\shuffle B,n)\,,
}
or equivalently \refs{\Ree,\BGschocker}, the vanishing of pure shuffles in $n{-}1$ legs,
\eqn\noshuff{
Z(P|A \shuffle B,n) = 0 \ \ \ \forall \ A,B \neq \emptyset \ .
}
The shuffle operation in \KKZint\ and \noshuff\ is defined recursively via \reutenauer
\eqn\Shrecurs{
\emptyset\shuffle A = A\shuffle\emptyset = A,\qquad
A\shuffle B \equiv a_1(a_2 \ldots a_{|A|} \shuffle B) + b_1(b_2 \ldots b_{|B|}
\shuffle A)\,,
}
and it acts linearly on the parental objects, e.g.\ $Z(123|1(2\shuffle3))=Z(123|123) + Z(123|132)$.  
Finally, integration by parts yields the same BCJ relations among permutations of $Z(P|Q)$ in $Q$ as
known from \BCJ\ for color-stripped SYM tree amplitudes \Polylogs
\eqn\QBCJ{
0 = \sum_{j=2}^{n-1}  k_{q_1} \cdot k_{q_2 q_3\ldots q_j}
Z(P| q_2q_3\ldots q_{j}q_1 q_{j+1} \ldots q_n) \ .
}
Note that neither \KKZint\ nor \QBCJ\ depends on the domain $P$, and they allow to expand any $Z(P|Q)$
in an $(n{-}3)!$-element basis $\{Z(P|Q_i) , \ i=1,2,\ldots,(n{-}3)!\}$ at fixed $P$ \BCJ. The symmetries \Qcyc, \KKZint\ 
and \QBCJ\ known from SYM interactions crucially support the interpretation of $Z(P|Q)$ as 
doubly partial amplitudes \abZtheory.

\subsubsec Symmetries of disk integrals in the domain

As a consequence of the form of the integration region $D(P)$ in \domain, disk integrals obey a cyclicity and 
parity property in the domain $P=p_1p_2 \ldots p_n$,
\eqn\symtrace{
Z(p_2p_3 \ldots p_n p_1|Q) = Z(p_1p_2 \ldots p_n|Q)\,,\qquad
Z(\tilde P|Q) = (-1)^\len{P} Z(P|Q)\,,
}
which tie in with the simplest symmetries \Qcyc\ of the integrand $Q$. However, the
Kleiss--Kuijf symmetry \KKZint\ and BCJ relations \QBCJ\ of the integrand do not hold
for the integration domain $P$ in presence of $\alpha'$-corrections. This can be seen from
the real and imaginary part of the monodromy relations
\refs{\BjerrumBohrRD, \StiebergerHQ} (see \TourkineBAK\ for a recent generalization
to loop level)
\eqn\PBCJ{
0 = \sum_{j=2}^{n-1} \exp\! \big[i\pi \alpha' k_{p_1} \cdot k_{p_2 p_3\ldots p_j}
 \big] Z(p_2p_3\ldots p_{j}p_1 p_{j+1} \ldots p_n | Q) \ .
}
Nevertheless, \PBCJ\ is sufficient to expand any $Z(P|Q)$ in an $(n{-}3)!$-element 
basis $\{Z(P_i|Q) , \ i=1,2,\ldots,(n{-}3)!\}$ at fixed $Q$ \refs{\BjerrumBohrRD, \StiebergerHQ}.

\subsec Open superstring disk amplitudes

The $n$-point tree-level amplitude $A^{\rm open}$ of the open superstring takes a
particularly simple form once the contributing disk integrals are cast into an
$(n{-}3)!$ basis via partial fraction \noshuff\ and integration by parts
\QBCJ\ \refs{\nptTreeI, \nptTreeII}:
\eqn\npttree{
A^{\rm open}(1,P,n{-}1,n) = \sum_{Q \in S_{n-3}} F_P{}^{Q} A^{\rm SYM}(1,Q,n{-}1,n)
}
While all the polarization dependence on the right hand side has been expressed
through the BCJ basis \BCJ\ of SYM trees $A^{\rm SYM}$, the entire reference to
$\alpha'$ stems from the integrals
\eqnn\defF
$$\eqalignno{
F_P{}^{Q} &\equiv (-\ap)^{n-3} \! \! \! \! \! \! \! \! \! \!  \! \! \! \! \! \! \!  \int\limits_{0\leq z_{p_2} \leq z_{p_3} \leq \ldots \leq z_{p_{n-2}} \leq 1}   \! \! \! \! \! \! \! \! \! \! \! \! \! \! \! \! \!  \dd z_2 \, \dd z_3\, \ldots \, \dd z_{n-2} \prod_{i<j}^{n-1} |z_{ij}|^{\alpha' s_{ij}} \, { s_{1 q_2} \over z_{1 q_2}}
\left(  { s_{1 q_3} \over z_{1 q_3}} + { s_{q_2 q_3} \over z_{q_2 q_3}}
\right) &\defF\cr 
& \ \ \ \ \ \times \left(  { s_{1 q_4} \over z_{1 q_4}} + { s_{q_2 q_4} \over z_{q_2 q_4}} + { s_{q_3 q_4} \over z_{q_3 q_4}}
\right) \ldots \left(  { s_{1 q_{n-2}} \over z_{1 q_{n-2}}} + { s_{q_2 q_{n-2}} \over z_{q_2 q_{n-2}}} +\ldots + { s_{q_{n-3} q_{n-2}} \over z_{q_{n-3} q_{n-2}}}
\right) \, ,
}$$
where $P=p_2 p_3\ldots p_{n-2}$ and $Q=q_2 q_3\ldots q_{n-2}$ are permutations of $23\ldots n{-}2$. 
The original derivation \refs{\nptTreeI, \nptTreeII} of \npttree\ and \defF\ has been performed
in the manifestly supersymmetric pure spinor formalism \psf, where the SYM amplitudes
$A^{\rm SYM}$ in \npttree\ have been identified from their Berends--Giele representation
in pure spinor superspace \nptMethod. Hence, \npttree\
applies to the entire ten-dimensional gauge multiplet in the external
states\foot{A bosonic-component check of the formula
\npttree\ at multiplicity $n\leq7$ within the RNS formalism has
been performed in \BarreiroDPA.}.

\subsubsec $Z$-theory

\noindent
After undoing the $SL(2,\Bbb R)$-fixing 
in \volsl, the integrals $F_P{}^{Q}$ can be identified as a linear combination of
disk integrals \Zintdef\ \Polylogs,
\eqn\KLTopen{
F_P{}^{Q} = \sum_{R \in S_{n-3}} S[Q|R]_1 Z(P|1,R,n,n{-}1) \ ,
}
where $P,Q$ and $R$ are understood to be permutations of $2,3,\ldots,n{-}2$.
The symmetric $(n{-}3)! \times (n{-}3)!$ matrix $S[Q|R]_1$ 
encodes the field-theory KLT relations \refs{\oldMomKer,\BohrMomKer} (see also
\MomKer\ for the $\ap$-corrections to $S[Q|R]_1$) and admits the following recursive
representation~\abZtheory,
\eqn\mker{
S[A,j|B,j,C]_i =
(k_{iB}\cdot k_{j}) S[A|B,C]_i,
\qquad S[\varnothing| \varnothing]_i \equiv 1\,,
}
in terms of multiparticle momenta \mandmom. Hence, the $n$-point open-superstring
amplitude \npttree\ with any domain $P$ can be obtained from the KLT formula,
\eqn\Zthyorigin{
A^{\rm open}(P) = \sum_{Q,R \in S_{n-3}} Z(P|1,R,n,n{-}1) S[R|Q]_1 A^{\rm SYM}(1,Q,n{-}1,n)
}
upon replacing the right-moving SYM trees via $\tilde A^{\rm SYM}(1,R,n,n{-}1)
\rightarrow Z(P|1,R,n,n{-}1)$ \Polylogs. The KLT form of \Zthyorigin\ reveals the
double-copy structure of the open-superstring tree-level S-matrix which in turn 
motivated the proposal of \abZtheory\ to interpret disk integrals as doubly partial amplitudes. 
The specification of disk integrals by two cycles $P,Q$ identifies the underlying particles to
be bi-colored scalars, and we collectively refer to their effective interactions that give rise
to tree amplitudes $Z(P|Q)$ as $Z$-theory.

Note that disk amplitudes of the bosonic
string are conjectured in \HuangTAG\ to also admit the form \npttree\ or \Zthyorigin, with
$\alpha'$-dependent kinematic factors $A^{\rm SYM}(1,Q,n{-}1,n) \rightarrow 
B(1,Q,n{-}1,n;\alpha')$ that also satisfy the KK- and BCJ relations.

\subsubsec $\alpha'$-expansion of disk amplitudes

\noindent
The $\alpha'$-expansion of disk amplitudes \npttree, i.e. their Taylor expansion in the
dimensionless Mandelstam invariants $\alpha' s_{ij}$, involves multiple zeta values (MZVs),
\eqn\MZVdef{
\zeta_{n_1,n_2,\ldots, n_r} \equiv \sum_{0<k_1<k_2<\ldots <k_r}^{\infty} k_1^{-n_1} k_2^{-n_2} \ldots k_r^{-n_r} 
\ , \ \ \ \ n_r \geq 2 \ .
}
The MZV in \MZVdef\ is said to have depth $r$ and weight $w=n_1+n_2+\ldots+n_r$ (which is
understood to be additive in products of MZVs). While the four-point instance of \defF,
\eqnn\Ffour
$$\displaylines{
F_2{}^2 = \exp \Big( \sum_{n=2}^{\infty} \frac{ \zeta_n}{n} (-\alpha')^n \big[
s_{12}^n +s_{23}^n - (s_{12}+s_{23})^n
\big]
\Big)  \hfil\Ffour\hfilneg \cr
 = 1 - \alpha'^2 \zeta_2 s_{12}s_{23} + \alpha'^3 \zeta_3 s_{12}s_{23}(s_{12}+s_{23}) - \alpha'^4 \zeta_4 s_{12}s_{23} \Big(s_{12}^2 +\frac{s_{12}s_{23}}{4} + s_{23}^2\Big)+ {\cal O}(\alpha'^5)\ ,
}
$$
boils down to a single entry with Riemann zeta values $\zeta_n$ of depth $r=1$ only, disk
integrals at multiplicity $n\geq 5$ generally involve MZVs of higher depth $r\geq 2$,
see \PuhlfuerstGTA\ for a recent closed-form solution at five points.
It has been discussed in the literature of both physics
\refs{\nptTreeII, \StiebergerRR, \SchlottererNY} and mathematics 
\refs{\mathMZV, \BrownmathMZV} that the disk integrals \Zintdef\
at any multiplicity exhibit uniform transcendentality: Their $\alpha'^w$-order is exclusively
accompanied by products of MZVs with total weight $w$. 

The basis of functions $F_P{}^{Q}$
in \KLTopen\ is particularly convenient to directly
determine the $\ap$-expansion of
the open-string amplitudes \npttree\ \BroedelAZA\
and to describe their pattern of MZVs\foot{After pioneering work in \OprisaWU,
the $\ap$-expansion of disk integrals at multiplicity $n\geq 5$ has later been
systematically addressed via all-multiplicity techniques based on polylogarithms
\Polylogs\ and the Drinfeld associator \BroedelAZA\ (see also \DrummondVZ).}
\refs{\SchlottererNY, \DrummondVZ}.
At multiplicities five, six and seven, explicit results for the leading orders
in the $\ap$-expansion of
$F_P{}^{Q}$ are available for download on \WWW.

\subsubsec Basis-expansion of disk integrals

In setting up the Berends--Giele recursion for the fundamental objects $Z(A|B)$ of this work,
it is instrumental to efficiently extract their $\ap$-expansion from the basis functions $F_P{}^Q$.
However, solving the mediating BCJ and monodromy relations can be very cumbersome, and the
explicit basis expansions spelled out in \Polylogs\ only address an $(n{-}2)!$ subset of integrands $B$.
These shortcomings are surpassed by the following formula,
\eqn\ZinttoF{
Z(1,P,n{-}1,n|R) = \sum_{Q\in S_{n-3}} F_P{}^Q m(  1,Q,n{-}1,n | R ) \,,
}
where $m(A|B)$ denote the doubly partial amplitudes of biadjoint $\phi^3$-theory which arise in the 
field-theory limit of disk integrals \DPellis
\eqn\defdp{
m(A|B) = \lim_{\alpha' \rightarrow 0}  Z(A|B)  \ .
}
Note the striking
resemblance of the formulas \ZinttoF\ and \npttree, which
further point out the similar roles played by the amplitudes
$A^{\rm open}(P)$ and $Z(P|Q)$ of string and $Z$-theory.


\subsec Berends--Giele recursion for the field-theory limit

The task we want to accomplish in this paper concerns the computation of the
$\ap$-expansion of the disk integrals \Zintdef\ in a recursive and efficient
manner. In the field-theory limit $\ap\to0$, all-multiplicity techniques have been
developed in \nptTreeII, and a relation to the inverse KLT matrix \mker\ has been
found in \Polylogs. The equivalent description of the $\ap\to0$ limit in terms of
doubly partial amplitudes \defdp\ \DPellis\ has inspired a recent Berends--Giele 
description \FTlimit\ via bi-adjoint scalars $\Phi^{(0)} \equiv \Phi^{(0)}_{a|b}
t^a \otimes \tilde t^b$. The latter take values in the tensor product of two gauge groups
with generators $t^a$ and $\tilde t^b$ as well as structure constants $f^{acd}$ and 
$\tilde f^{bgh}$, respectively.

The superscript of the biadjoint scalar $\Phi^{(0)}$ indicates that this is the $\ap\to0$ limit of 
the $Z$-theory particles $\Phi$ whose interactions give rise to the disk integrals $Z(P|Q)$
as their doubly partial amplitudes. The non-linear field equations in the low-energy limit
\eqn\nonabtwo{
\Box \Phi^{(0)}_{a|b} = f_{acd} \tilde f_{bgh} \Phi^{(0)}_{c|g} \Phi^{(0)}_{d|h}
}
with d'Alembertian $\Box \equiv \partial^2$ will later be completed such as to
incorporate the $\ap$-corrections in $Z(P|Q)$. One can solve \nonabtwo\ through a
perturbiner \Selivanov\ expansion\foot{See \refs{\moreSelivanov, \Selivanov} for perturbiner
solutions to self-dual sectors of four-dimensional gauge and gravity theories (see
also \Bardeen) and \Gauge\ for perturbiners in ten-dimensional SYM.} \FTlimit,
\eqnn\perturbiner
$$\eqalignno{
\Phi^{(0)} & = \sum_{a_1,b_1} \phi^{(0)}_{a_1|b_1} \,e^{k_{a_1}x} \, t^{a_1} \otimes \tilde t^{b_1}
+\!\!\!\!\! \sum_{a_1,a_2,b_1,b_2}
\phi^{(0)}_{a_1 a_2|b_1 b_2}   \, e^{k_{a_1a_2}x} \, t^{a_1} t^{a_2} \otimes \tilde t^{b_1}\tilde t^{b_2} \cr
& \ \ \ \ +\!\!\!\!\!\! \sum_{a_1,a_2,a_3,b_1,b_2,b_3}\!\!\!
\phi^{(0)}_{a_1 a_2 a_3|b_1 b_2 b_3} \, e^{k_{a_1a_2a_3}x} \, t^{a_1} t^{a_2} t^{a_3} \otimes \tilde t^{b_1}\tilde t^{b_2} t^{b_3} +\cdots \cr
&= \sum_{A,B} \phi^{(0)}_{A|B} \, e^{k_A x}\, t^{A}  \otimes\tilde t^B \ ,&\perturbiner
}$$
which resums tree-level subdiagrams and is compactly written as a sum over all
words $A,B$ with length $|A|, |B|\geq 1$ in
the last line. We are using the collective notation
\eqn\coll{
t^{A} \equiv t^{a_1}t^{a_2} \ldots t^{a_{\len{A}}}\,,
\qquad \tilde t^B \equiv t^{b_1}t^{b_2} \ldots t^{b_{\len{B}}}
}
for products of Lie-algebra generators associated with multiparticle label $A=a_1
a_2 \ldots a_\len{A}$ and $B=b_1 b_2 \ldots b_\len{B}$. The coefficients in \perturbiner\ are 
recursively determined by the non-linear
field equations \nonabtwo\ \FTlimit,
\eqn\FTphi{
 s_A \phi^{(0)}_{A|B} =  \sum_{A_1A_2=A\atop B_1B_2=B}
\bigl(\phi^{(0)}_{A_1|B_1} \phi^{(0)}_{A_2|B_2} - \phi^{(0)}_{A_1|B_2}
\phi^{(0)}_{A_2|B_1}\bigr)\,,
}
and referred to as Berends--Giele double
currents $\phi^{(0)}_{A|B}$. The notation
$\sum_{A_1A_2=A}$ and $\sum_{B_1B_2=B}$ instructs to sum over
deconcatenations $A = a_1a_2\ldots a_{|A|}$ into 
non-empty words $A_1 = a_1a_2\ldots a_j$ and $A_2=a_{j+1}\ldots a_{|A|}$
with $j=1,2,\ldots,|A|{-}1$ and to independently deconcatenate $B$ in the same manner. 
The initial conditions for the recursion in \FTphi,
\eqn\nonabfive{
\phi^{(0)}_{i|j} = \delta_{i,j}\, ,
}
guarantee that $\phi^{(0)}_{A|B}$ vanishes unless $A$ is a permutation
of $B$ and yield expressions such as
\eqn\exBGBG{
\phi^{(0)}_{12|12} = - \phi^{(0)}_{12|21} = {1\over s_{12}}\,,\quad
\phi^{(0)}_{123|123} = {1\over s_{12} s_{123}}+{1\over s_{23} s_{123}}\,, \quad
\phi^{(0)}_{123|312} = -{1\over s_{12} s_{123}}
}
at the two- and three-particle level. 

As shown in \FTlimit, the field-theory limits of the disk integrals \Zintdef\
and thereby the doubly partial amplitudes \defdp\ are given by the Berends--Giele double currents
$\phi^{(0)}_{A|B}$,
\eqn\nonabsix{
m(A,n|B,n)= s_A \phi^{(0)}_{A|B}\,.
}
Given the cyclic symmetry \Qcyc\ of $Z(P|Q)$ in the word $Q$, one can always choose the
last letter of the integrand $Q\equiv(B,n)$ to coincide with the last letter of the
integration region $P\equiv(A,n)$ as has been done in \nonabsix. The recursive
definition of $\phi^{(0)}_{A|B}$ in \FTphi\ gives rise to an efficient algorithm to
obtain the field-theory limit of disk integrals $Z(A,n|B,n)$ directly from the two words $A$,
$B$ encoding the integrand and integration domain, respectively.

Furthermore, the BG double currents allow
the inverse of the KLT matrix \mker\ to be obtained without any
matrix algebra \FTlimit,
\eqn\phimker{
 S^{-1}[P|Q]_1 = \phi^{(0)}_{1P|1Q}\,.
}

\subsubsec Example application of the Berends--Giele recursion

The computation of the field-theory limit of the five-point disk integral
\eqn\FTexamp{
m(13524|32451) = \lim_{\ap\to0}\ap^2 \! \! \!
\int \limits_{D(13524)}\! \! {\dz_1\dz_2\cdots \dz_5 \over {\rm vol}(SL(2,\Bbb R))}
\prod_{i<j}^5 |z_{ij}|^{\ap s_{ij}} {1\over z_{32}z_{24}z_{45}z_{51}z_{13}} \,
}
using the Berends--Giele formula \nonabsix\ proceeds as
follows. First, one exploits the cyclic symmetry of the integrand to rotate
its labels until the last leg matches
the last label of the integration region. After applying \nonabsix\ one
obtains,
\eqn\firstex{
m(13524|32451) = m(13524|51324) = s_{1352}\phi^{(0)}_{1352|5132} =
\phi^{(0)}_{135|513} \phi^{(0)}_{2|2}\,.
}
Terms such as $\phi^{(0)}_{1|5} \phi^{(0)}_{352|132}$ following from the
deconcatenation \FTphi\
have been dropped from the last equality because the condition \nonabfive\
implies that $\phi^{(0)}_{1|5}=0$. In addition,
the overall factor $s_{1352}$ from \nonabsix\ cancels the
propagator $1/s_{1352}$ in the current $\phi^{(0)}_{1352|5132}$.
Recursing the above steps until no factor
of $\phi^{(0)}_{A|B}$ remains yields,
\eqn\secex{
m(13524|32451) =  \phi^{(0)}_{135|513}
= {1\over s_{135}}( - \phi^{(0)}_{13|13} \phi^{(0)}_{5|5}) = -{1\over s_{135}}
\phi^{(0)}_{13|13} = - {1\over s_{135}s_{13}}\,,
}
in agreement with the expression for the doubly partial amplitude $m(13524|32451)$
that follows from the methods of \DPellis. In the next section this method will
be extended to compute the $\ap$-corrections of string disk integrals.

\newsec Berends--Giele recursion for disk integrals
\par\seclab\secthree

\noindent In this section, we develop a Berends--Giele recursion\foot{For a review
of the Berends--Giele recursion for gluon amplitudes \BerendsME\ which is adapted
to the current discussion, see section~2 of \FTlimit.} for the full-fledged disk
integrals $Z(P|Q)$ defined in \Zintdef. The idea is to construct $\ap$-dependent
Berends--Giele double currents $\phi_{A|B}$ such that the integrals $Z(P|Q)$
including $\ap$-corrections are obtained in the same manner as their field-theory
limit in \nonabsix,
\eqn\ZapBG{
Z(A,n|B,n) = s_A \phi_{A|B}\,.
}
And similarly, the $\ap$-corrected BG double currents 
$\phi_{A|B}$ in \ZapBG\ will be given by the coefficients 
of a perturbiner expansion analogous to \perturbiner,
\eqn\perturbinerA{
\Phi = \sum_{A,B}\phi_{A|B} e^{k_A\cdot x}\; t^A\otimes \tilde t^B\,,
}
that solves non-linear equations of motions which can be viewed as an augmentation
of \nonabtwo\ by $\ap$-corrections. The field equation obeyed by the perturbiner
\perturbinerA\ will be interpreted as the equation of motion of $Z$-theory, the
collection of effective theories involving bi-colored scalars encoding all the
$\ap$-corrections relevant to the open superstring \abZtheory. In addition, the BG
double currents above are subject to the initial and vanishing condition
\eqn\deltaphiAB{
\phi_{i|j} = \delta_{i,j} \ , \ \ \ \ \ \ 
\phi_{A|B}=0\,,
\quad\hbox{unless $A$ is a permutation of $B$.}
}
Given their role in equation \ZapBG, the words $A$ and $B$
on the BG double current $\phi_{A|B}$ will be referred to
as the {\it integration domain} $A$ and the {\it integrand} $B$,
respectively.

\subsec Symmetries of the full Berends--Giele double currents

In the representation \ZapBG\ of the disk integrals, their parity symmetries \KKZint\ and \symtrace\
can be manifested if the double currents $\phi_{A|B}$ satisfy
\eqn\BGsymmetriesA{
\phi_{A|B} = (-1)^{|A|-1} \phi_{\tilde A|B} = (-1)^{|B|-1} \phi_{A| \tilde B}\,,
}
upon reversal of either the integration domain $A$ or the integrand $B$.
Similarly, the Kleiss--Kuijf relations \noshuff\ of the disk integrals
follow from the shuffle symmetry\foot{In the
mathematics literature, objects $T_B$ satisfying the symmetry $T_{P\shuffle Q}=0$ 
for any $P,Q \neq \emptyset$ are known as ``alternal moulds'', see e.g.\ \chapoton.}
of $\phi_{A|B}$ within the integrand $B$,
\eqn\BGsymmetriesB{
\phi_{A|P\shuffle Q} = 0  \ \ \ \forall \ P,Q \neq \emptyset\,.
}
Note that $\phi_{A|B}$ does {\it not} exhibit shuffle symmetries in the integration
domain $A$: The $\alpha'$-correction in the monodromy relations
\refs{\BjerrumBohrRD, \StiebergerHQ}, more specifically in the real part of \PBCJ,
yields non-zero expressions\foot{Since the monodromy relations only differ from the
KK relations by rational multiples of $ \pi^{2n}$ or $(\zeta_2)^n$, the sub-sector
of $Z(A,n|B,n)$ without any factors of $\zeta_2$ still satisfies shuffle
symmetries, e.g.\ $\phi_{P\shuffle Q | B} \, \big|_{\zeta_{2n+1}} = 0$, also see
\StiebergerHBA\ for analogous statements for the heterotic string and
section~\secfive\ for implications for a Berends--Giele approach to closed-string
integrals.} ${\cal O}((\alpha' \pi)^2)$ for $\phi_{P\shuffle Q|B}$. As a
consequence, the perturbiner \perturbinerA\ is Lie-algebra valued w.r.t.\ the
$\tilde t^b$ generators \Ree\ but not w.r.t.\ the $t^a$ generators. That is why the
$Z$-theory scalar $\Phi$ is referred to as {\it bi-colored} rather than {\it
biadjoint}.

The symmetries \BGsymmetriesA\ and \BGsymmetriesB\ will play a fundamental role in
the construction of ansaetze for the $\ap$-corrections of the Berends--Giele double
currents, see appendices \appshuff\ and \appANS\ for further details.

\subsec The $\ap^2$-correction to Berends--Giele currents of disk integrals
\par\subseclab\secZetatwo

Assuming that the $\ap^2$-corrections of the integrals \Zintdef\ can be described
by Berends--Giele double currents as in \ZapBG, dimensional analysis admits two
types of terms at this order. They have the schematic form $k^2\phi^3$ and $\phi^4$
since $\phi$ has dimension of $k^2$, and the $\ap^2$-terms contain a factor of
$k^4$ compared to the leading contribution from $\phi^2$ in \FTphi.
Therefore, an ansatz for
$s_A \phi_{A|B}$ at this order must be based on a linear combination of
\eqn\AnsatzZeta{
\sum_{A_1 A_2 A_3=A\atop B_1 B_2 B_3=B}
 (k_{A_i} \cdot k_{A_j}) \phi_{A_1|B_k} \phi_{A_2|B_l}\phi_{A_3|B_m},\quad
\sum_{A_1 \ldots A_4=A\atop B_1 \ldots B_4=B} 
 \phi_{A_1|B_p} \phi_{A_2|B_q}\phi_{A_3|B_r} \phi_{A_4|B_s}\,,
}
see the explanation below \FTphi\ for the deconcatenations $A=A_1 A_2 A_3$ 
and $A=A_1 \ldots A_4$ into non-empty words. By the initial condition \deltaphiAB,
$\phi_{A|B}$ vanishes unless $A$ is a permutation of $B$, so there is no need to
consider momentum dependence of the form $(k_{A_i} \cdot k_{B_j})$ or $(k_{B_i}
\cdot k_{B_j})$.

The most general linear combination of the terms \AnsatzZeta\ contains $36+24=60$
parameters. Imposing the symmetries \BGsymmetriesA\ and \BGsymmetriesB\ reduces
them to $6+4=10$ parameters, see appendix \appshuff\ for the implementation of
the shuffle symmetry.  Then, matching the outcomes of \ZapBG\ with the known
$\ap^2$-order of various integrals at four and five points fixes six parameters,
leaving a total of four free parameters. The $\ap^2$-order of $(n\geq 6)$-point
integrals does not provide any further input: As we have checked with all the known
$(n\leq 9)$-point data \WWW, they are automatically reproduced for any choice of
the four free parameters. This is where the predictive power of the Berends--Giele
setup kicks in: A finite amount of low-multiplicity data -- the coefficients of
$k^2 \phi^3$- and $\phi^4$-terms \AnsatzZeta\ at the $\ap^2$-order -- determines
the relevant order of disk integrals at {\it any} multiplicity.

\subsubsec Free parameters versus $Z$-theory equation of motion

It is not surprising that the ansatz based on \AnsatzZeta\ is not completely fixed (yet) by matching the
data. The reason for this can be seen from the interpretation of the Berends--Giele
recursion method as the perturbiner solution \perturbiner\ to the $Z$-theory
equation of motion with the schematic form $\Box \Phi = \Phi^2 + {\cal O}(\Phi^3)$.
Self-contractions $(k_{A_i}\cdot k_{A_i})$ signal the appearance of $\Box \Phi
=\Phi^2+\alpha'^2 \zeta_2 \Phi^2 \Box \Phi+\ldots$ on the right hand side, where
$\Box \Phi$ along with $\alpha'^2 \zeta_2 \Phi^2$ can be replaced by the entire
right hand side. The result $\Box \Phi =\Phi^2+\alpha'^2 \zeta_2 \Phi^2
(\Phi^2+\alpha'^2 \zeta_2 \Phi^2 \Box \Phi)+\ldots$ in turn leads to another
appearance of $\Box \Phi$ at higher orders in $\alpha'$ and the fields.  In order
to obstruct an infinite iteration of the field equations, we fix three additional
parameters by demanding absence of $(k_{A_i}\cdot k_{A_i})$ with $i=1,2,3$ and
thereby leave one free.

The last free parameter reflects the freedom to perform field redefinitions. Terms
of the form $\alpha'^2 \zeta_2\Box( \Phi^3)$ on the right hand side of $\Box \Phi$
can be absorbed via $\Phi' \equiv \Phi - \alpha'^2 \zeta_2 \Phi^3$, i.e.\ the
right-hand side of $\Box \Phi'$ will no longer contain the term $\ap^2\zeta_2\Box
(\Phi^3)$ in question. This leftover freedom can be fixed by requiring the absence
of the dot product $(k_{A_1} \cdot k_{A_3})$ among the leftmost and the rightmost
slot-momentum\foot{In general, in a $p$-fold deconcatenation $\sum_{A=A_1 \ldots
A_p} \sum_{B=B_1 \ldots B_p}$, the dot product $(k_{A_1} \cdot k_{A_p})$ among the
leftmost and the rightmost momentum will not be included into an ansatz for $s_A
\phi_{A|B}$ at given order in $\alpha'$. This freezes the freedom to perform field
redefinitions while preserving the manifest parity property \BGsymmetriesA\ in
$A$.} in the deconcatenation $A=A_1 A_2 A_3$ in \AnsatzZeta.  Like this,
ambiguities to shift $\Box \Phi$ by a total d'Alembertian $\Box(\ldots)$ are
systematically avoided while preserving the manifest parity property
\BGsymmetriesA\ in $A$.

At the end of the above process, one finds the unique recursion
that generates the $\ap^2$ terms in the low-energy expansion of disk integrals
at any multiplicity via \ZapBG:
\eqnn\uptozetatwo
$$\eqalignno{
&\qquad\qquad\qquad s_A \phi_{A|B} =
\sum_{A_1A_2=A\atop B_1B_2=B} (\phi_{A_1|B_1} \phi_{A_2|B_2}
- \phi_{A_1|B_2} \phi_{A_2|B_1}) &\uptozetatwo\cr
&\qquad{} + \ap^2\zeta_2 \sum_{A_1 \ldots A_3=A\atop B_1 \ldots B_3=B}
 \Big[(k_{A_1}\cdot k_{A_2}) \bigl(
 \phi_{A_{1}|B_{1}}\phi_{A_{2}|B_{3}}\phi_{A_{3}|B_{2}}
- \phi_{A_{1}|B_{1}}\phi_{A_{2}|B_{2}}\phi_{A_{3}|B_{3}} \cr
&\qquad\qquad\qquad\qquad\qquad\qquad\qquad{}
+ \phi_{A_{1}|B_{3}}\phi_{A_{2}|B_{1}}\phi_{A_{3}|B_{2}}
- \phi_{A_{1}|B_{3}}\phi_{A_{2}|B_{2}}\phi_{A_{3}|B_{1}}
\bigr) \cr
&\qquad\qquad\qquad\qquad\quad{}+  (k_{A_2}\cdot k_{A_3}) \bigl(
\phi_{A_{1}|B_{2}}\phi_{A_{2}|B_{1}}\phi_{A_{3}|B_{3}}
 - \phi_{A_{1}|B_{1}}\phi_{A_{2}|B_{2}}\phi_{A_{3}|B_{3}}\cr
&\qquad\qquad\qquad\qquad\qquad\qquad\qquad{}
+ \phi_{A_{1}|B_{2}}\phi_{A_{2}|B_{3}}\phi_{A_{3}|B_{1}}
 - \phi_{A_{1}|B_{3}}\phi_{A_{2}|B_{2}}\phi_{A_{3}|B_{1}} \bigr) \Bigr] \cr
\noalign{\smallskip}
&\quad\quad{} + \ap^2\zeta_2 \sum_{A_1 \ldots A_4=A\atop B_1 \ldots B_4=B} 
          \Bigl[\phi_{A_{1}|B_{1}}\phi_{A_{2}|B_{2}}\phi_{A_{3}|B_{4}}\phi_{A_{4}|B_{3}}
         - \phi_{A_{1}|B_{1}}\phi_{A_{2}|B_{2}}\phi_{A_{3}|B_{3}}\phi_{A_{4}|B_{4}}\cr
&{}         + \phi_{A_{1}|B_{1}}\phi_{A_{2}|B_{3}}\phi_{A_{3}|B_{2}}\phi_{A_{4}|B_{4}}
         - \phi_{A_{1}|B_{1}}\phi_{A_{2}|B_{4}}\phi_{A_{3}|B_{2}}\phi_{A_{4}|B_{3}}
         + \phi_{A_{1}|B_{2}}\phi_{A_{2}|B_{1}}\phi_{A_{3}|B_{3}}\phi_{A_{4}|B_{4}}
          \cr
&{}         - \phi_{A_{1}|B_{2}}\phi_{A_{2}|B_{1}}\phi_{A_{3}|B_{4}}\phi_{A_{4}|B_{3}}
         - \phi_{A_{1}|B_{2}}\phi_{A_{2}|B_{3}}\phi_{A_{3}|B_{1}}\phi_{A_{4}|B_{4}}
         + \phi_{A_{1}|B_{2}}\phi_{A_{2}|B_{4}}\phi_{A_{3}|B_{1}}\phi_{A_{4}|B_{3}}
          \cr
&{}         - \phi_{A_{1}|B_{3}}\phi_{A_{2}|B_{1}}\phi_{A_{3}|B_{4}}\phi_{A_{4}|B_{2}}
         + \phi_{A_{1}|B_{3}}\phi_{A_{2}|B_{2}}\phi_{A_{3}|B_{4}}\phi_{A_{4}|B_{1}}
         + \phi_{A_{1}|B_{3}}\phi_{A_{2}|B_{4}}\phi_{A_{3}|B_{1}}\phi_{A_{4}|B_{2}}
          \cr
&{}         - \phi_{A_{1}|B_{3}}\phi_{A_{2}|B_{4}}\phi_{A_{3}|B_{2}}\phi_{A_{4}|B_{1}}
         + \phi_{A_{1}|B_{4}}\phi_{A_{2}|B_{1}}\phi_{A_{3}|B_{3}}\phi_{A_{4}|B_{2}}
         - \phi_{A_{1}|B_{4}}\phi_{A_{2}|B_{2}}\phi_{A_{3}|B_{3}}\phi_{A_{4}|B_{1}}
          \cr
&{}         - \phi_{A_{1}|B_{4}}\phi_{A_{2}|B_{3}}\phi_{A_{3}|B_{1}}\phi_{A_{4}|B_{2}}
         + \phi_{A_{1}|B_{4}}\phi_{A_{2}|B_{3}}\phi_{A_{3}|B_{2}}\phi_{A_{4}|B_{1}}
	 \Bigr]
	 + {\cal O}(\alpha'^3) \ .
}$$
For example, applying the above recursion to the disk integral $Z(13524|32451)$ whose
field-theory limit was computed in \secex\ leads
to the following result up to $\ap^2$:
\eqn\secexap{
Z(13524|32451) = - {1\over s_{13}s_{135}} + \ap^2\zeta_2\Bigl(
{s_{35}\over s_{135}}
+ {s_{25}\over s_{13}}
-1
\Bigr) + {\cal O}(\alpha'^3) \,.
}
It is important to emphasize that, while only four- and five-point data entered
in the derivation of \uptozetatwo, this recursion allows the computation of
$\ap^2$ terms of disk integrals at arbitrary multiplicity. The eleven-point example
\eqn\elevenex{
Z(134582679ba|123456789ab) =
%
	   {-\ap^2 \zeta_2\over s_{19ab} s_{ab} s_{345} s_{67}}\Bigl(
	   {1\over s_{34} } {+} {1\over s_{45}}\Bigr)\Bigl({1\over s_{1ab}} {+} {1\over s_{9ab}}\Bigr) + {\cal O}(\ap^3)
}
with the shorthands $a=10$ and $b=11$
was computed within two seconds on a regular laptop with
the program available in \BGapwww.

\subsubsec Manifesting the shuffle symmetries of BG currents

The length of the recursion in \uptozetatwo\ at the $\alpha'^2\zeta_2$ order
calls for a more efficient representation. In this subsection, we identify the sums of
products of $\phi_{A_i|B_j}$ which satisfy the shuffle symmetries \BGsymmetriesB\ in the
$B_j$-slots. This allows to rewrite the recursion \uptozetatwo\ in a compact form which inspires the generalization 
to higher orders and clarifies the commutator structure in the $Z$-theory equation of motion
upon rewriting the results in the language of perturbiners \perturbinerA.

In order to do this, recall from the theory of free Lie algebras
that all shuffle products are annihilated by a linear map $\rho$ acting on words $(B_1,B_2, \ldots,B_n)$
of $n$ letters $B_i$ which is defined by $\rho(B_i) \equiv B_i$ and \Ree
\eqn\rhomap{
\rho(B_1, B_2, \ldots ,B_n) \equiv \rho(B_1,B_2, \ldots ,B_{n-1}),B_n
- \rho(B_2,B_3, \ldots ,B_{n}),B_1 \, .
}
For example, it is easy to see that $\rho(B_1,B_2) = (B_1,B_2) - (B_2,B_1)$ and
\eqn\rhoexpl{
\rho(B_1,B_2,B_3) = (B_1, B_2, B_3) -
(B_2,B_1,B_3) - (B_2,B_3,B_1) + (B_3,B_2,B_1)
}
imply the vanishing of $\rho(B_1 \shuffle B_2)$
and $\rho((B_1,B_2) \shuffle B_3)$.
Therefore, after defining
\eqn\tensP{
T^{\rm dom}_{A_1,A_2, \ldots, A_n}\otimes T^{\rm int}_{B_1, B_2, \ldots, B_n} \equiv \phi_{A_1|B_1} \phi_{A_2|B_2} \ldots \phi_{A_n|B_n} \ .
}
it is straightforward to check
that the following linear combinations
\eqnn\TBdef
$$\eqalignno{
T^{B_1,B_2, \ldots, B_n}_{A_1,A_2, \ldots, A_n}&\equiv
T_{A_1,A_2, \ldots, A_n}^{\rm dom}\otimes T_{\rho(B_1,B_2, \ldots, B_n)}^{\rm int} &\TBdef\cr
&= T^{B_1, B_2, \ldots, B_{n-1}}_{A_1,A_2, \ldots,  A_{n-1}}\, \phi_{A_n|B_n}
- T^{B_2, B_3, \ldots, B_{n}}_{A_1, A_2, \ldots A_{n-1}}\, \phi_{A_n|B_1}\,,
}$$
with $T^B_A \equiv \phi_{A|B}$ satisfy the
shuffle symmetries
on the $B_j$-slots \Ree\foot{The parenthesis around the $B$ labels signifies that
the shuffle product treats the (multiparticle) labels $B_j$ as single entries, e.g.\ $(B_1,B_2)\shuffle(B_3)=(B_1,B_2,B_3) + (B_1,B_3,B_2) + (B_3,B_1,B_2)$.},
\eqn\shuffleB{
T^{(B_1,B_2, \ldots,B_i)\shuffle(B_{i+1}, \ldots,B_n)}_{A_1, A_2,\ldots,A_n}= 0 \ , \ \ \ \ i=1,2,\ldots,n-1 \,.
}
The first few examples of \TBdef\ read as follows,
\eqnn\TABex
$$\eqalignno{
T^{B_1, B_2}_{A_1,A_2} &\equiv \phi_{A_1 | B_1} \phi_{A_2 | B_2}
- \phi_{A_1 | B_2} \phi_{A_2 | B_1}\,,  &\TABex\cr
T^{B_1, B_2, B_3}_{A_1, A_2, A_3}  &\equiv
  \phi_{A_1 | B_1} \phi_{A_2 | B_2} \phi_{A_3 | B_3}
- \phi_{A_1 | B_2} \phi_{A_2 | B_3} \phi_{A_3 | B_1}\cr
&{}
- \phi_{A_1 | B_2} \phi_{A_2 | B_1}  \phi_{A_3 | B_3}
+ \phi_{A_1 | B_3} \phi_{A_2 | B_2} \phi_{A_3 | B_1}\,,\cr
T^{B_1, B_2, B_3, B_4}_{A_1, A_2, A_3, A_4} &\equiv
  \phi_{A_1 | B_1} \phi_{A_2 | B_2} \phi_{A_3 | B_3} \phi_{A_4 | B_4}
- \phi_{A_1 | B_2} \phi_{A_2 | B_1} \phi_{A_3 | B_3}  \phi_{A_4 | B_4}\cr
&{}
- \phi_{A_1 | B_2} \phi_{A_2 | B_3} \phi_{A_3 | B_1} \phi_{A_4 | B_4}
+ \phi_{A_1 | B_3} \phi_{A_2 | B_2} \phi_{A_3 | B_1} \phi_{A_4 | B_4}\cr
&{}
- \phi_{A_1 | B_2} \phi_{A_2 | B_3} \phi_{A_3 | B_4} \phi_{A_4 | B_1}
+ \phi_{A_1 | B_3} \phi_{A_2 | B_2} \phi_{A_3 | B_4} \phi_{A_4 | B_1}\cr
&{}
+ \phi_{A_1 | B_3} \phi_{A_2 | B_4} \phi_{A_3 | B_2} \phi_{A_4 | B_1}
- \phi_{A_1 | B_4} \phi_{A_2 | B_3} \phi_{A_3 | B_2} \phi_{A_4 | B_1}\,,
}$$
and their shuffle symmetries \shuffleB\ are easy to verify, starting with 
\eqn\simpshuff{
T^{B_1, B_2}_{A_1,A_2} =-T^{B_2, B_1}_{A_1,A_2}  \ , \ \ \
T^{B_1, B_2,B_3}_{A_1,A_2,A_3} + T^{B_1, B_3,B_2}_{A_1,A_2,A_3} + T^{B_3,B_1, B_2}_{A_1,A_2,A_3} = 0
\ .
}
Moreover, the $\rho$-map in \rhomap\ exhausts all tensors of the type \tensP\ subject to
shuffle symmetry in the $B_j$-slots it acts on \refs{\Ree, \Griffing}. Hence, a BG recursion which
manifests the shuffle symmetry in the $B_j$-slots is necessarily expressible in terms 
of $T^{B_1,B_2, \ldots ,B_n}_{A_1,A_2, \ldots ,A_n}$ in \TBdef.

Rather surprisingly, it turns out
that the definition \TBdef\ not only manifests the shuffle symmetries
on the $B_j$-slots but also implies {\it generalized Jacobi identities}
with respect to the $A_j$-slots. In other words, the above $T^{B_1,B_2, \ldots ,B_n}_{A_1,A_2, \ldots ,A_n}$
satisfy the same symmetries as the nested commutator $[[ \ldots[[A_1,A_2],A_3]
\ldots],A_n]$, see appendix~A.2 for a proof.

\subsubsec Simplifying the $\alpha'^2$-correction to BG currents

As discussed in the previous subsection,
the BG double current can always be written
in terms of $T_{A_1, A_2,\ldots,A_n}^{B_1,B_2, \ldots,B_n}$ from 
the definition \TBdef. For example, the expression
\uptozetatwo\ becomes
\eqnn\cztwo
$$\eqalignno{
s_A &\phi_{A|B} = \! \! \sum_{A=A_1A_2\atop B=B_1B_2}\! \! T^{B_1, B_2 }_{A_1,A_2}
- \ap^2\zeta_2 \!\!\! \! \sum_{A=A_1 \ldots A_3\atop B=B_1 \ldots B_3} \! \!\!\!\!
\Big[(k_{A_2}\cdot k_{A_3}) T^{B_1, B_2, B_3}_{A_1, A_2, A_3} +
(k_{A_1}\cdot k_{A_2}) T^{B_1, B_2, B_3}_{A_3,A_2,A_1} \Big] &\cztwo \cr
&{}+ \ap^2\zeta_2 \!\!\! \sum_{A=A_1 \ldots A_4\atop B=B_1 \ldots B_4}\!\!\!\! \Big(
T^{B_1, B_2, B_3, B_4}_{A_1,A_2, A_4,A_3} - T^{B_1, B_2, B_3, B_4}_{A_1,A_2, A_3,A_4}
- T^{B_1, B_2, B_3, B_4}_{A_1, A_3, A_4, A_2} + T^{B_1, B_2, B_3, B_4}_{A_1, A_3, A_2, A_4}\Big) + {\cal O}(\ap^3)\,. 
}$$
From a practical perspective, it could be a daunting task to convert a huge expression 
in terms of $\phi_{A_i|B_j}$ such as \uptozetatwo\ into linear combinations of
$T_{A_1, A_2,\ldots,A_n}^{B_1,B_2, \ldots,B_n}$ on the right-hand side of \cztwo. 
Fortunately, since both the BG double current
and $T_{A_1, A_2,\ldots,A_n}^{B_1,B_2, \ldots,B_n}$ satisfy
generalized Jacobi identities in the $A_j$-slots,
an efficient algorithm due to Dynkin, Specht and Wever \thibon\ can be used to accomplish
this at higher orders in $\ap$.
See the appendix~A.3 for more details.

\subsec The perturbiner description of $\alpha'$-corrections
\par\subseclab\secpertdesc

\noindent The recursion \cztwo\ for the coefficients $\phi_{A|B}$ of the perturbiner
\perturbinerA\
can be rewritten in a more compact form by defining
the shorthand
\eqn\commNot{
[[ \ldots [[\Phi_{i_1} ,\Phi_{i_2}] , \Phi_{i_3}] , \ldots ,\Phi_{i_{p-1}}], \Phi_{i_p}] 
\equiv \! \! \! \! \!\!\! \sum_{A_1,A_2,\ldots,A_p \atop{B_1,B_2,\ldots,B_p}}
\! \! \! \! \! \! \! \! e^{k_{A_1\ldots A_p} \cdot x}
\,T_{A_{i_1},A_{i_2}, \ldots, A_{i_p}}^{B_1, B_2, \ldots, B_p}\;t^{A_1A_2 \ldots A_p}\otimes
\tilde t^{B_1B_2 \ldots B_p}\,,
}
which exploits the generalized Jacobi symmetry of the $A_j$-slots in
$T^{B_1,B_2, \ldots, B_p}_{A_{i_1},A_{i_2}, \ldots ,A_{i_p}}$.
That is, the
numeric indices $i_1,i_2, \ldots,i_p$ of the various {\it formal} perturbiners
$\Phi_i$ in
the commutator match the ordering of the labels within the $A$-slots in
$T^{B_1,B_2, \ldots, B_p}_{A_{i_1},A_{i_2},
\ldots ,A_{i_p}}$, while the ordering
of the $B$-slots is always the same. Finally,
the color degrees of freedom enter
in a global multiplication order;
$t^{A_1A_2 \ldots A_p}\otimes \tilde t^{B_1B_2 \ldots B_p}$.

The above definition implies that the Berends--Giele recursion \cztwo\
condenses to,
\eqnn\pertEOM
$$\eqalignno{
\half\Box \Phi &= [\Phi_1, \Phi_2] -
\ap^2 \zeta_2 \big(
\partial_{23} [ [\Phi_1,\Phi_2],\Phi_3]
- \partial_{12} [ \Phi_1, [\Phi_3,\Phi_2]]
 \big)\cr
&{} +\ap^2 \zeta_2 \big( [[\Phi_1,\Phi_2] , [ \Phi_4,\Phi_3]  ] -
[  [\Phi_1,\Phi_3] , [ \Phi_4,\Phi_2]  ]
\big) + {\cal O}(\alpha'^3)\,, &\pertEOM
}$$
with the following shorthand for the derivatives:
\eqn\partder{
\partial_{ij} \equiv (\partial_i \cdot \partial_j) \ .
}
The convention for the derivatives $\partial_j$ is to only act on
the position of $\Phi_j$, e.g.\ the perturbiner expansion
of $\p_{12} [ [\Phi_3,\Phi_2],\Phi_1]$
reproduces $\sum_{A=A_1A_2 A_3} \sum_{B=B_1B_2 B_3} (k_{A_1}\cdot k_{A_2})
T^{B_1, B_2, B_3}_{A_3 ,A_2, A_1}$. 

In view of the increasing number of $\Phi$-factors at higher order in $\alpha'$, 
we will further lighten the notation and translate the commutators
into multiparticle labels $\Phi_P\equiv \Phi_{i_1 i_2 \ldots i_p}$,
\eqn\nested{
\Phi_{i_1 i_2 \ldots i_p} \equiv [[ \ldots [[\Phi_{i_1} ,\Phi_{i_2}] , \Phi_{i_3}] ,
\ldots ,\Phi_{i_{p-1}}], \Phi_{i_p}]\,,
}
which exhibit generalized Jacobi symmetries by construction\foot{
These are the same symmetries in $P=i_1i_2\ldots i_p$ obeyed by contracted structure
constants $f^{i_1 i_2 a} f^{a i_3 b} \ldots f^{x i_p y}$ as well as the local
multiparticle superfields $V_P$ \EOMBBs\ in pure spinor superspace.}. Hence, any
subset of the nested
commutators of \pertEOM\ can be separately expressed in terms of $\Phi_P$;
e.g.\ $[[\Phi_1,\Phi_2],[\Phi_3,\Phi_4]] = 
[\Phi_{12} ,\Phi_{34}] = \Phi_{1234} - \Phi_{1243}$.
In this language, the $Z$-theory equation of motion \pertEOM\ becomes
\eqn\zetatwopert{
\half \Box \Phi
=[\Phi_{1}, \Phi_{2}]
- \ap^2\zeta_2 \Bigl(
\p_{23} [\Phi_{12}, \Phi_3]
-\p_{12} [\Phi_{1}, \Phi_{32}]
- [\Phi_{12}, \Phi_{43}] + [\Phi_{13}, \Phi_{42}]\Bigr) + {\cal O}(\ap^3)\,.
}
As will be explained below, this form of the $Z$-theory
equation of motion provides the essential clue for
proposing the Berends--Giele recursion to arbitrary orders of $\ap$.

As a reformulation of \pertEOM\ which does not rely on the notion of perturbiners,
one can peel off the $t^a$ generators\foot{In view of the $\ap$-corrections to KK
relations from \PBCJ, the $Z$-theory scalar $\Phi$ is {\it not} Lie-algebra valued
in the gauge group of the $t^a$ but instead exhibits an expansion in the universal
enveloping algebra spanned by $t^A = t^{a_1} t^{a_2}\ldots t^{a_{\len{A}}}$.} from
the bi-colored fields $\Phi = \sum_A t^A \Phi_{A}$. The coefficients $\Phi_{A}$ are
still Lie-algebra valued with respect to the $\tilde t^b$, and this is where the
nested commutators act in the following rewriting of \pertEOM:
\eqnn\altEOM
$$\eqalignno{
\half\Box \Phi&= \! \!\! \sum_{A_1, A_2} \! \! t^{A_1 A_2}[\Phi_{A_1}, \Phi_{A_2}] - \! \! \! \! \! \!
\sum_{A_1, A_2,A_3} \! \! \! \! \! \! t^{A_1 A_2 A_3}
\ap^2 \zeta_2 \big(
\partial_{23} [ [\Phi_{A_1},\Phi_{A_2}],\Phi_{A_3}]
- \partial_{12} [ \Phi_{A_1}, [\Phi_{A_3},\Phi_{A_2}]]
 \big)\cr
&{} \hskip-20pt +\ap^2 \zeta_2 \! \! \! \! \! \! \! \! \! \sum_{A_1, A_2,A_3,A_4} \! \! \! \!  \! \! \! \! \! t^{A_1 A_2 A_3 A_4} \big( [[\Phi_{A_1},\Phi_{A_2}] , [ \Phi_{A_4},\Phi_{A_3}]  ] -
[  [\Phi_{A_1},\Phi_{A_3}] , [ \Phi_{A_4},\Phi_{A_2}]  ]
\big) + {\cal O}(\alpha'^3)\,. &\altEOM
}$$
Upon comparison with \zetatwopert, the notation in \nested\ can be understood as a compact way 
to track the relative multiplication orders of the $t^a$ and $\tilde t^b$ generators.

\subsubsec Perturbiners at higher order in $\ap$

The procedure of subsection \secZetatwo\ to determine the Berends--Giele recursion
that reproduces the $\ap^2$-corrections to the disk integrals was also
applied to fix the recursion
at the orders $\ap^3$ and $\ap^4$ (see appendix \appANS\ for more details).
Luckily, the analogous ansaetze at orders $\alpha'^{w\geq 5}$ could be bypassed 
since the general pattern of the field equations became apparent from the leading orders $\alpha'^{w\leq 4}$.
To see this, it is instructive to spell out
the $Z$-theory equation of motion up to the $\ap^3$-order:
\eqnn\zetatwoT
$$\eqalignno{
\half \Box \Phi &=[\Phi_{1}, \Phi_{2}]
+  \Bigl(\ap^2\zeta_2 \p_{12} - \ap^3\zeta_3 \p_{12}(\p_{12} + \p_{23}) \Bigr)  [\Phi_{{1}},\Phi_{{3}{2}}]&\zetatwoT\cr
&- \Bigl(\ap^2\zeta_2 \p_{23} - \ap^3\zeta_3 \p_{23} (\p_{12} + \p_{23}) \Bigr)  [\Phi_{{1}{2}},\Phi_{{3}}]\cr
& + \Bigl( \ap^2\zeta_2 - \ap^3\zeta_3 \big(\p_{21}
+ 2 \p_{31} + 2 \p_{32}
+ 2 \p_{42}  +  \p_{43}\big)  \Bigr)
[\Phi_{{1}{2}}, \Phi_{{4}{3}}] 
\cr
& - \Bigl(\ap^2\zeta_2 - \ap^3\zeta_3\big(2 \p_{21}
+  \p_{31} + 3 \p_{32}
 +  \p_{42} + 2 \p_{43} \big) \Bigr)
  [\Phi_{{1}{3}}, \Phi_{{4}{2}}] \cr
&+2 \ap^3\zeta_3 \bigl( \p_{42} +  \p_{43} \bigr)
[\Phi_{{1}{2}{3}}, \Phi_{{4}}] 
-\ap^3\zeta_3  \bigl( 3 \p_{42} +  \p_{43} \bigr)
[\Phi_{{1}{3}{2}}, \Phi_{{4}}]   \cr
&+2  \ap^3\zeta_3 \bigl( \p_{31} +  \p_{21} \bigr) [\Phi_{{1}}, \Phi_{{4}{3}{2}}] 
-  \ap^3\zeta_3  \bigl( 3 \p_{31}  +  \p_{21} \bigr) [\Phi_{{1}}, \Phi_{{4}{2}{3}}] \cr
&+  \ap^3\zeta_3   \Bigl(
-  [\Phi_{{1}{2}}, \Phi_{{5}{3}{4}}] 
{} +  2 [\Phi_{{1}{2}}, \Phi_{{5}{4}{3}}] 
{} -  2 [\Phi_{{1}{2}{3}}, \Phi_{{5}{4}}] 
{} -  2 [\Phi_{{1}{3}}, \Phi_{{5}{2}{4}}]
{} +  [\Phi_{{1}{3}{2}}, \Phi_{{5}{4}}]\cr
&{} +  2 [\Phi_{{1}{3}{4}}, \Phi_{{5}{2}}]
{} +  3 [\Phi_{{1}{4}}, \Phi_{{5}{2}{3}}]
{} -  2 [\Phi_{{1}{4}}, \Phi_{{5}{3}{2}}]
{} +  2 [\Phi_{{1}{4}{2}}, \Phi_{{5}{3}}]
{} -  3 [\Phi_{{1}{4}{3}}, \Phi_{{5}{2}}]
\Bigr) + {\cal O}(\ap^4)\,.
}$$
After identifying $s_{ij}\leftrightarrow\partial_{ij}$,
the coefficients of $[\Phi_{12}, \Phi_3]$ and
$[\Phi_{1}, \Phi_{32}]$ in \zetatwoT\ are identical to
the first {\it regular} terms in the expansion of
the four-point disk integrals considered in \Polylogs:
\eqnn\regAB
$$\eqalignno{
 {\rm reg} \int_0^1 \frac{\dd z_2}{z_{12}}  &\sum_{m,n=0}^{\infty} \! \! { (\ap s_{12} \ln |z_{12}|)^m \over m!}\,
 { (\ap s_{23} \ln |z_{23}|)^n \over n!}
= \ap \zeta_2 s_{23} -\ap^2 \zeta_3 s_{23}(s_{12} {+} s_{23}) + {\cal O}(\ap^3)  \cr
 {\rm reg} \int_0^1 \frac{\dd z_2}{z_{32}}  &\sum_{m,n=0}^{\infty} \! \! { (\ap s_{12} \ln |z_{12}|)^m \over m!} \,
{ (\ap s_{23} \ln |z_{23}|)^n \over n!} =
-\ap \zeta_2 s_{12} + \ap^2\zeta_3 s_{12}(s_{12} {+} s_{23})+ {\cal O}(\ap^3) \cr
&&\regAB
}$$
The endpoint divergences of these integrals as $z_2 \rightarrow z_1=0$ and $z_2
\rightarrow z_3=1$ require a regularization prescription denoted by ``reg'' and
explained in section \regpoly. The infinite sums in the above integrands arise from
the Taylor expansion of a $SL(2,\Bbb R)$-fixed four-point Koba--Nielsen factor via
\eqn\KNtaylor{
|z_{ij}|^{\ap s_{ij}} =
\sum_{n=0}^{\infty} \frac{1}{n!} (\ap s_{ij} \ln|z_{ij}|)^n\,,
}
which removes the kinematic poles from the full disk integrals and yields their
non-singular counterparts \Polylogs\ upon regularization. Comparing the expansion
of \regAB\ at the next order in $\ap$ with the expression for the BG current
obtained from an ansatz confirms the pattern, and we will later on see that the
terms of order $\Phi^4$ and $\Phi^5$ in \zetatwoT\ can be traced back to
regularized five- and six-point integrals.

\subsec All-order prediction for the BG recursion

From the observations in the previous subsection, we propose 
a closed form for the $\Phi^3$ contributions to the
$Z$-theory equations of motion for $\Box \Phi$, to all orders in $\alpha'$: 
\eqnn\closedthree
$$\eqalignno{
\half \Box \Phi &=[\Phi_{1}, \Phi_{2}] -
\ap\,{\rm reg} \int_0^1 \dd z_2 \, \sum_{m=0}^{\infty} { (\ap \p_{12} \ln |z_{12}|)^m \over m!} 
\, \sum_{n=0}^{\infty} { (\ap \p_{23} \ln |z_{23}|)^n \over n!}   \cr
& \ \ \ \ \ \ \ \ \ \ \ \ \ \ \  \ \ \ \ \ \ \ \ \ \ \times
\Big(
\frac{ [\Phi_{12}, \Phi_3] }{z_{12}}+\frac{ [\Phi_{1}, \Phi_{32}]  }{z_{32}}
\Big) + {\cal O}(\Phi^4) \ . &\closedthree
}$$
The integrand in the second line bears a strong structural similarity to
the correlation function in the
four-point open string amplitude \refs{\nptTreeI,\fourptids}
\eqn\fourdisk{
A^{\rm open}(1,2,3,4) = -\ap \int^1_0\dd z_2 \,
\prod_{i<j}^3 |z_{ij}|^{\alpha's_{ij}}
\Big\langle {  V_{12} V_3 V_4  \over z_{12}}
+ {  V_{1} V_{32} V_4  \over z_{32}} \Big\rangle \ ,
}
with $\langle V_P V_Q V_n \rangle$ denoting certain kinematic factors in pure
spinor superspace.  The precise correspondence between \closedthree\ and \fourdisk\
maps multiparticle vertex operators $V_P$ \EOMBBs\ to perturbiner commutators
$\Phi_P$ defined in \nested. Moreover, since $V_P$ is fermionic and satisfies
generalized Jacobi symmetries \EOMBBs, the all-multiplicity mapping
\eqn\correspV{
\langle V_P V_Q V_n \rangle \longleftrightarrow [\Phi_P,\Phi_Q]\,,\quad \len{P}+\len{Q} = n-1
}
preserves all the symmetry properties of its constituents. Finally, the Koba--Nielsen
factor $\prod_{i<j}^3 |z_{ij}|^{\alpha's_{ij}}$ with $s_{ij} \rightarrow \p_{ij}$
has been Taylor expanded according to \KNtaylor\ in converting \fourdisk\ to
\closedthree.  This projects out the kinematic poles of the integrals to ensure
locality of the $Z$-theory equation of motion, but requires a regularization of
the endpoint divergences at $z_2 \rightarrow 0$ and $z_2 \rightarrow 1$ as
discussed in section \regpoly.

It is easy to see that the correspondence \correspV\ correctly ``predicts''
the first term in the right hand side of \closedthree\ from the well-known \psf\
expression $A^{\rm open}(1,2,3)=\langle V_1V_2V_3\rangle$ of the three-point
massless disk amplitude under the mapping \correspV;
$\langle V_1 V_2 V_3\rangle \longleftrightarrow [\Phi_1,\Phi_2]$.

Extrapolating the above pattern,
a natural candidate for the higher-order contributions $\Phi^4, \Phi^5,\ldots$ to
the $Z$-theory equation of motion emerges from the
integrand of the $(n{-}2)!$-term
representation of the $n$-point disk amplitude \refs{\nptTreeI},
\eqnn\VVVn
$$\eqalignno{
&A^{\rm open}(1,2,\ldots,n) = (-\ap)^{n-3} 
\! \! \! \! \! \! \! \! \! \!\! \! \! \! \! \!    \int\limits_{0\leq z_{2} \leq z_{3} \leq \ldots \leq z_{n-2} \leq 1} 
\! \! \! \! \! \! \! \! \! \! \! \! \! \! \! \!  \dd z_2 \, \dd z_3 \, \ldots \, \dd z_n
\prod_{i<j}^{n-1} |z_{ij}|^{\ap s_{ij}}  &\VVVn \cr
& \ \ \ \times \, \Big\langle
 \sum_{l=1}^{n-2} \frac{  V_{12\ldots l}  V_{n-1,n-2,\ldots, l+1} V_n }{(z_{12}z_{23}\ldots z_{l-1,l})
(z_{n-1,n-2} z_{n-2,n-3} \ldots z_{l+2,l+1})} + \perm(2,3,\ldots,n{-}2) \Big\rangle \, ,
}
$$
which appeared in an intermediate step towards the minimal $(n{-}3)!$-term
expression \npttree.
This expression leads us to propose the following $Z$-theory equation of motion
to all orders in the fields and their derivatives (with $SL(2,\Bbb R)$-fixing $z_1=0$ and $z_p=1$):
\eqnn\fullBG
$$\displaylines{
\frac{1}{2} \Box \Phi =
\sum_{p=2}^{\infty}(-\ap)^{p-2}  \int^\eom \prod_{i<j}^{p} |z_{ij}|^{\ap
\partial_{ij}}  \hfil\fullBG\hfilneg\cr
\times\Bigl(\,
\sum_{l=1}^{p-1}
\frac{ [\Phi_{12\ldots l} , \Phi_{p,p-1\ldots l+1}] }{(z_{12}z_{23}\ldots z_{l-1,l})
 (z_{p,p-1} z_{p-1,p-2} \ldots z_{l+2,l+1})}
+ \perm(2,3,\ldots,p{-}1)\, \Bigr)\,.
}$$
Apart from the correspondence \correspV\ which
settles the perturbiner commutators suggested by \VVVn, we introduce a formal operator
$\int^\eom$ that maps the accompanying disk integrals to local expressions. 
The precise rules for the map $\int^\eom$ to be explained in the next section 
include a Taylor expansion \KNtaylor\ of the Koba--Nielsen factor as seen in \closedthree. Also, $\int^\eom$
incorporates a regularization along with particular parameterization of the ubiquitous domain 
$0 \leq z_2 \leq z_3\leq \ldots  \leq z_{p-1} \leq 1$
for the $p{-}2$ integration variables $z_2,z_3,\ldots,z_{p-1}$ which is left implicit
in \fullBG\ for ease of notation. The shorthands $\Phi_{i_1 i_2\ldots i_k}$ in \fullBG\
explained in section \secpertdesc\ compactly track the relative multiplication order of the 
gauge-group generators $t^a$ and $\tilde t^b$ which govern the color structure of $\Phi$.

For example, the equation of motion up to $\Phi^4$-order following from \fullBG\ reads
\eqnn\pertFive
$$\eqalignno{
\half \Box \Phi &=[\Phi_{1}, \Phi_{2}] -
\ap\, \int^{\rm eom}  \prod_{i<j}^{3} |z_{ij}|^{\ap \partial_{ij}}
\Big(
\frac{ [\Phi_{12}, \Phi_3] }{z_{12}} + \frac{ [\Phi_{1}, \Phi_{32}]  }{z_{32}}
\Big) &\pertFive\cr
&{} +\ap^2 \int^\eom  \prod_{i<j}^{4} |z_{ij}|^{\ap \partial_{ij}}\Bigl(
 \frac{ [\Phi_{123} , \Phi_{4}] }{z_{12}z_{23}}
 +\frac{ [\Phi_{12} , \Phi_{43}] }{z_{12}z_{43}}
 +\frac{ [\Phi_{1} , \Phi_{432}] }{z_{43}z_{32}}
 +(2\leftrightarrow 3)\Bigr) + {\cal O}(\Phi^5)\,,
}$$
and the low-energy expansion of the five-point integrals in the second line
spelled out in appendix \appfivereg\ reproduces
the $\zeta_2  \Phi^4$- and $\zeta_3 \partial^2 \Phi^4$-orders of the $Z$-theory equation of motion \zetatwoT.
Using the rules explained in the next section
for obtaining the local terms indicated by $\int^\eom$, we
have made an explicit form of the Berends--Giele double current from \fullBG\ up to
$\ap^7$ publicly available on \BGapwww.

The Berends--Giele recursion for the $\ap$-expansion of disk integrals is
particularly advantageous over previous methods when computing the $\ap^w$-order of
disk integrals at {\it high multiplicities} $n> w{+}3$. That is because only a {\it
finite} number of terms up to $\Phi^{w+2}$ in the field equation \fullBG\ is
required to obtain terms of order $\ap^w$ in the disk integrals to {\it all
multiplicities} (by simple deconcatenation of words as seen in \uptozetatwo). This
bypasses the manual pole subtractions in the polylogarithm-based method of
\Polylogs\ and the increasingly expensive matrix algebra involving matrices of
dimension $(n{-}2)!\times(n{-}2)!$ in the Drinfeld associator method of
\BroedelAZA.

\newsec Local disk integrals in the $Z$-theory equation of motion
\par\seclab\regpoly

\noindent In the previous section, we have proposed the $Z$-theory equation of motion \fullBG\
which determines the Berends--Giele double currents of disk integrals \ZapBG. The proposed field equations 
are inspired by the form \VVVn\ of the open-superstring disk amplitude and rely on a
formal operator $\int^\eom$ which converts the associated iterated integrals into local expressions. 
The purpose of this section is to give a precise definition of the map 
$\int^\eom$ in \fullBG\ which incorporates a regularization of the disk integrals' 
endpoint divergences along with a prescription to settle the resulting ambiguities.

In section~\secfourC, we will briefly review the definition and properties of
polylogarithms that are used to perform the integrals that appear in \fullBG.
Already the examples at the four-point level \regAB\ will be seen to yield endpoint
divergences, for which we will specify a suitable regularization scheme.
Consequently, the results of iterated integrals at the $(n\geq 5)$-point level will
depend on the order of integration\foot{Note that Fubini's theorem stating the
equivalence of integration orders for iterated integrals does not apply to the
divergent integrals and their regularized values under discussion.}. The first
non-trivial example is given by \noparity, where different regularized values may
arise for the two orders of integration $\int^1_0 \dd z_3 \int^{z_3}_0 \dd z_2$ and
$\int^1_0 \dd z_2 \int^{1}_{z_2} \dd z_3$. A priori, it is not clear that any
choice will lead to their regularized values required by the $Z$-theory equation of
motion. But, on empirical grounds, we find a prescription that gives the correct
answers: we identify a new basis for the integrands in \fullBG\ under partial fraction relations 
along with the integration orders for each of its elements. The recursive algorithms implementing these rules
are described in sections \secfourA\ and \secfourB.

The above prescription was derived by trial and error through
comparison with known data for $Z(P|Q)$ at low number of points, and
its consequences were extrapolated to arbitrary multiplicity.
It remains an open question to find its rigorous mathematical
justification.

\subsec Multiple polylogarithms and their regularization
\par\subseclab\secfourC

\noindent In this section, we review selected aspects of the polylogarithm-based
setup of \Polylogs\ to extract local terms (also called regular terms) from the disk
integrals\foot{There is a vast body of literature related to
iterated integrals on moduli spaces of genus-zero curves with $n$ ordered
marked points, see e.g.\ \refs{\BrownmathMZV, \BrownUM, \PanzerIDA,\BognerMHA}
and references therein. Moreover, their symbolic computation
have been recently implemented in computer programs \refs{
\hyperint,\BognerNDA}.}
in \VVVn. The requirement that the $\int^\eom$ map 
must reproduce the correct $Z$-theory equation of motion induces systematic
departures from \Polylogs\ which will be highlighted in the subsequent discussion.

\subsubsec Polylogarithms and MZVs

We recall that multiple polylogarithms $G(A;z)$ 
with $A=a_1,a_2,\ldots,a_n$ and $a_j,z \in \Bbb C$ are defined by\foot{
Our conventions for polylogarithms agree with the work \GoncharovIEA\ of Goncharov as well as for instance 
reference \DuhrZQ. See e.g.\ \AblingerCF\ for other aspects of multiple polylogarithms.}
\eqn\Gdef{
G(a_1,a_2,\ldots,a_n;z) \equiv \int^z_0 \frac{ \dd t}{t-a_1} \, G(a_2,\ldots,a_n;t),\quad
G(\emptyset;z)\equiv 1,\quad  \forall z\neq0\,,
}
setting $G(\emptyset ;0)\equiv 0$. The variables $a_j$ and $z$ on the left and right of the semicolon
are referred to as the {\it labels} and the {\it argument} of the polylogarithm, respectively, and the
number $n$ of labels $a_j$ is called the {\it weight}.
Their recursive definition \Gdef\ as iterated integrals
endows polylogarithms with a shuffle algebra
 \eqn\shufalg{
G(A;z)G(B;z) = G(A\shuffle B;z)\,,
}
and the regularization prescription discussed in the sequel is designed to preserve \shufalg.
After repeated application of the recursion \Gdef, disk integrals ultimately boil down to $G(\ldots;1)$ at unit 
argument \Polylogs. In the framework of the $Z$-theory equation of motion \fullBG, this
follows from the endpoint $z_{p}=1$ for the uppermost integration variable
$z_{p-1}$ and reproduces the integral representation of MZVs \MZVdef,
\eqn\Maxim{
\zeta_{n_1,n_2,\ldots,n_r} = (-1)^r G( \underbrace{ 0,0,\ldots,0,1}_{n_r} , \ldots, \underbrace{ 0,0,\ldots,0,1}_{n_1};1) \ , 
}
see appendix D.1 for examples and extensions to regularized values of divergent integrals.

\subsubsec Polylogarithms and the Koba--Nielsen factor

Using the special cases of the multiple polylogarithms \Gdef,
\eqn\Gexa{
 G(\underbrace{0,0,\ldots,0}_{w};z) \equiv \frac{1}{w!} \big[ \ln(z)\, \big]^w\,,\quad
G(\underbrace{a,a,\ldots,a}_{w};z) = \frac{1}{w!} \Big[ \ln\Big(1-\frac{z}{a}
\Big)\, \Big]^w \quad a \neq 0\,,
}
see \regwone\ for the regularization involved in the
convention for $G(0,0,\ldots,0;z)$,
the Taylor expansion \KNtaylor\ of the Koba--Nielsen factor
with the $SL(2,\Bbb R)$-fixing $z_1=0$ and $z_p=1$
can be written as \Polylogs
\eqn\knfinal{
 \prod_{i<j}^{p}|z_{ij}|^{\ap \p_{ij}} 
=  \prod_{i=2}^{p-1} \sum_{n_i=0}^{\infty} \left(\sum_{l=1}^{i-1} \ap \p_{il}
\right)^{\mkern-8mu n_i}
G( \underbrace{ 0,\ldots,0}_{n_i} ; z_i)
\prod_{2\leq j<k}^p \sum_{n_{jk}=0}^{\infty} (\ap \p_{jk})^{n_{jk}}
 G(\underbrace{ z_k,\ldots,z_k}_{n_{jk}};z_j)\,.
}
Therefore,
the leading orders of the regularized four-point integrals in \regAB\ can be traced back to
 \eqnn\kobafour
 $$\eqalignno{
\int^\eom \prod_{i<j}^{3}|z_{ij}|^{\ap \p_{ij}} &\;  {1 \over z_2-a} =\reg \int^0_1\frac{ \dd z_2}{z_2-a}\, \Big\{ 1
 	   + \ap \Bigl(G(0;z_2)\p_{12} 
 	   +  G(1;z_2) \p_{23}\Bigr) 
 	   &\kobafour\cr
 &{}  \!\!\!\!\!\!\!\!\!\!   \!\!\!\!\!\!\!\!\!\!       + \ap^2\Bigl( G(0,0;z_2)\p_{12}^2
           +  G(0\shuffle 1;z_2) \p_{12} \p_{23}
           +  G(1,1;z_2) \p_{23}^2\Bigr)+ {\cal O}(\ap^3) \Big\}\,,
 }$$
with $a \in \{0,1\}$, using \Gdef\ to perform the $z_2$-integral as well as \Maxim\ to convert the 
results to MZVs. Divergent cases
as exemplified in \Gregs\ are addressed by the regularization scheme which is denoted by ``reg'' in \kobafour\ 
and will be the subject of the next subsection.

\subsubsec Regularization of endpoint divergences

It follows from their definition \Gdef\ that multiple polylogarithms diverge at the endpoints of the 
integration domains whenever $a_1=z$ or $a_n=0$, and therefore they need to be
regularized. The convention for $G(0,0,\ldots,0;z)$ in \Gexa\ is part of the regularization procedure
of interest to this work and can be understood in terms of a cutoff $\epsilon$: The left hand side of
\eqn\cutoff{
\int_\epsilon^z {\dd t \over t} = \ln|t|\, \big|^{t=z}_{t=\epsilon} = \ln|z| - \ln|\epsilon| 
}
formally tends to $G(0;z)$ in the $\epsilon \rightarrow 0$ limit, and its regularized value $\ln|z|$
can be obtained from the right hand side by manually discarding (the source of
divergences) $\ln|\epsilon|$.
Together with a similar reasoning for divergences from the upper integration limit, we specify
the following regularized values for divergent integrals at weight one\foot{We are indebted to Erik Panzer for
suggesting this regularization to us.}:
\eqn\regwone{
\reg \int_0^z {\dd t \over t} = G(0;z) \equiv \ln|z| \ , \ \ \ \  
\reg \int_0^z {\dd t \over t-z}  \equiv -\ln|z| = -G(0;z) \ , \ \ \ \ z >0 
}
Further subtleties arise in situations where the endpoint divergence
as $t\rightarrow z$ is approached from above. In this case, one defines
\eqnn\regsubtle
 $$\eqalignno{
\reg \int_z^w {\dd t \over t-z}  &
\equiv G(z;w)+ G(0;z) - i \pi \ , \ \ \ \ \ \ w> z \, , &\regsubtle
}$$
where the occurrence of imaginary parts is an artifact of the decomposition of the 
integration domains in later sections. The choice of sign along with $i \pi$ in \regsubtle\
is a convention, and the cancellation of imaginary parts in the $Z$-theory equation of motion
serves as a consistency check of our integration setup.

One can combine \regwone\ with \regsubtle\ such as to define the regularized value of
$G(z;z)$ via
\eqn\EriksReg{
G(z;z) \equiv - G(0;z) + i\pi\d\,,
}
where $\d=0$ and $\d=1$ if $G(z;z)$ is obtained after
integration over $t$ such that $t<z$ and $t> z$, respectively.

Since the regularization scheme in this work is defined to preserve the shuffle
algebra \shufalg, the regularized values at weight one in \regwone\ and \EriksReg\ determine
endpoint divergences at higher weight, see appendix D.2 for more
details. For instance, the special cases $G(0;1)=G(1;1)=0$ of \regwone\ along with
the shuffle algebra allow to extract finite linear combinations of MZVs from
$G(1,\ldots;1)$ and $G(\ldots,0;1)$ with labels $\in \{0,1\}$, see \Gregs.

In contrast to the regularizations
\regwone\ and \EriksReg\ of this work
which are selected by the $Z$-theory equation of motion,
the regularization scheme of \Polylogs\
preserves the scaling property of polylogarithms and implies a vanishing
regularized value for $G(z;z)$.

\subsubsec Dependence on the integration order

As a subtle consequence of the shuffle-preserving regularization scheme based on
\regwone\ and \EriksReg, regularized values of disk integrals relevant to the
$Z$-theory equation of motion \fullBG\ depend on the integration order. A simple
example where the two integration orders $\int^1_0 \dd z_3 \int^{z_3}_0 \dd z_2$
and $\int^1_0 \dd z_2 \int^{1}_{z_2} \dd z_3$ for the integration domain $0\leq z_2
\leq z_3\leq 1$ yield inequivalent results stems from the five-point integral over
${\ln|z_{23}|\over z_{12} z_{13}}$ which arises from the partial-fraction identity
\eqn\partfrac{
{1 \over z_{12} z_{13}} = 
{1 \over z_{12} z_{23}} 
+ {1 \over z_{13} z_{32}} 
}
along with the Koba--Nielsen expansion
\knfinal\ at linear order in $\ap$. Using $\ln|z_{23}| = G(0;z_3) + G(z_3;z_2)$ and
$G(0,z_3;z_3) = -\zeta_2$ (see \zrmlenthree) as well as \genzrm\ to render
$G(0,z_2;1)$ suitable for integration over $z_2$, one finds the two different
results\foot{In order to evaluate the second integral of \noparity\ through the
definition \Gdef\ of polylogarithms, the integration limits are rearranged
according to
$$
\int^{1}_{z_2} \dd z_3 \, f(z_3) = \int^{1}_{0} \dd z_3 \, f(z_3)  - \int^{z_2}_{0} \dd z_3 \, f(z_3)  \, .
$$}
\eqn\noparity{
{\rm reg} \int^1_0 \frac{ \dd z_3}{z_3} \int^{z_3}_0 \frac{ \dd z_2}{z_2} \; \ln|z_{23}| = 0
\ , \ \ \ \ \
{\rm reg} \int^1_0 \frac{ \dd z_2}{z_2} \int^{1}_{z_2} \frac{ \dd z_3}{z_3} \; \ln|z_{23}|= \zeta_3 \, .
}
The task to rewrite $G(0,z;z)$ and $G(0,z;1)$ in a form suitable for integration
over $z$ via \Gdef\ is ubiquitous to regularized $(n\geq 5)$-point disk integrals
\Polylogs. The systematics of such ``$z$-removal identities'' is discussed in
appendix D.3. It is worth noting that
the symbolic program {\tt HyperInt} \hyperint\
contains routines that automate this task.

It turns out that between the two orders of integration displayed in \noparity, the
$Z$-theory equation of motion \zetatwoT\ (obtained from an ansatz for the equivalent BG 
recursion) is reproduced at the $\ap^3\zeta_3$ order only if the regularized integral of
$\ln|z_{23}|/(z_{12}z_{13})$ vanishes. Therefore $z_2$ must be integrated prior to
$z_3$ in presence of $(z_{12}z_{13})^{-1}$ in a five-point integrand. By worldsheet parity $z_j \rightarrow
z_{5-j}$, the integral over $(z_{24} z_{34})^{-1}$ must then follow the converse
order where $z_3$ is integrated first. The conclusion here is that different integrands require different
orders of integration. Adapting the integration order to each integrand will be
part of the map $\int^\eom$ to be elaborated below.

Similarly, we identified the appropriate integration orders for the
$4!$ six-point integrals in \fullBG\ at $p=5$ by matching with the $\ap^3$ and
$\ap^4$ order of the Berends--Giele recursion obtained from an ansatz. Moreover, an
alternative method to determine the desired outcome of regularized integrals to
arbitrary orders in $\ap$ is presented in appendix \Jregapp\ which closely follows
the handling of poles in \Polylogs. An all-multiplicity algorithm to determine the
integration orders which are observed to reproduce the $Z$-theory
equation of motion will be described in section \secfourB. As a preparation for
this, however, a systematic change of integral bases via repeated use of
partial-fraction identities will be introduced in the next section.

\subsec Towards the simpset basis
\par\subseclab\secfourA

\noindent Our investigations showed that the $(n{-}2)!$ integrals in the
open-superstring amplitude \VVVn\ need to be rewritten in a very particular basis
to define the $\int^\eom$ prescription in the $Z$-theory equation of motion
\fullBG. In this section, we will introduce a basis where the $\int^\eom$
prescription can be associated with appropriate integration orders for the
regularized integrals such as to settle the ambiguity seen in \noparity. In order
to explain this change of basis\foot{The ``basis'' of dimension $(n{-}2)!$ refers
to the minimum elements under partial-fraction identities; integration by parts
further reduce their number to $(n{-}3)!$ \refs{\nptTreeI,\nptTreeII}. The
reduction of products of $z_{ij}^{-1}$ via partial fractions to a
$(n{-}2)!$-dimensional basis is also described in appendix A of \StiebergerHZA.} it
will be convenient to introduce the following chain of worldsheet propagators
\eqn\calZ{
\cZ_A\equiv {1\over z_{a_1a_2}z_{a_2 a_3} \ldots z_{a_{\len{A}-1},a_{\len{A}}}}\ , \ \ \ \ \ \ |A|\geq 2
 \,,
}
in which two consecutive $z_{ij}$ factors in the denominator always share a label, with a formal extension
$\cZ_A\equiv1$ to words of length $|A|=1$. One can check that $z_{ij}=-z_{ji}$ and 
partial-fraction identities \partfrac\ imply the shuffle symmetry \cresson
\eqn\shufcalZ{
\cZ_{A\shuffle B} = 0\,,\quad\forall A,B\neq\emptyset\,.
}
Using the above definition, the $(n{-}2)!$ {\it chain basis} integrals in the
amplitude \VVVn\ can be distinguished by their chain factors of $\cZ_{1P}\cZ_{(n-1)Q}$,
with $\len{P}+\len{Q}=n{-}3$. As a part of the prescription
for the map $\int^\eom$, the integrals from the chain basis in the
$Z$-theory equation of motion \fullBG\ are rewritten in another basis which is
referred to as the {\it simpset basis}.

\subsubsec Description of the algorithm

At generic multiplicity, the elements of the simpset basis are obtained from the
chain basis $\cZ_{1P}\cZ_{(n-1)Q}$ by recursively stripping off factors of
$\cZ_{ij}=z_{ij}^{-1}$. At each step, the shuffle symmetry \shufcalZ\ is applied to
$\cZ_{1P}$ and $\cZ_{(n-1)Q}$ to factor out $\cZ_{ij}$, where $i$ and $j$ are the
labels in $1P$ which are maximally apart (i.e.\ at highest value of $|i-j|$). This
procedure is repeated for the coefficient $\cZ_{R}$ in the decomposition $\cZ_{1P}
= \cZ_{ij} \cZ_{R}$, leading to a recursive algorithm.

In a factor of $\cZ_{1243}$ relevant at six points, the labels~$1$ and $4$ constitute the
pair which is maximally apart with a separation of $|1-4|=3$. Therefore, to arrive
at the elements in the simpset basis, one needs to rewrite $\cZ_{1243}$ in such a way
as to contain the factor $z_{14}^{-1}$. In this case it is easy to show using
partial-fraction identities that
\eqn\exampsimp{
\cZ_{1243} = -\cZ_{14}\cZ_{12}\cZ_{34} -\cZ_{14}\cZ_{24}\cZ_{34}\,,
}
in which the factor $\cZ_{14}=z_{14}^{-1}$ has been stripped off from the chain
$\cZ_{1243}$. The two integrals on the right-hand side of \exampsimp\ 
belong to the simpset basis since $\cZ_{12}\cZ_{34}$ cannot be written as a
single chain factor $\cZ_R$ and the maximally separated labels $2,4$ in 
$\cZ_{24}\cZ_{34}=-\cZ_{243}$ are already factored out. 

The following recursive algorithm implements the change of basis required by the
$\int^\eom$ map. For each factor of $\cZ_{R}$ one identifies the pair of labels $i$
and $j$ that are maximally separated and recursively applies the following corollaries of
\shufcalZ\ and \calZ,
\eqn\changebasis{
\cZ_{iAaj} = - \cZ_{(iA\shuffle j) a},\qquad \cZ_{iAjB} = \cZ_{iAj}\cZ_{jB} \, ,
}
which eventually stops at $\cZ_{ija}=\cZ_{ij}\cZ_{ja}$ where the factor
$\cZ_{ij}$ is singled out.

In order to illustrate the algorithm \changebasis, consider the seven-point
integral characterized by the factor $\cZ_{63425}$ with five labels. Since the labels~$2$ and $6$ are
maximally separated, the second identity in \changebasis\ rewrites it as
$\cZ_{63425} = \cZ_{6342}\cZ_{25}$. The first factor now contains only four labels
and iterating the application of the identities in \changebasis\ yields,
\eqnn\longex
$$\eqalignno{
\cZ_{6342} &= - \cZ_{6324} - \cZ_{6234} - \cZ_{2634} = -\cZ_{632}\cZ_{24} -
\cZ_{62}\cZ_{234} - \cZ_{26}\cZ_{634} &\longex\cr
&= (\cZ_{623} + \cZ_{263})\cZ_{24} +
\cZ_{62}(\cZ_{243}+\cZ_{423})-\cZ_{26}\cZ_{63}\cZ_{34}\cr
&= (\cZ_{62}\cZ_{23} + \cZ_{26}\cZ_{63})\cZ_{24}
+ \cZ_{62}(\cZ_{24}\cZ_{43} + \cZ_{42}\cZ_{23})
-\cZ_{26}\cZ_{63}\cZ_{34}\cr
&=\cZ_{26}\cZ_{63}\cZ_{24} + \cZ_{62}\cZ_{24}\cZ_{43}
- \cZ_{26}\cZ_{63}\cZ_{34}\,.
}$$
In order to arrive at the second line, the factor $\cZ_{234}$ was manipulated w.r.t
the maximally-separated labels~$2$ and $4$ (with similar considerations for the
other factor $\cZ_{632}$). Therefore,
\eqn\eomover{
{1\over z_{63}z_{34}z_{42}z_{25}} = {1\over z_{26}z_{63}z_{24}z_{25}} +
{1\over z_{62}z_{24}z_{43}z_{25}}
- {1\over z_{26}z_{63}z_{34}z_{25}}\,,
}
is the transformation from the chain to the simpset basis.

The first non-trivial application of the above algorithm
leads the five-point simpset basis
\eqn\simpsetfive{
\left\{ 
{1\over z_{12} z_{13}} , \ 
{1\over z_{13} z_{23}} , \ 
{1\over z_{12} z_{43}} , \ 
{1\over z_{13} z_{42}} , \ 
{1\over z_{42} z_{43}} , \ 
{1\over z_{32} z_{42}} 
\right\} \,.
}
The complete set of denominators in the six-point simpset basis can be found in
\sixvuf\ (upon adjoining their parity images under $z_j \rightarrow z_{6-j}$),
while the appendix \orderapp\ contains
an overview of the seven-point simpset basis.

\subsubsec Back to the chain basis

For completeness, it is straightforward to exploit the shuffle symmetry \shufcalZ\ to
obtain a recursive algorithm to expand the simpset basis elements back in the chain
basis\foot{This algorithm summarizes the discussion of the appendix A of
\StiebergerHZA\ after noticing that simpset and chain basis elements
can be described via Cayley graphs and Hamilton paths, respectively.}.
To motivate the algorithm below, consider the following example:
To rewrite $\cZ_{12}\cZ_{13}\cZ_{14}$ in the chain basis
note that $\cZ_{12}\cZ_{13}\ = -\cZ_{213}$.
Next, to make a chain out of $\cZ_{213}\cZ_{14}$ one
uses the identity $\cZ_{A1B} =
(-1)^\len{A}\cZ_{1(\tilde A \shuffle B)}$ in the first factor to allow it to
be prefixed by $\cZ_{14}=-\cZ_{41}$, yielding $\cZ_{12}\cZ_{13}\cZ_{14} =
-\cZ_{4123} - \cZ_{4132}$. Then, $\cZ_{41ij}=- \cZ_{14ij}-\cZ_{1i4j}-\cZ_{1ij4}$ 
completes the basis change to $\cZ_{12}\cZ_{13}\cZ_{14}= \cZ_{1234}+{\rm perm}(2,3,4)$. 

Hence, the general algorithm to expand the simpset basis elements in the chain basis is
based on the recursive application of the following two identities,
\eqn\ToHamiltonBasis{
\cZ_{Pi}\cZ_{iQ}= \cZ_{PiQ}\,,\qquad \cZ_{AiB} =
(-1)^\len{A}\cZ_{i(\tilde A \shuffle B)}\,.
}
The second identity follows from \shufcalZ\ and implies that the basis
dimension of Hamilton paths $\cZ_{P}$ is $(\len{P}-1)!$.

\subsec Integration orders for the simpset elements
\par\subseclab\secfourB

\noindent In the simpset basis of integrals attained through the algorithm
\changebasis, we can now complete the definition of the $\int^\eom$ map in \fullBG.
For each simpset element, the algorithm to be described in this section identifies
at least one integration order for which the regularized\foot{When the disk
integrals do not contain any kinematic poles, the Taylor expansion of the
Koba--Nielsen factor results in convergent integrals where all integration orders
are equivalent.} integrals involving the Koba--Nielsen factor \knfinal\ are
observed to yield the correct $Z$-theory equation of motion.

It should be emphasized once more that the order of integration must not be
confused with the integration domain in \fullBG\ which is always fixed to be $0\leq
z_2\leq z_3\leq \ldots \leq z_{p-1} \leq 1$. Instead, ``order of integration''
refers to the decision whether an iterated integral over $z_2,z_3$ subject to
$0\leq z_2\leq z_3\leq 1$ is represented as $\int^1_0 \dd z_3 \int^{z_3}_0 \dd z_2$
or as $\int^1_0 \dd z_2 \int^{1}_{z_2} \dd z_3$. In the first case, the integration
over $z_2$ is performed first, and we will write $23$, whereas the opposite
integration order will be referred to through the shorthand $32$, with obvious
generalization to higher multiplicity.

\subsubsec Description of the algorithm

Let us introduce a formal operator ``$\ord$'' that takes as input a product of
$z_{ij}$ from the denominators in the simpset basis and outputs a combination of
words encoding the admissible integration orders. For example, $\ord(z_{12}z_{13}) = 23$ 
for the integrand in \noparity\ means that $\int^\eom$ requires the integral over $z_2$ to be
performed first, followed by $z_3$.

In order to describe a recursive algorithm to determine the order of integration, 
we associate a graph to each element in the simpset basis where each factor of
$z_{ij}$ contributes an edge between vertices $i$ and $j$. Then, $\ord(\ldots)$ for 
a given element of the simpset basis can be obtained by repeated application of
two steps:
\medskip
\item{1.}
If the graph of $z_{a_1a_2} \ldots z_{a_na_{n+1}}= (z_{b_1b_2} \ldots z_{b_pb_{p+1}})(z_{c_1c_2} \ldots z_{c_qc_{q+1}})$ 
is not connected (i.e.\ if $b_i\neq c_j$ $\forall \ i,j$), apply $\ord( \ldots)$ to each of its
connected subgraphs representing $z_{b_1b_2} \ldots z_{b_pb_{p+1}}$ as well as $z_{c_1c_2} \ldots z_{c_qc_{q+1}}$
and shuffle the resulting words,
\eqn\secord{
\ord( z_{a_1a_2} \ldots z_{a_na_{n+1}}) = \ord(z_{b_1b_2} \ldots z_{b_pb_{p+1}} ) \shuffle \ord( z_{c_1c_2} \ldots z_{c_qc_{q+1}} )\,.
}
The shuffle between the ordered sequences $ijk \ldots$ generated by the individual
$\ord(\ldots)$ operators indicates that the associated integrations commute, e.g., $23\shuffle 4$ means
that any integration order among $234, 243, 423$ is allowed.
\item{2.}
If the element
$z_{a_1a_2} \ldots z_{a_n a_{n+1}} z_{ij}$ is represented by a connected graph where
$z_{ij}$ is the factor with maximal separation $|i-j|$, and $j$ corresponds
to the integration variable that has not yet been pulled out of $\ord(\ldots)$, then
\eqn\firstord{
\ord(z_{a_1a_2} \ldots z_{a_n a_{n+1}} z_{ij}) = \ord(z_{a_1a_2} \ldots z_{a_n
a_{n+1}})j\,.
}
By design of the algorithm, only one of $i$ or $j$ can correspond
to an integration variable that has not yet been pulled out of $\ord(\ldots)$.

\medskip
\noindent For example, consider the element $z_{12}z_{35}z_{36}z_{46}$ from the
seven-point simpset basis where $z_{12}$ is associated with a disconnected subgraph. 
The first step splits $\ord( \ldots)$
according to its connected components, and iterating the algorithm above yields,
\eqnn\ordexample
$$\eqalignno{
\ord(z_{12}z_{35}z_{36}z_{46}) &= \ord(z_{12})\shuffle
\bigl(\ord(z_{35}z_{36}z_{46})\bigr)
= 2 \shuffle \bigl(\ord(z_{35}z_{46})3\bigr) &\ordexample\cr
&= 2\shuffle \Bigl(\bigl(\ord(z_{35})\shuffle \ord(z_{46})\bigr)3\Bigr)
= 2\shuffle \bigl((5\shuffle 4)3\bigr) \,.
}$$
The above ordering means that any permutation of $2345$ such that
$4$ and $5$ appear before $3$ (e.g.\ $5243$) defines a viable integration order,
while the position of $2$ is arbitrary.

\subsubsec Examples

Let us list the outcomes of the above algorithm for a few elements. At four
points, the two-dimensional basis has a unique order:
\eqn\fourR{
\ord(z_{12}) = 2 \, ,\quad \ord(z_{23}) = 2\,.
}
At five points, the six-dimensional simpset basis requires the following integration orders:
\medskip
\eqnn\fiveR
\settabs \+ \hskip 4.1cm \hfill & \hskip 4.1 cm \hfill & \hskip 6.1cm & \cr
\+  $\ord(z_{12}z_{13}) = 23$ ,&
    $\ord(z_{12}z_{34}) = 2\shuffle 3$ ,&
    $\ord(z_{23}z_{24}) = 32$ , \hskip1.8cm\fiveR\cr
\+  $\ord(z_{13}z_{23}) = 23$ ,&
    $\ord(z_{13}z_{24}) = 2\shuffle 3$ ,&
    $\ord(z_{24}z_{34}) = 32$ .\cr
\medskip
\noindent
At six points, the order for twelve simpset basis elements is given by
\medskip
\eqnn\sixvuf
\settabs \+ \hskip 4.3cm \hfill & \hskip 4.5 cm \hfill & \hskip 6.1cm & \cr
\+ $\ord(z_{12}z_{13}z_{14}) = 234$ ,&
   $\ord(z_{12}z_{34}z_{14}) = (2\shuffle 3)4$ ,&
   $\ord(z_{23}z_{24}z_{14}) = 324$ , \hskip0.55cm\sixvuf\cr
\+ $\ord(z_{13}z_{23}z_{14}) = 234$ ,&
   $\ord(z_{13}z_{24}z_{14}) = (2\shuffle 3)4$ ,&
   $\ord(z_{24}z_{34}z_{14}) = 324$ , \cr
\+ $\ord(z_{12}z_{13} z_{45}) = 23\shuffle 4$ ,&
   $\ord(z_{14}z_{24} z_{35}) = 24\shuffle 3$ ,&
   $\ord(z_{13}z_{14} z_{25}) = 34\shuffle 2$ ,\cr
\+ $\ord(z_{13}z_{23} z_{45} ) = 23\shuffle 4$ ,&
   $\ord(z_{12}z_{14} z_{35}) = 24\shuffle 3$ ,&
   $\ord(z_{14}z_{34} z_{25}) = 34\shuffle 2$ ,\cr
\medskip
\noindent while the integration order for the remaining twelve integrals are
obtained from worldsheet parity $z_j \rightarrow z_{6-j}$, e.g.\
$\ord(z_{25}z_{35}z_{45}) = 432$. The integration orders of the simpset basis at
seven points are explicitly listed in appendix \orderapp.

\subsubsec Iterated integrals and integration order

The above algorithm generates the allowed integration orders for all
the $(n{-}2)!$ elements in the simpset basis, with $p=n{-}1$ in the $Z$-theory 
equation of motion \fullBG. Since the integration domain is always 
$0\le z_2\le z_3 \ldots\le z_{p-1}\le 1$, one can show that the resulting words
$\ord(z_{i_1 j_1} z_{i_2 j_2}\ldots z_{i_k j_k}) = a_1 a_2\ldots a_k$ translate
into the following iterated integrals
\eqn\orddomain{
\int^\eom\! \! {1 \over z_{i_1 j_1} z_{i_2 j_2} \ldots
z_{i_k j_k}} =
\reg \int \limits^1_0 \dd z_{a_k} \int \limits^{c_{k-1}}_{b_{k-1}} \dd z_{a_{k-1}} \ldots \int \limits^{c_2}_{b_2}
\dd z_{a_2} \int \limits^{c_1}_{b_1} \dd z_{a_1}  \, {1 \over z_{i_1 j_1} z_{i_2 j_2} \ldots
z_{i_k j_k}}\, ,
}
with lower limits $b_j \equiv {\rm max}\{x \in \{0,z_{a_{j+1}},z_{a_{j+2}},\ldots ,z_{a_k}\}\, | \, x\leq z_{a_j}\}$
as well as upper limits $c_j \equiv {\rm min}\{x \in \{1,z_{a_{j+1}},\ldots ,z_{a_k}\}\, | \, x\geq z_{a_j}\}$.

\subsec Summary and overview example
\par\subseclab\secfourD

As discussed in the previous subsections, the $\int^\eom$ map converts the integrands
in the $Z$-theory equation of motion \fullBG\ to series expansions in derivatives and MZVs by:
\medskip
\item{(i)} changing the basis of integrals to the simpset basis through the
algorithm in \changebasis

\item{(ii)} determining the integration orders $\ord(\ldots)$ for simpset
denominators through the algorithm in \secord\ and \firstord

\item{(iii)} applying the regularization techniques of section \secfourC\
to perform the integrals \orddomain\ with Koba--Nielsen insertions \knfinal

\medskip
\noindent
The above steps will be illustrated through a simple yet representative example
\eqn\longex{
\prod^5_{i<j}|z_{ij}|^{\ap s_{ij}} \langle V_1V_{5243}V_6\rangle\,{\cal Z}_{5243} \ \longleftrightarrow \
\prod^5_{i<j}|z_{ij}|^{\ap \partial_{ij}} [ \Phi_1,\Phi_{5243} ] \,{\cal Z}_{5243}
}
taken from the six-point open superstring amplitude and the $\Phi^5$-order of the
$Z$-theory equation of motion \fullBG, respectively. We focus on the term
proportional to $\ap \partial_{12}G(0;z_2)$ in the expansion \knfinal\ of the
Koba--Nielsen factor to order $\ap$. This example was chosen because it touches all
the subtle points of the regularization prescription in section~\secfourC. The
calculations are long and tedious to perform by hand, but they are straightforward
to automate in a computer\foot{We are releasing our code that performs this task
via \BGapwww. The evaluation of the $8!=40.320$ integrals in the $10$-point simpset
basis to their leading order $\sim \zeta_7 , \, \zeta_2 \zeta_5 , \, \zeta_2^2
\zeta_3$ takes about two hours on a laptop. The program is written in FORM \FORM,
and improvements to the code are certainly possible and highly welcomed.}.

The chain basis element ${\cal Z}_{5243}$ under discussion also belongs to the simpset basis 
with $\ord(z_{52}z_{24}z_{43}) = 342$. Hence, the $\int^\eom$ map instructs to 
evaluate the regularized integral
\eqn\needEval{
\ap^4  \partial_{12}  [ \Phi_1,\Phi_{5243} ] \ \longleftrightarrow \
\reg \int_0^1 \dd z_2\int_{z_2}^1\dd z_4\int_{z_2}^{z_4}\dd z_3
{G(0;z_2)\over z_{52}z_{24}z_{43}}\,,
}
where the reference to the shuffle regularization scheme \regwone\
and \EriksReg\ via ``reg'' will be left implicit in the
remainder of this section. The integration limits in \needEval\ associated
to the order $\ord(z_{52}z_{24}z_{43}) = 342$ follow from \orddomain.
Rewriting $\int_{z_i}^{z_j}=\int_0^{z_j} - \int_{0}^{z_i}$ yields four
integrals, where integration over $z_3$ leads to
\eqnn\fourints
$$\eqalignno{
+\int_0^1 {\dd z_2 \over z_{52}}\int_{0}^1 { \dd z_4 \over z_{24} }  \int_{0}^{z_4}\dd z_3 {G(0;z_2)\over
 z_{43}} &=
-\int_0^1 {\dd z_2 \over z_{52}} \int_{0}^1 { \dd z_4 \over z_{24} } G(0;z_2)G(z_4;z_4)  &\fourints\cr
-\int_0^1{\dd z_2 \over z_{52}}\int_{0}^1{ \dd z_4 \over z_{24} }\int_{0}^{z_2}\dd z_3{G(0;z_2) \over
 z_{43}}
&= +\int_0^1{\dd z_2 \over z_{52}}\int_{0}^1{ \dd z_4 \over z_{24} } G(0;z_2)G(z_4;z_2) \cr
- \int_0^1{\dd z_2 \over z_{52}}\int_{0}^{z_2}{ \dd z_4 \over z_{24} }\int_{0}^{z_4}\dd z_3{G(0;z_2) \over
 z_{43}}
&= +\int_0^1{\dd z_2 \over z_{52}}\int_{0}^{z_2}{ \dd z_4 \over z_{24} } G(0;z_2)G(z_4;z_4) \cr
+ \int_0^1{\dd z_2 \over z_{52}}\int_{0}^{z_2}{ \dd z_4 \over z_{24} }\int_{0}^{z_2}\dd z_3{G(0;z_2) \over
 z_{43}}
&=-\int_0^1{\dd z_2 \over z_{52}}\int_{0}^{z_2}{ \dd z_4 \over z_{24} } G(0;z_2)G(z_4;z_2) \,.
}$$
We stress that the shuffle regularization to use
in the first and third integrals is \EriksReg\ with $\d=0$ since $G(z_4;z_4)$ is obtained after
integration over $z_3$ (subject to $z_3<z_4$),
$$
G(z_4;z_4) = - G(0;z_4)\,.
$$
In addition, in order to integrate over $z_4$, the polylogarithm $G(z_4;z_2)$
needs to be rewritten using the general $z$-removal identities, in particular
\zrmCtetwo,
\eqn\genzrm{
G(z_4;z_2) = G(z_2;z_4) + G(0;z_2) - G(0;z_4) - i\pi\,.
}
After the above considerations, the integrals over $z_4$ yield
\eqnn\firzf
\eqnn\seczf
\eqnn\thrzf
\eqnn\fthzf
$$\eqalignno{
-\int_0^1 {\dd z_2 \over z_{52}} \int_{0}^1 { \dd z_4 \over z_{24} } G(0;z_2)G(z_4;z_4) 
&= -\int_0^1 {\dd z_2\over z_{52}} G(0;z_2)G(z_2,0;1) &\firzf\cr
+\int_0^1{\dd z_2 \over z_{52}}\int_{0}^1{ \dd z_4 \over z_{24} } G(0;z_2)G(z_4;z_2) 
&=\int_0^1{\dd z_2\over z_{52}} G(0;z_2) &\seczf\cr
&\hskip-80pt \times \big(G(z_2,0;1) - G(z_2,z_2;1)
+ i\pi G(z_2;1) - G(0;z_2)G(z_2;1)\big)\cr
+\int_0^1{\dd z_2 \over z_{52}}\int_{0}^{z_2}{ \dd z_4 \over z_{24} } G(0;z_2)G(z_4;z_4) 
&=\int_0^1{\dd z_2\over z_{52}} G(0;z_2) &\thrzf\cr
&\hskip-80pt \times \bigl( i\pi\, G(0;z_2) - G(0;z_2)G(0;z_2) - G(0,z_2;z_2)\bigr)\cr
- \int_0^1{\dd z_2 \over z_{52}}\int_{0}^{z_2}{ \dd z_4 \over z_{24} } G(0;z_2)G(z_4;z_2) 
&=\int_0^1{\dd z_2 \over z_{52}} G(0;z_2)  &\fthzf\cr
&\hskip-80pt \times \bigl(
G(z_2,z_2;z_2) - i\pi\,G(z_2;z_2) - G(z_2,0;z_2) + G(0;z_2)G(z_2;z_2)
\bigr)\,.
}$$
As an important distinction from the previous integration (over $z_3$), the present
divergent polylogarithms of the form $G(z_2, \ldots;z_2)$ were generated after
integration over $z_4$, where the endpoint divergence is approached from above by
$z_4> z_2$.  Hence, the shuffle regularization in this case requires $\d=1$ in
\EriksReg, and the techniques of appendix D.3 imply
\eqnn\deltaOne
$$\eqalignno{
G(z_2,z_2;z_2) &= \half\big(-G(0;z_2) + i\pi\big)\big(-G(0;z_2) + i\pi\big) &\deltaOne\cr
G(z_2,0;z_2) &=\big(-G(0;z_2) + i\pi\big)G(0;z_2) +\zeta_2 \cr
G(z_2;z_2) &= -G(0;z_2) + i\pi \, ,
}$$
using $G(0,z_2;z_2)=- \zeta_2$ by \zrmlenthree.
It is interesting to observe that the last line of \fthzf\ becomes
$\half G(0;z_2)^2 + G(0,z_2;z_2)+\half\pi^2$, where the term
$\pi^2$ can be traced back to an interplay between two subtle factors of $i\pi$
from very distinct sources: one from the general $z$-removal identity \genzrm\
and the other from the $\d=1$ shuffle regularization \EriksReg\foot{Fortunately,
the independent proposal for the regularized value for the integral \needEval\
inspired by the methods of \Polylogs\ and described in the appendix~\Jregapp\
allowed us to fix all these subtleties.
This ultimately led us to our final regularization prescription that has
ever since passed many tests at much higher order in $\ap$.}.

In addition to the above shuffle regularizations, the following $z$-removal identities
based on $G(0;1) = G(0,0;1) = 0$ are needed to perform the final integration over $z_2$:
\eqnn\finalids
$$\eqalignno{
G(z_2,z_2;1) &=
\half\Big(G(0;z_2)^2 + G(1;z_2)^2 -\pi^2 \Big)
- G(0;z_2)G(1;z_2) + i\pi \Big( G(1;z_2) - G(0;z_2)\Bigr) \cr
G(z_2,0;1) &=
2\zeta_2 + i\pi G(0;z_2) - G(0,0;z_2) + G(0,1;z_2)  \cr
G(0,z_2;1) &=
- 2\zeta_2 -i\pi G(0;z_2) + G(0,0;z_2)- G(0,1;z_2) &\finalids\cr
G(z_2;1) &=
G(1;z_2) - G(0;z_2) + i\pi\ .
}$$
In combination with the shuffle algebra \shufalg, the identities in \finalids\ yield the
following results for the remaining integral over $z_2$ (setting $z_5=1$):
\eqnn\intresults
$$\eqalignno{
\firzf
&= \half \zeta_2^2
       - 2 i\pi \zeta_3\,,\qquad
\seczf = {17\over 10} \zeta_2^2\,, &\intresults\cr
\thrzf & = {7\over 5}\zeta_2^2
       + 2 i\pi \zeta_3\,,\qquad
\fthzf  =- {16\over 5} \zeta_2^2\,.
}$$
Finally, summing the above results yields the regularized value
of the integral \needEval,
\eqn\finally{
\reg
\int_0^1 \dd z_2\int_{z_2}^1\dd z_4\int_{z_2}^{z_4}\dd z_3{G(0;z_2)\over
z_{52}z_{24}z_{43}}= {2 \over 5} \zeta_2^2 = \zeta_4\,.
}
Using the prescription \fullBG, this implies that the $Z$-theory equation of
motion contains the term $-\ap^4 \zeta_4\p_{12}[\Phi_1,\Phi_{5243}]$,
in agreement with the Berends--Giele recursion at order $\ap^4$ previously
obtained from an ansatz.

\newsec Closed-string integrals
\par\seclab\secfive

\noindent Our results have a natural counterpart for closed-string scattering,
where tree-level amplitudes involve integrals over worldsheets of sphere topology.
Similar to the characterization of disk integrals \Zintdef\ via two cycles $P$ and
$Q$, any sphere integral in tree-level amplitudes of the type II
superstring\foot{The same kind of organization in terms of \ZZclosed\ is expected
to be possible in tree-level amplitudes of the heterotic string and the bosonic
string. This would imply the universality of gravitational tree-level
interactions in these theories whenever their order of $\ap$ ties in with the weight of
the accompanying MZV \HuangTAG.} \SchlottererNY\ boils down to
\eqn\ZZclosed{
W(P|Q) \equiv 
\left({\ap \over \pi} \right)^{n-3}
\int \limits_{\Bbb C^n} {\dd^2 z_1 \, \dd^2 z_2 \cdots \dd^2 z_n \over {\rm vol}(SL(2,\Bbb C))}
\prod_{i<j}^n |z_{ij}|^{\ap s_{ij}} \, C(P) \, \bar C(Q) \,.
}
The inverse volume of the conformal Killing group $SL(2,\Bbb C)$ of the sphere
generalizes \volsl\ in an obvious manner, and
$\bar C(Q)$ denotes the complex conjugate of the chain \Zposdef\ of
worldsheet propagators with $z_{ij}\rightarrow \bar z_{ij}$.

While the field-theory limit of the sphere integrals \ZZclosed\ yields the
same doubly partial amplitudes as the
corresponding disk integrals \StiebergerHBA,
\eqn\closedA{
m(A,n|B,n)=\lim_{\alpha' \rightarrow 0}  W(A,n|B,n) \, ,
}
only a subset of the $\ap$-corrections in $Z(P|Q)$ can be found in the closed
string \ZZclosed. These selection rules obscured by the KLT relations \KLT\ have
been identified to all orders in \SchlottererNY\ and realize the single-valued
projection ``sv'' \SchnetzHQA\ of the MZVs in the disk integrals
\refs{\StiebergerWEA, \StiebergerHBA}
\eqn\closedB{
W(P|Q) = {\rm sv} \big[ Z(P|Q) \big] \ .
}
The single-valued map projects Riemann zeta values to their representatives
of odd weights, ${\rm sv}(\zeta_{2n})=0$ 
and ${\rm sv}(\zeta_{2n+1})=2\zeta_{2n+1}$, and acts on MZVs \MZVdef\ of
depth $r\geq 2$ in a manner explained in \SchnetzHQA. As an immediate consequence of \closedB, the
Berends--Giele representation
\eqn\closedC{
W(A,n|B,n) 
 = s_A  \, {\rm sv} \big[\phi_{A|B} \big]\,,
}
of closed-string integrals can be derived from the same currents $\phi_{A|B}$
which govern the disk integrals via \ZapBG. Hence, any tentative ``single-valued
$Z$-theory'' defined by reproducing the closed-string integrals \ZZclosed\ as its
doubly partial amplitudes is necessarily contained in the non-abelian $Z$-theory
of this paper. 

Note that reality of the sphere integrals $W(P|Q)$ along with the phase-space constraint
$s_A=0$ for $n$ on-shell particles with $P=(A,n)$ implies that single-valued currents obey the following on-shell properties
\eqn\onshell{
{\rm sv} \big[\phi_{A|B} \big]={\rm sv} \big[\phi_{B|A} \big] + {\cal O}(s_A)
\ , \ \ \ \
{\rm sv} \big[\phi_{P\shuffle Q|B} \big]=  {\cal O}(s_A)\, .
} 
Hence, one can perform field redefinitions such as to render the associated perturbiner 
${\rm sv}[\Phi]$ Lie-algebra valued in both gauge groups.


\newsec Conclusions and outlook
\par\seclab\secsix

\noindent We have proposed a recursive method to calculate the
$\ap$-expansion of disk integrals present in the massless $n$-point tree-level
amplitudes of the open superstring \refs{\nptTreeI,\nptTreeII}. As a backbone of this
method, the disk integrals themselves are interpreted as the tree amplitudes in an
effective field theory of bi-colored scalars $\Phi$, dubbed as $Z$-theory in
previous work \abZtheory. Its equation of motion \fullBG\ furnishes the central
result of this work and compactly encodes the Berends--Giele
recursions that elegantly compute the $\ap$-expansions of the disk
integrals at arbitrary multiplicity. More precisely, the $Z$-theory equation of motion
\fullBG\ is satisfied by the perturbiner series of the
Berends--Giele currents, and its structure is shared by an $(n{-}2)!$-term
representation of the $n$-point open-string tree amplitude derived in \nptTreeI.

As a practical result of this work, the BG recursion relations for disk integrals
$Z(P|Q)$ with any given words $P$ and $Q$ of arbitrary multiplicity is
made publicly available up to order $\ap^7$ in a {\tt FORM}
\FORM\ program called {\tt BGap}. In order to ease replication, the auxiliary
computer programs used in the derivation of the BG recursion via regularized
polylogarithms are also available to download on the website \BGapwww.

As a conceptual benefit of this computational achievement, the Berends--Giele
description of disk integrals sheds new light on the double-copy structure of the
open-string tree-level S-matrix \Polylogs. As manifested by \Zthyorigin, disk
amplitudes exhibit a KLT-like factorization into SYM amplitudes and disk integrals
$Z(P|Q)$. Following the interpretation of $Z(P|Q)$ as $Z$-theory amplitudes \abZtheory, the perturbiner
description of the Berends--Giele recursion for disk integrals pinpoints the field
equation \fullBG\ of $Z$-theory. Hence, our results give a more precise definition of 
$Z$-theory, the second double-copy component of open superstrings.

\bigskip
\noindent{\it 6.1 Further directions}
\medskip

\noindent To conclude, we would like to mention an incomplete selection of the
numerous open questions raised by the results of this work.

The non-linear equation of motion \fullBG\ of $Z$-theory gives rise to wonder about
a Lagrangian origin.  Moreover, the form of \fullBG\ is suitable for
(partial) specialization to abelian generators in gauge group of the integration domain.  Hence,
we will explore the implications of our results for the $\ap$-corrections to the
NLSM \abZtheory\ as well as mixed $Z$-theory amplitudes involving both bi-colored
scalars and NLSM pions in future work \inprog.

Do worldsheet integrals over higher-genus surfaces admit a similar interpretation
as $Z$-theory amplitudes? It might be rewarding to approach the low-energy
expansion of superstring loop amplitudes at higher multiplicity with 
Berends--Giele methods. At the one-loop order, this concerns annulus integrals
involving elliptic multiple zeta values \BroedelVLA\ and torus integrals involving
modular graph functions \DHokerLBK.

Is there an efficient BCFW description of $Z$-theory amplitudes? Given that BCFW on-shell 
recursions \BrittoFQ\ can in principle be applied string amplitudes \BoelsBV,
it would be interesting to relate the Berends--Giele recursion for $Z$-theory
amplitudes to BCFW methods.

Furthermore, what are the non-perturbative solutions to
the full $Z$-theory equation of motion \fullBG?
A non-perturbative solution to the field equation $\Box\Phi=\Phi^2$
of bi-adjoint scalars (obtained from the field-theory
limit $\ap\to0$) has been recently found \white\
in an attempt to understand the non-perturbative regime of the double-copy
construction.

In addition, is it possible to obtain field equations or effective actions for 
massless open- or closed-superstring states along similar lines of \fullBG? In
order to approach the $\ap$-corrections to the SYM action, the
resemblance of such an equation of motion with the Berends--Giele description of
superfields in pure spinor superspace \refs{\EOMBBs, \Gauge} is intriguing. This
parallel might for instance be useful in generating the $\ap$-corrections to the 
on-shell constraint $\{\nabla_{\a},\nabla_{\b}\}-\gamma^m_{\alpha \beta}
\nabla_m=0$ of ten-dimensional SYM \howeSYM. 

Related to this, it would be desirable to express the $Z$-theory equation of motion
and tentative corollaries for superstring effective actions in terms of the
Drinfeld associator.  Given that disk integrals in a basis \defF\ of $F_{P}{}^Q$
have been recursively computed from the associator \BroedelAZA, we expect that
suitable representations of its arguments allow to cast the $\ap$-expansion of the
Berends--Giele recursion into a similarly elegant form. One could even envision to
generate the tree-level effective action of the open superstring from the SYM
action by acting with appropriate operator-valued arguments of the associator.

Finally, a rigorous mathematical justification for the various prescriptions
used in ``converting'' the open string amplitude \VVVn\ to the $Z$-theory
equation of motion was not the subject of this paper but clearly deserves
further investigation. In particular, it seems mysterious to us at this point why the $Z$-theory setup
selects the regularization scheme for $G(0;z),\, G(z;z)$, the
integration orders, and the change of basis presented in section \regpoly.

\bigskip \noindent{\bf Acknowledgements:} We are grateful to Johannes Br\"odel,
John Joseph Carrasco and Ellis Yuan for combinations of valuable discussions and fruitful 
collaboration on related
topics. We are indebted to Erik Panzer for indispensable email exchange and
numerous enlightening discussions, in particular for guidance on the subtleties of
polylog regularization and different orders of integration.
Also, we would like to thank Johannes Br\"odel and Erik Panzer
for helpful comments on an initial draft. The authors would like to thank IAS at Princeton where this work was
initiated as well as Nordita and in particular Paolo Di Vecchia and Henrik Johansson for providing
stimulating atmosphere, support and hospitality through the ``Aspects of
Amplitudes'' program. CRM thanks the Albert-Einstein-Institut in Potsdam
for hospitality during the final stages of this work.
CRM is supported by a University Research
Fellowship from the Royal Society, and gratefully acknowledges
support from NSF
grant number PHY 1314311 and the Paul Dirac Fund during the initial phase
of this work.

\appendix{A}{Symmetries of Berends--Giele double currents}
\applab\appshuff\

\noindent In this appendix we discuss the symmetries obeyed by the Berends--Giele
double currents.

\medskip
\noindent{\it A.1 Shuffle symmetry}
\medskip

\noindent In order to make sure that our ansaetze for BG currents \ZapBG\ for disk integrals satisfy the
shuffle-symmetry $\phi_{A|P\shuffle Q}=0$, we will need the generalization
of the result proven in the appendix of \Gauge. That is,
in a deconcatenation (into non-empty words $X_i$) of the form
\eqn\phideconc{
\phi_P = \sum_{X_1X_2=P} H_{X_1,X_2} + \sum_{X_1X_2X_3=P} H_{X_1,X_2,X_3}
+ \sum_{X_1X_2X_3X_4=P} H_{X_1,X_2,X_3,X_4} + \cdots \ ,
}
if $H_{X_1,X_2, \ldots, X_n}$ satisfies shuffle symmetries within each individual slot
and collectively on all the slots
(treating each $X_i$ as a single letter)
\eqn\indiv{
H_{X_1,X_2, \ldots, A\shuffle B, \ldots, X_n} = 0\,,\qquad
H_{(X_1,X_2, \ldots,X_j) \shuffle (X_{j+1}, \ldots,X_n)} = 0 \,,\qquad j=1,2,\ldots,n-1\,,
}
then $\phi_P$ in \phideconc\ is expected to satisfy the shuffle symmetry for words of arbitrary length,
\eqn\theclaim{
\phi_{R\shuffle S} = 0, \quad \forall \ R,S\neq\emptyset \, .
}
It would be interesting to rigorously derive the symmetry in \theclaim\ from the properties \indiv\ of the
deconcatenations in \phideconc, possibly along the lines of the appendix of \Gauge.

\medskip
\noindent{\it A.2 Generalized Jacobi symmetry}
\medskip

\noindent
The definition of $T^{B_1,B_2, \ldots ,B_n}_{A_1,A_2, \ldots ,A_n}$
in \TBdef\ implies the shuffle symmetries \shuffleB\
in the $B_j$-slots at fixed ordering of the
$A_j$-slots. This raises the question about the dual
symmetry properties when the $A_j$-slots are permuted
at a fixed ordering of the $B_j$-slots. For this purpose
it is convenient to use
the left-to-right Dynkin bracket mapping $\ell$ defined by $\ell(A_1)=A_1$
and \refs{\Ree,\reutenauer},
\eqn\Dynkindef{
\ell(A_1,A_2, \ldots ,A_n) = \ell(A_1,A_2, \ldots, A_{n-1}),A_n - A_n,\ell(A_1,A_2, \ldots ,A_{n-1})
}
such as $\ell(A_1,A_2) = (A_1,A_2) - (A_2,A_1)$ and $\ell(A_1,A_2,A_3) = (A_1,A_2,A_3) - (A_2,A_1,A_3)
 - (A_3,A_1,A_2) + (A_3,A_2,A_1)$. One can show that \Dynkindef\ projects to the symmetries of
 nested commutators with
\eqn\genJac{
\ell([[ \ldots[[A_1,A_2],A_3] \ldots ],A_n]) = n [[ \ldots[[A_1,A_2],A_3] \ldots ],A_n]\,.
}

\smallskip
\proclaim Lemma 1. The object $T^{B_1,B_2, \ldots ,B_n}_{A_1,A_2, \ldots ,A_n}$ defined by \TBdef\ satisfies
the generalized Jacobi symmetries in the $A_j$-slots, i.e.\ the symmetries of nested commutators
\eqn\equivSym{
T^{B_1,B_2, \ldots, B_n}_{A_1, A_2, \ldots, A_n}\quad \longleftrightarrow\quad [ \ldots
[[A_1,A_2],A_3], \ldots, A_n]
}
such as $T^{B_1,B_2}_{A_1,A_2 }=-T^{B_1,B_2}_{A_2,A_1}$
and $T^{B_1,B_2,B_3}_{A_1,A_2,A_3}+T^{B_1,B_2,B_3}_{A_2,A_3,A_1}+T^{B_1,B_2,B_3}_{A_3,A_1,A_2}=0$.

\proof According to \genJac\ it suffices to show that
\eqn\needshow{
T_{\ell(A_1,A_2, \ldots,A_n)}^{B_1,B_2, \ldots,B_n} = n
T_{A_1,A_2, \ldots,A_n}^{B_1,B_2, \ldots,B_n}\,,
}
which in turn follows from 
\eqn\firstshow{
T^{B_1,B_2, \ldots, B_n}_{A_1,A_2, \ldots, A_n} = T^{\rm dom}_{\ell(A_1,A_2, \ldots, A_n)}\otimes T^{\rm int}_{B_1,B_2, \ldots, B_n} \ ,
}
since the Dynkin bracket satisfies $\ell^2 (A_1,\ldots,A_n) = n \ell(A_1,\ldots,A_n)$ \reutenauer.
One can conveniently verify \firstshow\ by induction:
\eqnn\TAdef
$$\eqalignno{
&T^{\rm dom}_{\ell(A_1,A_2, \ldots, A_n)}\otimes T^{\rm int}_{B_1,B_2, \ldots, B_n}
= \bigl(
T^{\rm dom}_{\ell(A_1,A_2, \ldots,  A_{n-1}),A_n}
- T^{\rm dom}_{A_n,\ell(A_1, A_2, \ldots A_{n-1})}\bigr)
\otimes T^{\rm int}_{B_1, B_2, \ldots, B_{n}}\cr
& \ \ \ \ =  \phi_{A_n|B_n} 
T^{\rm dom}_{\ell(A_1,A_2, \ldots,  A_{n-1})}
\otimes T^{\rm int}_{B_1, B_2, \ldots, B_{n-1}}
- \phi_{A_n|B_1}  T^{\rm dom}_{\ell(A_1, A_2, \ldots A_{n-1})} 
\otimes T^{\rm int}_{B_2, \ldots, B_{n}}
\cr
& \ \ \ \ =
 \phi_{A_n|B_n} T^{B_1,B_2, \ldots, B_{n-1}}_{A_1,A_2, \ldots, A_{n-1}}
- \phi_{A_n|B_1} T^{B_2,B_3, \ldots, B_{n}}_{A_1,A_2, \ldots, A_{n-1}}\,.
&\TAdef
}$$
In the first line, we apply the recursive definition \Dynkindef\ of the Dynkin bracket operator, followed by the
definition \tensP\ of the tensor product $T^{\rm dom}_{\ldots}
\otimes T^{\rm int}_{\ldots}$ in the second line. In passing to the third line, we have used
the inductive assumption, i.e.\ \firstshow\ at $n\rightarrow n-1$, and the resulting expression
can be identified with the recursive definition \TBdef\ of $T^{B_1,B_2, \ldots, B_n}_{A_1,A_2, \ldots, A_n}$
which finishes the proof.\hfill\qed

\medskip
Note that $\rho^2(A_1, \ldots,A_n) = n\rho(A_1,\ldots,A_n)$ \Ree\ and \genJac\ imply a duality
between the shuffle symmetry of the $B_j$ slots and the generalized Jacobi symmetry of the $A_j$ slots,
\eqn\Symduality{
T^{\rho(B_1,B_2, \ldots, B_n)}_{A_1,A_2, \ldots, A_n} = T^{B_1,B_2, \ldots,
B_n}_{\ell(A_1,A_2, \ldots, A_n)}\,.
}

\medskip
\noindent{\it A.3 Berends--Giele double current and nested commutators}
\medskip

\noindent As discussed above, the BG double current satisfies generalized Jacobi symmetries
within the $A_j$ slots. This means
that its expansion in terms of products of $\phi_{A_i|B_j}$ can be
written as linear combinations of $T_{A_1,\ldots,A_n}^{B_1, \ldots,B_n}$ as,
according to Lemma~1,
they encode the symmetries of nested commutators.
For example, the following
terms of order $\ap^2$ that multiply the factor $(k_{A_1}\cdot k_{A_2})$ in
\uptozetatwo
\eqn\dyex{
 \phi_{A_{1}|B_{1}}\phi_{A_{2}|B_{3}}\phi_{A_{3}|B_{2}}
- \phi_{A_{1}|B_{1}}\phi_{A_{2}|B_{2}}\phi_{A_{3}|B_{3}} 
+ \phi_{A_{1}|B_{3}}\phi_{A_{2}|B_{1}}\phi_{A_{3}|B_{2}}
- \phi_{A_{1}|B_{3}}\phi_{A_{2}|B_{2}}\phi_{A_{3}|B_{1}}
}
are equal to $T_{A_2,A_3,A_1}^{B_1,B_2,B_3}$.
This is easy to verify but hard to obtain when the expressions are large.
Fortunately, one can use an efficient algorithm due to
Dynkin, Specht and Wever (for a pedagogical account, see \thibon)
to find the linear combinations of $T_{A_1, \ldots,A_n}^{B_1, \ldots,B_n}$
that capture the products of $\phi_{A_i|B_j}$.
The solution exploits the fact that the Dynkin bracket $\ell$ gives
rise to a Lie idempotent; $\t_n\equiv \frac{1}{n}\ell(A_1, \ldots,A_n)$.
Therefore, rewriting each word of length $n$ within a Lie polynomial
as $\frac{1}{n}\ell(P)$ leads to the answer, e.g.,
$ab-ba = \half\ell(ab) - \half\ell(ba) = \ell(ab)$.

In order to apply this algorithm to products of $\phi_{A_i|B_j}$,
first rewrite its products such that the $B_j$ labels
are always in the same order $B_1 B_2 B_3$. For example, \dyex\ becomes,
\eqnn\dynext
$$\displaylines{
\phi_{A_{1}|B_{1}}\phi_{A_{3}|B_{2}}\phi_{A_{2}|B_{3}}
- \phi_{A_{1}|B_{1}}\phi_{A_{2}|B_{2}}\phi_{A_{3}|B_{3}}
+ \phi_{A_{2}|B_{1}}\phi_{A_{3}|B_{2}}\phi_{A_{1}|B_{3}}
- \phi_{A_{3}|B_{1}}\phi_{A_{2}|B_{2}}\phi_{A_{1}|B_{3}}\cr
\equiv L_1L_3L_2 - L_1L_2L_3 + L_2L_3L_1
- L_3L_2L_1\,, \hfil\dynext\hfilneg
}$$
where in the second line we used the shorthand notation
$\phi_{A_i|B_1}\phi_{A_j|B_2}\phi_{A_k|B_3} \equiv L_iL_jL_k$
with non-commutative variables $L_{\ldots}$.
Applying the idempotent operator $\t_n$ one obtains
$$\eqalignno{
\dynext &= {1\over 3}\ell(L_1,L_3,L_2)
-{1\over 3}\ell(L_1,L_2,L_3)
+ {1\over 3}\ell(L_2,L_3,L_1)
- {1\over 3}\ell(L_3,L_2,L_1)\cr
&=
-{1\over 3}\ell(L_1,\ell(L_2,L_3))
- {1\over 3}\ell(L_1,\ell(L_2,L_3))
+ {1\over 3}\ell(L_1,\ell(L_3,L_2))\cr
& = - \ell(L_1,\ell(L_2,L_3)) = \ell(L_2,L_3,L_1)\equiv T_{A_2,A_3,A_1}^{B_1,B_2,B_3} \, ,
}$$
where we used the property $\ell(a_1,a_2,i) = - \ell(i,\ell(a_1,a_2))$ \reutenauer.
This algorithm has been used to cast the $\ap$-expansion of the
BG double current in terms of the definition \TBdef.

\appendix{B}{Ansatz for the Berends--Giele recursion at higher order in $\ap$}
\applab\appANS

\noindent As explicitly tested up to and including order $\ap^4$, one arrives at a
unique recursion for the Berends--Giele double current $\phi_{A|B}$ that
reproduces, via \ZapBG, the disk integrals at various $\ap^{w\geq 2}$-orders by
imposing the following constraints on an ansatz of the form in \AnsatzZeta:
\medskip
\item{1.} adjusting the powers of momenta and fields to the mass dimensions
of the $\ap^w$-order
\item{2.} reflection symmetry in both slots $A$ and $B$ as well as shuffle symmetry
in the $B$ slot
\item{3.} absence of dot products $(k_{A_i} \cdot k_{B_j})$,
$(k_{B_i} \cdot k_{B_j})$ and $k_{A_i}^2$
\item{4.} absence of dot products $(k_{A_1} \cdot k_{A_p})$ referring to the outermost slots in $\sum_{A=A_1A_2 \ldots A_p}$
\item{5.} matching the order-$\ap^w$ recursion with known $n$-point disk integrals for all $n\le w+3$
\medskip
\noindent
By dimensional analysis and triviality of the three-point integral, the BG recursion
of the disk integrals at a given order is captured by the following number
of fields and derivatives,
$$
(\hbox{order $\ap^w$})  \ \leftrightarrow \ 
(k_{A_i} \cdot k_{A_j})^p \phi_{A_1|B_{i_1}} \phi_{A_2|B_{i_2}} \ldots
\phi_{A_{w+2-p} |B_{i_{w+2-p}}} \ , \ \ \ p=0,1,\ldots,w-1 \,,
$$
e.g.\ the ansatz of the form \AnsatzZeta\ for the $\alpha'^2 \zeta_2$-order generalizes to
three types of terms with schematic form $k^4 \phi^3,\ \, k^2 \phi^4,\ \, \phi^5$ along with $\alpha'^3 \zeta_3$,
$$
(\hbox{order $\ap^3$})  \ \leftrightarrow \ 
(k_{A_p}\cdot k_{A_q})(k_{A_r}\cdot k_{A_s}) \prod_{j=1}^3\phi_{A_j|B_{i_j}} \, , \ \ \ 
(k_{A_p}\cdot k_{A_q})\prod_{j=1}^4\phi_{A_j|B_{i_j}} \, , \ \ \ 
\prod_{j=1}^5\phi_{A_j|B_{i_j}} \, .
$$

\appendix{C}{Regular parts of five-point integrals}
\applab\appfivereg

\noindent The contributions to $\Box \Phi$ of order $\Phi^4$ in the fields are
governed by the $\alpha'$-expansion of regularized five-point integrals, see \pertFive. 
In the regularization scheme explained in section \regpoly, the relevant
leading orders are given by
\eqnn\fivereg
$$\eqalignno{
 \int^\eom  \prod_{i<j}^{4} |z_{ij}|^{\ap \partial_{ij}}\frac{ 1}{z_{12}z_{23}} &=  2 \ap\zeta_3(\p_{24}+\p_{34}) +{\cal O}(\ap^2) &\fivereg
\cr
 \int^\eom  \prod_{i<j}^{4} |z_{ij}|^{\ap \partial_{ij}}\frac{ 1}{z_{13}z_{32}}&=  -\ap \zeta_3 (3 \p_{24} + \p_{34}) +{\cal O}(\ap^2)
\cr
 \int^\eom  \prod_{i<j}^{4} |z_{ij}|^{\ap \partial_{ij}} \frac{ 1 }{z_{12}z_{34}}&=   -  \zeta_2 +\ap \zeta_3(\p_{12}+ 2 \p_{13} + 2 \p_{23} + 2 \p_{24} + \p_{34}) +{\cal O}(\ap^2)
\cr
 \int^\eom  \prod_{i<j}^{4} |z_{ij}|^{\ap \partial_{ij}} \frac{ 1 }{z_{13}z_{24}}&=    \zeta_2 + \ap \zeta_3( -2 \p_{12} - \p_{13} - 3 \p_{23} - \p_{24} - 2 \p_{34}) +{\cal O}(\ap^2)
\cr
 \int^\eom  \prod_{i<j}^{4} |z_{ij}|^{\ap \partial_{ij}}\frac{ 1}{z_{42}z_{23}} &=- \ap \zeta_3(\p_{12} + 3 \p_{13}) +{\cal O}(\ap^2)
\cr
 \int^\eom  \prod_{i<j}^{4} |z_{ij}|^{\ap \partial_{ij}}\frac{ 1}{z_{43}z_{32}} &= 2\ap \zeta_3(\p_{12} +  \p_{13})+{\cal O}(\ap^2) \ ,
}$$
while the terms at higher orders in $\ap$ can be found in the ancillary
files. Note that the integrals over $(z_{12}z_{23})^{-1}$ and
$(z_{43}z_{32})^{-1}$ have been assembled from the simpset basis \simpsetfive.

\appendix{D}{Multiple polylogarithm techniques}
\applab\appvarious

\medskip
\noindent{\it D.1 Polylogarithms and MZVs}
\medskip

\noindent Polylogarithms $G(a_1,a_2,\ldots,a_n;1)$ at unit argument with labels $a_i \in
\{0,1\}$ can be converted to MZVs via \Maxim\ provided that $a_1=0$ and $a_n=1$
prevent endpoint divergences. Divergent iterated integrals $G(1,\ldots;1)$ and
$G(\ldots,0;1)$ in this work will be shuffle-regularized based on the special cases
$G(1;1) = G(0;1) = 0$ of \regwone. At weight two and three, the appearance of
$\zeta_2$ and $\zeta_3$ in \regAB\ can be traced back to
\eqnn\Gregs
$$\displaylines{
 G(1,0;1) = + \zeta_2,\quad G(0,1;1) = - \zeta_2 \hfil\Gregs\hfilneg\cr
 G(1,0,0;1) = - \zeta_3,\quad G(0,1,0;1) = + 2 \zeta_3,\quad G(0,0,1;1) = - \zeta_3 \, \phantom{.}\cr
 G(1,1,0;1) = + \zeta_3,\quad G(1,0,1;1) = - 2 \zeta_3,\quad G(0,1,1;1) = + \zeta_3 \, .
}$$
The analogous higher-weight relations follow from \Maxim, while several identities
among MZVs can be found in \datamine\ (obtained using harmonic polylogarithms \harmonicP).

\medskip
\noindent{\it D.2 Methods for shuffle regularization}
\medskip

\noindent
By the shuffle algebra \shufalg, the regularized values \regwone\ and \EriksReg\ for weight-one cases $G(0;z)$
and $G(z;z)$ propagate to divergent multiple polylogarithms at higher weight, e.g.
\eqnn\lowershufreg
\eqnn\uppershufreg
$$\eqalignno{
G(A,a_{n-1},0;z) &= G(A,a_{n-1};z)G(0;z) - G(A\shuffle 0,a_{n-1};z) \ , \ \ \ \  a_{n-1} \neq 0
&\lowershufreg \cr
G(z,a_2,A;z) &= G(z;z)G(a_2,A;z) - G(a_2, z\shuffle A;z) \ , \ \ \ \ \ \ \ \ \ \ \ \ \ \ a_{2} \neq z \ .
&\uppershufreg
}$$
In case of \lowershufreg, $a_{n-1} \neq 0$ implies that $G(0;z)\equiv \ln|z|$ captures the entire endpoint divergence from the
lower integration limit. The same kind of shuffle operations including $G(0,0;z)={1\over 2} G(0;z)^2$ allows
to reduce cases with multiple terminal labels $0$ such as $G(A,a_{n-2},0,0;z)$ with $a_{n-2}\neq 0$ to
convergent polylogarithms and polynomials in $G(0;z)$ \PanzerIDA. Analogous statements based on
a regularization prescription for $G(z;z)$ can be made
for upper-endpoint divergences in integrals like $G(z,z,\ldots,z,a_k, \ldots,a_n;z)$ with $a_k\neq z$.

\medskip
\noindent{\it D.3 $z$-removal identities}
\medskip

\noindent 
The definition \Gdef\ of polylogarithms applies to situations where the
integration variable $z$ only appears on the right of the semicolon in
$G(a_1,a_2,\ldots,a_n;z)$, i.e.\ to labels $a_j \neq z$. This appendix is devoted to
integration techniques for polylogarithms with more general arguments, i.e.\ with
multiple appearances of the integration variable $z$ as $G(\ldots,z,\ldots;z)$ or
$G(\ldots,z,\ldots;b)$ with $b\neq z$. These techniques rely on rewritings such as
\Polylogs,
\eqnn\diffG
$$\eqalignno{
G(a_1, \ldots,a_{i-1},z,a_{i+1}, \ldots,a_n;z) &= \int_0^z
\dd t \, {\dd\over \dd t} \, G(a_1, \ldots,a_{i-1},t,a_{i+1}, \ldots,a_n;t)\cr
&{}  \ \ \ \ \ \ \ \ + c(a_1, \ldots,a_{i-1},\hat z,a_{i+1}, \ldots,a_n)\, ,  &\diffG
}$$
with appropriate initial value $c(a_1, \ldots,a_{i-1},\hat z,a_{i+1}, \ldots,a_n)$
at $z=0$. The total derivative 
in \diffG\ can be evaluated through the differential equations ($\hat a_j$ means
that $a_j$ is omitted)
\eqnn\diffone
$$\eqalignno{
  \frac{\p}{\p z}G(\vec{a};z)&=\frac{1}{z-a_1}G(a_2,\ldots,a_n;z)&\diffone\cr
\frac{\p}{\p a_i}G(\vec{a};z)&=
  \frac{1}{a_{i-1}-a_i}G(\ldots,\hat{a}_{i-1},\ldots;z)
  + \frac{1}{a_i-a_{i+1}}G(\ldots,\hat{a}_{i+1},\ldots;z)\cr
&{}+\Bigl({1\over a_i - a_{i-1}}
-{1\over a_i - a_{i+1}}\Bigr)G(\ldots,\hat{a}_i,\ldots;z)\,, \qquad i \neq 1,n \cr
\frac{\p}{\p a_n}G(\vec{a};z)&=\frac{1}{a_{n-1}-a_n}G(\ldots,\hat{a}_{n-1},a_n;z)
+ \Bigl({1\over a_n - a_{n-1}} - {1\over a_n}\Bigr)G(\ldots,a_{n-1};z) \ .
}$$

\medskip
\noindent{\it D.3.1 Simple $z$-removal identities}
\medskip

\noindent 
Let us first address the simpler subset of $z$-removal identities, where the integration variable is
present on both sides of the semicolon, i.e.\ cases of the schematic form $G(\ldots,z,\ldots;z)$. Inserting
the differential equations \diffone\ into \diffG\ recursively eliminates the variable $z$
from the labels \Polylogs,
\eqnn\recursG
$$\eqalignno{
&G(a_1,\ldots, a_{i-1},z,a_{i+1},\ldots,a_n;z) =c(a_1,\ldots,a_{i-1},\hat
z,a_{i+1},\ldots,a_n)&\recursG\cr
&{} +  G(a_{i-1},a_1,\ldots,a_{i-1},\hat{z},a_{i+1},\ldots,a_n;z)  -\int_0^z\frac{\dd t}{t-a_{i-1}} \,G(a_1,\ldots,\hat{a}_{i-1},t,a_{i+1},\ldots,a_n;t)\cr
&{} -G(a_{i+1},a_1,\ldots,a_{i-1},\hat{z},a_{i+1},\ldots,a_n;z)  +\int_0^z\frac{\dd t}{t-a_{i+1}}\,G(a_1,\ldots,a_{i-1},t,\hat{a}_{i+1},\ldots,a_n;t)\cr
&{} +\int_0^z\frac{\dd t}{t-a_1}\,G(a_2,\ldots,a_{i-1},t,a_{i+1},\ldots,a_n;t), \qquad i\neq 1,n\, ,
}$$
with the following specialization for when $z$ is the rightmost label (with $n\neq 1$):
\eqnn\espec
$$\displaylines{
G(a_1,\ldots,a_{n-1},z;z)=c(a_1,\ldots,a_{n-1},\hat z)
+  G(a_{n-1},a_1,\ldots,a_{n-1};z) -G(0,a_1,\ldots,a_{n-1};z)\cr
-\int_0^z\frac{\dd t}{t-a_{n-1}}\,G(a_1,\ldots,a_{n-2},t;t)
+\int_0^z\frac{\dd t}{t-a_1}\,G(a_2,\ldots,a_{n-1},t;t)\,.\hfil\espec\hfilneg
}$$
Similar recursions for repeated appearance of $z$ among the labels as in 
$G(\ldots,z,z,\ldots;z)$ can be derived from
\diffone\ and \diffG\ in exactly the same manner.

The integration constants $c(\ldots,\hat z,\ldots)$ in \diffG\ are generically zero unless
the labels are exclusively formed from letters $a_j \in \{0,\hat z\}$, in which
case they yield MZVs \Maxim:
\eqn\Gregstwo{
c(a_1,a_2,\ldots,a_n) = \cases{
0 &: $\exists \ a_j \notin \{0,\hat z\}$ \cr
G({a_1\over \hat z},{a_2\over\hat z},\ldots,{a_n\over\hat z};1) &:
$a_j \in \{0,\hat z\}$
}}
The simplest nonzero applications of \Gregstwo\ at weight two and three are
\eqn\cteex{
c(0,\hat z) = -\zeta_2  ,\quad c(\hat z,0) = +\zeta_2 ,\quad c(0,0,\hat z)= c(\hat z,0,0)=  - \zeta_3,\quad c(0,\hat z,0)=
2\zeta_3
}
and follow from \Gregs. For example, the above steps lead to the $z$-removal identities\foot{Note
that the identities (E.1) in reference \Polylogs\ exclude $a_j=0$ and therefore do not exhibit the constant terms of
\zrmlenthree\ in the analogous identities.}
\eqnn\zrmlenthree
$$\eqalignno{
G(a_1,z;z) &= G(a_1,a_1;z) - G(0,a_1;z) - \d_{a_1,0}\zeta_2\,, &\zrmlenthree \cr
G(a_1,a_2,z;z) &= G(a_2,0,a_1;z) - G(a_2,a_1,a_1;z) + G(a_1, a_2,a_2;z)    \cr
&{} - G(a_1,0,a_2;z) + G(a_2,a_1,a_2;z) - G(0,a_1,a_2;z) \cr
&{}- \delta_{a_2,0} G(a_1;z) \zeta_2 + \delta_{a_1,0} G(a_2;z) \zeta_2 - \delta_{a_1,0} \delta_{a_2,0} \zeta_3 \cr
G(a_1,z,a_2;z) &= G(a_1,a_1,a_2;z) - G(a_2,0,a_1;z) +G(a_2,a_1,a_1;z) \cr
&{}  - G(a_2,a_1,a_2;z) - \d_{a_1,0} G(a_2;z) \zeta_2 + 2 \d_{a_1,0}\d_{a_2,0}\zeta_3 \cr
G(a,z,z;z) &= G(0,0,a;z) - G(0,a,a;z) - G(a,0,a;z)+ G(a,a,a;z) + \delta_{a,0} \zeta_3\,.
}$$
Note that analogous $z$-removal identities for $G(z,a_1;z)$, $G(z,a_1,a_2;z) $ and
other divergent cases follow from the shuffle relation \shufalg, see \EriksReg\ for the regularized 
values of $G(z;z)$ that differ from the choice in \Polylogs.

\medbreak
\medskip
\noindent{\it D.3.2 General $z$-removal identities}
\medskip

\noindent
As exemplified by \noparity, some of the regularized integrals require different orders of integration 
over the variables $z_2,z_3,\ldots,z_{n-2}$. In these situations
it can happen that polylogarithms such as $G(0,z_4;z_3)$ need to be converted to
$G(\ldots;z_4)$ with no additional instance of $z_4$ in the ellipsis in order to
integrate over $z_4$ first. This requires a generalization
of the techniques in the previous subsection. As before, the starting
point for a recursion is the differential equation \diffG\ for derivatives in
the labels of polylogarithms. The recursion is supplemented by the initial
condition
\eqn\zrmCtetwo{
G(z_1;z_2) = G(z_2;z_1)+G(0;z_2)-G(0;z_1) - i\pi \, \sign(z_2,z_1)\,,
}
where
\eqn\defsign{
\sign(z_i,z_j)\equiv
 \cases{
\phantom{+}1 &: $z_i<z_j$ \cr
-1 &: $z_i>z_j$} \ .
}
For example, the first identity in \zrmlenthree\ generalizes to
\eqnn\genzrm
$$\eqalignno{
G(a_1,z_1;z_2) &= G(a_1,0;z_2) - G(a_1,z_2;z_1) - G(0;z_2) G(a_1;z_1) + G(a_1,0;z_1)\cr
&{}  + G(a_1;z_2) \big[ G(a_1;z_1)-G(0;z_1) \big] -2 \delta_{a_1,0} \zeta_2\cr
&{}	+ i\pi \, \sign(z_2,z_1) (G(a_1;z_1) - G(a_1;z_2))\,.  &\genzrm
}$$
Note that the polylogarithms on the right hand side are suitable for integration over $z_1$
since there are no instances of $z_1$ among their labels.

The use of $z$-removal identities represents the most expensive
step in the computation of regularized integrals as they tend to increase the
number of terms considerably. An overview of the weights of the identities
required at a given order of the Berends--Giele recursion is given in Table 1. 
For example, terms at the order of $\ap^6\zeta_6 \Phi^5$ in the $Z$-theory equation
of motion \fullBG\ arise from integrating the third subleading order $\sim \ap^3$
of the Koba--Nielsen factor \knfinal\ -- the offset is due to the factor
$(-\ap)^{(n{-}3)}$ in \VVVn\ -- and
require $z$-removal identities for $G(P;z)$ at weight $\len{P}=5$.

\moveright2.6cm\vbox{\offinterlineskip
\halign{
\strut\vrule\hskip3pt\hfil #\hfil &%
\vrule\hskip3pt\hfil #\hfil &%
\vrule\hskip3pt\hfil #\hfil &%
\vrule\hskip3pt\hfil #\hfil &%
\vrule\hskip3pt\hfil #\hfil\hskip3pt \vrule\cr
\noalign{\hrule}
$n$-pts & MZVs & BG current & $z$-removal & Koba--Nielsen\cr
\noalign{\hrule}
    & $\zeta_2$ & $\phi^4$ & $w=1$ & $\ell=0$\cr
    & $\zeta_3$ & $k^2\phi^4$ & $w=2$ & $\ell=1$\cr
$5$ & $\zeta_4$ & $k^4\phi^4$ & $w=3$ & $\ell=2$\cr
    & $\zeta_5$ & $k^6\phi^4$ & $w=4$ & $\ell=3$\cr
    & $\zeta_6$ & $k^8\phi^4$ & $w=5$ & $\ell=4$\cr
    & $\zeta_7$ & $k^{10}\phi^4$ & $w=6$ & $\ell=5$\cr
\noalign{\hrule}
    & $\zeta_3$ & $\phi^5$ & $w=2$ & $\ell=0$\cr
    & $\zeta_4$ & $k^2\phi^5$ & $w=3$ & $\ell=1$\cr
$6$ & $\zeta_5$ & $k^4\phi^5$ & $w=4$ & $\ell=2$\cr
    & $\zeta_6$ & $k^6\phi^5$ & $w=5$ & $\ell=3$\cr
    & $\zeta_7$ & $k^8\phi^5$ & $w=6$ & $\ell=4$\cr
\noalign{\hrule}
    & $\zeta_4$ & $\phi^6$ & $w=3$ & $\ell=0$\cr
$7$ & $\zeta_5$ & $k^2\phi^6$ & $w=4$ & $\ell=1$\cr
    & $\zeta_6$ & $k^4\phi^6$ & $w=5$ & $\ell=2$\cr
    & $\zeta_7$ & $k^6\phi^6$ & $w=6$ & $\ell=3$\cr
\noalign{\hrule}
    & $\zeta_5$ & $\phi^7$ & $w=4$ & $\ell=0$\cr
$8$ & $\zeta_6$ & $k^2\phi^7$ & $w=5$ & $\ell=1$\cr
    & $\zeta_7$ & $k^4\phi^7$ & $w=6$ & $\ell=2$\cr
\noalign{\hrule}
   $9$ & $\zeta_6$ & $\phi^8$ & $w=5$ & $\ell=0$\cr
    & $\zeta_7$ & $k^2\phi^8$ & $w=6$ & $\ell=1$\cr
    \noalign{\hrule}
   $10$ & $\zeta_7$ & $\phi^9$ & $w=6$ & $\ell=0$\cr
\noalign{\hrule}
}}
\smallskip
{\leftskip=0pt\rightskip=20pt\noindent\ninepoint\baselineskip=11pt
{\bf Table 1.} Summary of the contributions from regularized $n$-point integrals, the
order of MZVs, the schematic form of the Berends--Giele double current, the
required weight $w$ of $z$-removal identities ($G(a_1, \ldots,a_w;z)$)
and the order $\ap^\ell$ of the Koba--Nielsen expansion \knfinal.\par}

\appendix{E}{Alternative description of regularized disk integrals}
\applab\Jregapp

\noindent
In this appendix, we present a method to determine the $\ap$-expansions for regularized disk integrals
selected by the $Z$-theory equation of motion from the $(n{-}3)!\times (n{-}3)!$ 
basis $F_{P}{}^Q$ defined in \defF. This approach has been very useful to 
constrain the required regularization scheme via explicit data at high orders of $\alpha'$, without the need to obtain the 
Berends--Giele recursion from an ansatz at these orders. However, we only understand this method as an intermediate
tool to determine the appropriate regularization scheme selected by the $Z$-theory equation of motion: The ultimate goal and 
achievement of this work is to compute $\alpha'$-expansions of disk integrals at multiplicities and orders where no prior knowledge of $F_{P}{}^Q$ is available.

Closely following the lines of \Polylogs, the basic idea is to divide disk integrals\foot{For the sake of simplicity, the 
discussion of \Polylogs\ and the current appendix is restricted to
linear combinations of disk integrals $Z(I|P)$ with the canonical domain $I=12\ldots n$, where the choices of $P$ 
only leave a single pole channel in the field-theory limit.} $Z(I|P)$ into a singular and a regular part
with respect to region variables $s_{i,i+1\ldots j}$ in \mandmom. The singular parts associated with the
propagators of the field-theory limits can be subtracted with residues given by lower-multiplicity data,
and the leftover local expression is identified with the regularized integrals in \fullBG. However, there 
are ambiguities in the subtraction scheme by shifting the numerator $N \rightarrow N+{\cal O}(s)$ in the 
subtracted singular expression $N/s$ by polynomials in the associated Mandelstam invariant $s\equiv s_{i,i+1\ldots j}$. 
Five-point examples suggest that changes in the regularization scheme or the integration order
can be compensated by the choice of subtraction scheme when reproducing the associated local expressions from
regularized integrals over Taylor-expanded Koba--Nielsen factors.

In the setup of \Polylogs, the regularization scheme for divergent integrals was fixed and designed to preserve the shuffle 
algebra and scaling relations of polylogarithms such that $G(z;z) \equiv 0$ instead of \EriksReg. 
Moreover, the integration orders were globally chosen as $23\ldots n{-}2$ (i.e.\ integrating over $z_2$ first and 
over $z_{n-2}$ in the last step). In all examples under consideration 
in \Polylogs, it was possible to choose a scheme for pole subtraction such that the resulting regular parts could 
be reproduced by integration in the canonical order $23\ldots n{-}2$ within the given scaling-preserving regularization. 
In these adjustments of the subtraction scheme, certain regular admixtures were 
incorporated by systematically shifting the arguments of the lower-point integrals in the above numerators $N$.

Here, by contrast, we work with a fixed (or ``minimal'') subtraction scheme for the poles of 
$Z(I|P)$. The resulting regular parts -- to be denoted by $J_{\ldots}^\reg(\ldots)$ in the sequel -- 
turn out to exactly reproduce the desired $Z$-theory equation of motion upon insertion into \fullBG. As 
will become clear from the following examples, 
this subtraction scheme is canonical in the sense that the aforementioned regular admixtures of \Polylogs\ are 
completely avoided, reflecting the different choices of regularization scheme and integration orders 
between this work and \Polylogs.

We will regard $SL(2,\Bbb R)$-fixed combinations of disk integrals $Z(P|Q)$ in the notation
\eqn\Jnonreg{
J_{u_1v_1,u_2 v_2,\ldots,u_{n-3} v_{n-3}}(k_1,k_2,\ldots,k_{n-1}) \equiv \ap^{n-3}
\! \! \! \! \! \! \! \! \! \! \!   \! \! \! \!  \int\limits_{0\leq z_{2} \leq z_{3} \leq \ldots \leq z_{n-2} \leq 1}
\! \! \! \! \! \! \! \! \! \! \! 
{ \dd z_2 \, \dd z_3 \, \ldots \, \dd z_{n-2}  \, \prod_{i<j}^{n-1}
|z_{ij}|^{\ap s_{ij}} \over z_{u_1, v_1}z_{u_2, v_2}\ldots z_{u_{n-3}, v_{n-3} }}
 }
as functions of $n{-}1$ massless momenta $k_j$ which determine the $s_{ij}$ on the right hand side
through their independent dot products. The product $k_1\cdot k_{n-1}$ can be eliminated by momentum conservation
and is absent in \Jnonreg\ by the $SL(2,\Bbb R)$-fixing $z_1=0$ and $z_{n-1}=1$. This reflects the choice of ansatz in
appendix \appANS, where $(k_{A_1}\cdot k_{A_p})$ referring to the outermost slots $A_1,A_p$ in a deconcatenation 
$\sum_{A=A_1A_2\ldots A_p}$ is excluded.

In the four-point case, the field-theory limit of \Jnonreg, which follows from the rules in section 4 
of \Polylogs\ or from \nonabsix, already exhausts the singular part. Hence, the
expressions
\eqn\fourJreg{
J_{21}^\reg(k_1,k_2,k_3) = J_{21}(k_1,k_2,k_3) - \frac{1}{s_{12}} \ , \ \ \ \ 
J_{32}^\reg(k_1,k_2,k_3) = J_{32}(k_1,k_2,k_3) - \frac{1}{s_{23}} 
}
are analytic in $s_{ij}$ and coincide with the regularized integrals \regAB\ \Polylogs\ in any regularization 
scheme of our awareness. Their $\ap$-expansion is straightforwardly determined by $F_2{}^2$ in \Ffour\ 
(also see \BoelsJUA\ for a neat representation in terms of $G(0,\ldots,0,1,\ldots,1;1)$),
\eqn\fourJregA{
J_{21}(k_1,k_2,k_3) = {F_2{}^2 \over s_{12}} \ , \ \ \ \
J_{32}(k_1,k_2,k_3) = {F_2{}^2 \over s_{23}} \, .
}
The regular parts $J_{ij}^\reg(\ldots)$ in \fourJreg\ are by themselves functions of three
light-like momenta
under $s_{pq} \rightarrow k_p\cdot k_q$ and can later on be promoted to massive momenta $k_P$ provided that
no reference to $k_P^2$ is expected.

\medskip
\noindent{\it E.1 Five-point pole subtraction}
\medskip

\noindent 
At five points, generic field-theory limits of $Z(P|Q)$ yield two simultaneous propagators, and by factorization 
on four-point integrals, the residue on single poles in $s_{ij}$ still involves all orders in $\ap$. As elaborated 
in \Polylogs, the $\ap$-dependence of the singular pieces can be removed using the regular four-point expressions 
in \fourJreg\ with composite momenta $k_{ij}\equiv k_i{+}k_j$,
\eqnn\fiveJreg
$$\eqalignno{
J_{21,43}^\reg(k_1,k_2,k_3,k_4) &= J_{21,43}(k_1,k_2,k_3,k_4)  
 -  \frac{J_{21}^\reg(k_1,k_2,k_{34})}{s_{34}}  - \frac{J_{32}^\reg(k_{12},k_3,k_4)}{s_{12}} 
 -  \frac{1}{s_{12} s_{34}}\cr
J_{31,42}^\reg(k_1,k_2,k_3,k_4) &= J_{31,42}(k_1,k_2,k_3,k_4) &\fiveJreg
\cr
J_{21,31}^\reg(k_1,k_2,k_3,k_4)&= J_{21,31}(k_1,k_2,k_3,k_4)   -   \frac{J_{21}^\reg(k_1,k_2,k_3)}{s_{123}}  -   \frac{J_{21}^\reg( k_{12},k_3,k_4 )}{s_{12}}   -  \frac{1}{s_{12} s_{123}} 
 \cr
J_{32,31}^\reg(k_1,k_2,k_3,k_4)&= J_{32,31}(k_1,k_2,k_3,k_4)  -  \frac{J_{32}^\reg(k_1,k_2,k_3)}{s_{123}}  -  
 \frac{J_{21}^\reg(k_1,k_{23},k_4)}{s_{23}}   -  \frac{1}{s_{23} s_{123}} \ .
}$$
Following the dot products of momenta, arguments $k_{12},k_3,k_4$ in the above $J_{ij}^\reg$ 
instruct to replace any $s_{12}$ and $s_{23}$ in their expansion from \fourJreg\ and \fourJregA\ by 
$s_{13}+s_{23}$ and $s_{34}$, respectively \Polylogs. Note that the counterpart of 
$J_{21}^\reg(k_1,k_{23},k_4)$ in \Polylogs\ required a different replacement $s_{12}\rightarrow s_{123}$
instead of the prescription $s_{12}\rightarrow s_{12}+s_{13}$ in \fiveJreg. This kind of dependence on 
$k_{23}^2=2s_{23}$ was inevitable to accommodate with the regularization scheme of the \Polylogs\ with $G(z;z) \equiv 0$.

In the same way as the $\ap$-dependence of the local four-point expressions $J_{ij}^\reg(\ldots)$ is accessible 
from $F_2{}^2$, their five-point counterparts $J_{ij,pq}^\reg(\ldots)$ can be expanded as soon as the right hand 
side of \fiveJreg\ is expressed in terms of the basis functions $\{F_{23}{}^{23},F_{23}{}^{32}\}$,
\eqnn\fiveJc
$$\eqalignno{
J_{21,43}(k_1,k_2,k_3,k_4)  &= \frac{F_{23}{}^{23}}{s_{12}s_{34}} \ , \ \ \ \
J_{21,31}(k_1,k_2,k_3,k_4) = 
\frac{F_{23}{}^{23}}{s_{12} s_{123}}  +  \frac{F_{23}{}^{32}}{s_{13} s_{123}}  &\fiveJc \cr
J_{31,42}(k_1,k_2,k_3,k_4)  &= \frac{F_{23}{}^{32}}{s_{13}s_{24}}
\ , \ \ \ \
J_{32,31}(k_1,k_2,k_3,k_4) =  \frac{F_{23}{}^{23} }{s_{23} s_{123}}  -  \Big( \frac{1}{s_{13}} + \frac{1}{s_{23} } \Big) \, \frac{F_{23}{}^{32}} {s_{123}} \ .
}$$
Explicit results on the $\ap$-expansion of $\{F_{23}{}^{23},F_{23}{}^{32}\}$ as pioneered in \OprisaWU\ are 
available from the all-multiplicity methods based on polylogarithms \Polylogs\ and the Drinfeld associator \BroedelAZA. 
Moreover, recent advances based on their hypergeometric-function representation \refs{\BoelsJUA, \PuhlfuerstGTA} 
render even higher orders in $\ap$ accessible, also see \PuhlfuerstGTA\ for a closed-form solution. Once we adjoin the 
parity images
\eqnn\fiveJd
$$\eqalignno{
J_{43,42}^\reg(k_1,k_2,k_3,k_4) &= J_{21,31}^\reg(k_1,k_2,k_3,k_4) \big|_{k_j \rightarrow k_{5-j}} &\fiveJd
\cr
J_{42,32}^\reg(k_1,k_2,k_3,k_4) &= J_{32,31}^\reg(k_1,k_2,k_3,k_4) \big|_{k_j \rightarrow k_{5-j}} \ ,
}$$
one can extract valuable all-weight information on the regularization scheme for
five-point integrals in \fullBG\ by
demanding the $\ap$-expansion of \fiveJreg\ and \fiveJd\ to match with
\eqn\matchJreg{
J_{pq,rs}^\reg(k_1,k_2,k_3,k_4) = \ap^2 \int^{{\rm eom}} \prod_{i<j}^4 |z_{ij}|^{\ap s_{ij}} \, {1\over z_{pq} z_{rs}} \ .
}
Again, the arguments $s_{ij} \rightarrow k_i\cdot k_j$ of $J_{pq,rs}^\reg$ can be promoted to massive momenta 
$k_i\rightarrow k_P$ as we will now see in the pole subtractions at higher-multiplicity.

\medskip
\noindent{\it E.2 Six and seven-point pole subtraction}
\medskip

\noindent 
The above five-point examples shed light on various aspects of the regularization scheme selected by the $Z$-theory
equation of motion including the integration orderings and the $z$-removal identities in appendix D.3. 
However, the appearance of $i\pi$ in \EriksReg\ cannot be seen from integrals below multiplicity six, so the 
$J^\reg_{\ldots}(\ldots)$ at $(n\geq 6)$-points have been an instrumental window to infer these particularly subtle 
ingredients of the regularization scheme. In this section, we present one example each at multiplicity six and seven:
\eqnn\sixJreg
$$\eqalignno{
J^\reg_{31,32,54}(k_1,k_2,\ldots,k_5) &=J_{31,32,54}(k_1,k_2,\ldots,k_5) - \frac{ J_{31,32}^\reg(k_1,k_2,k_3,k_{45}) }{s_{45}}  \cr
&\! \! \! \! \! \! \! \! \! \! \! \! \! \! \! \! \! \! \! \! \! \! \! \! - \frac{ J_{32}^\reg(k_1,k_2,k_3) J_{32}^\reg(k_{123},k_4,k_5)  }{s_{123}} - \frac{ J_{21,43}^\reg(k_1,k_{23},k_4,k_5) }{s_{23}}   - \frac{ J_{32}^\reg(k_1,k_2,k_3) }{s_{123} s_{45}}\cr
&\! \! \! \! \! \! \! \! \! \! \! \! \! \! \! \! \! \! \! \! \! \! \! \! -  \frac{ J_{21}^\reg(k_1,k_{23},k_{45}) }{s_{23}s_{45}}  - \frac{J_{32}^\reg(k_{123},k_4,k_5) }{s_{23} s_{123} } - \frac{1}{s_{23}s_{123}s_{45}}  &\sixJreg
}$$
Note that also the counterparts of $J_{21}^\reg(k_1,k_{23},k_{45})$ and
$J_{21,43}^\reg(k_1,k_{23},k_4,k_5) $ seen in \Polylogs\ exhibit additional
contributions $\sim s_{23}$ in their arguments. In the $J_{\ldots}^\reg(\ldots) $
under discussion, however, the argument $s_{23}={1\over 2}k_{23}^2$ is by
construction absent in $k_1\cdot k_{23}=s_{12}+s_{13}$.

At seven points, the local integral used in \Polylogs\ to generate the expansion
of $F_P{}^Q$ up to and including the $\ap^7$-order stored
on the website \WWW\ matches with
\eqnn\sevenJreg
$$\eqalignno{
&J^\reg_{21,31,41,65}  =  - \frac{1}{s_{12}s_{123}s_{1234}s_{56}}
- \frac{ J_{21}^\reg(k_1,k_2,k_3) }{s_{123}s_{1234}s_{56}}
 - \frac{ J^\reg_{21}(k_{12},k_3,k_{4}) }{s_{12}s_{1234}s_{56} }
  - \frac{ J^\reg_{21}(k_{123},k_4,k_{56}) }{s_{12}s_{123}s_{56}} \cr
&- \frac{ J^\reg_{32}(k_{1234},k_5,k_6) }{s_{12}s_{123}s_{1234}}
- \frac{ J_{21,31}^\reg(k_1,k_2,k_3,k_{4}) }{s_{56} s_{1234}} - \frac{  J_{21}^\reg(k_1,k_2,k_3) J^\reg_{21}(k_{123},k_4,k_{56})  }{s_{123}s_{56}} &\sevenJreg\cr
 &- \frac{  J_{21}^\reg(k_1,k_2,k_3)  J^\reg_{32}(k_{1234},k_5,k_6) }{s_{123}s_{1234}} - \frac{ J_{21,31}^\reg(k_{12},k_3,k_4,k_{56} )}{s_{12}s_{56}} - \frac{ J^\reg_{21,43}(k_{123},k_4,k_5,k_6) }{s_{12}s_{123}} \cr
 &- \frac{ J^\reg_{21}(k_{12},k_3,k_{4})  J^\reg_{32}(k_{1234},k_5,k_6)  }{s_{12}s_{1234}}
-  \frac{ J^\reg_{21,31,41}(k_1,k_2,k_3,k_4,k_{56}) }{s_{56}} - \frac{J^\reg_{21,31,54}(k_{12},k_3,k_4,k_5,k_6)}{s_{12}}  \cr
 &- \frac{ J_{21}^\reg(k_1,k_2,k_3) J^\reg_{21,43}(k_{123},k_4,k_5,k_6) }{ s_{123}} - \frac{ J_{21,31}^\reg(k_1,k_2,k_3,k_{4}) J^\reg_{32}(k_{1234},k_5,k_6) }{s_{1234}} + J_{21,31,41,65} \ .
}$$
The $\ap$-expansion of the right hand sides of \sixJreg\ and \sevenJreg\ is available
from the following decompositions into basis functions $F_P{}^Q$:
\eqnn\sixseven
$$\eqalignno{
J_{31,32,54}&= {F_{234}{}^{234} \over s_{23} s_{45} s_{123} } - 
 {F_{234}{}^{324} \over s_{123} s_{45}} \Big( {1 \over s_{13}} + {1 \over s_{23}} \Big) &\sixseven \cr
 J_{21,31,41,65} &= {1\over s_{1234} s_{56} } \Big( { F_{2345}{}\!\!\!^{2345}  \over s_{12} s_{123}} 
 + { F_{2345}{}\!\!\!^{2435}  \over s_{12} s_{124}} 
  + { F_{2345}{}\!\!\!^{3245}  \over s_{13} s_{123}} 
   + { F_{2345}{}\!\!\!^{3425}  \over s_{13} s_{134}} 
    + { F_{2345}{}\!\!\!^{4235}  \over s_{14} s_{124}} 
     + { F_{2345}{}\!\!\!^{4325}  \over s_{14} s_{134}} 
\Big) 
      }$$

\medskip
\noindent{\it E.3 The general strategy}
\medskip

\noindent 
The choice of labels and momenta for the $J^\reg_{\ldots}(k_{A_1},k_{A_2},\ldots,k_{A_{m-1}})$ in the above pole subtractions follows from an 
algorithm explained in section 4.3 of \Polylogs. This algorithm applies to integrals $J_{\ldots}(\ldots)$ of the 
form \Jnonreg\ with a single cubic diagram in their field-theory limit. Each factor of $z_{ij}^{-1}$ in the integrand is 
associated with one of the $n{-}3$ propagators of the field-theory diagram, and the pole subtraction exhausts 
all $2^{n-3}$ possibilities to relax a subset of these propagators. The residue of diagrams with less than $n{-}3$ 
propagators is a $J^\reg_{\ldots}(\ldots)$ labeled by the $z_{ij}^{-1}$-factors associated with the relaxed 
propagators, i.e.\ each relaxed propagator increases the multiplicity of the associated $J^\reg_{\ldots}(\ldots)$ by 
one. The massive momenta in its arguments can be read off from the structure of the leftover propagators in the 
diagram. The reader is referred to \Polylogs\ for further details, examples and diagrammatic illustrations.

From these rules, it is straightforward to extract the local parts of integrals at arbitrary multiplicity. We have checked up to 
and including the $5!$ integrals at seven points that these $J^\reg_{\ldots}(\ldots)$ at $s_{ij} \leftrightarrow k_i \cdot k_j$ 
are compatible with the integrals \fullBG\
in the regularization scheme and integration orders of this work,
\eqn\matchallN{
J_{u_1v_1,u_2 v_2,\ldots,u_{p-2} v_{p-2}}^\reg(k_1,k_2,\ldots,k_p) =(\ap)^{p-2} \int^{{\rm eom}}  \, {\prod_{i<j}^p |z_{ij}|^{\ap s_{ij}} \over 
z_{u_1, v_1}z_{u_2, v_2}\ldots z_{u_{p-2}, v_{p-2} }
} \ .
}
A variety of alternative regularization schemes and 
integration orders including those of \Polylogs\ are expected to correspond to a modified choice of arguments for 
$J^\reg_{\ldots}(k_{A_1},k_{A_2},\ldots,k_{A_{n-1}})$, where selected dot products $k_{A_p}\cdot k_{A_{q}}$ are 
shifted by (half of) $k_{A_{i}}^2$.

\appendix{F}{Integration orders for the seven-point integrals}
\applab\orderapp

\noindent
In this appendix, we explicitly list the results of section \secfourB\ on the integration orders for
regularized seven-point integrals in the simpset basis (see section \secfourA). The first topology
of seven-point integrals is spanned by single factors of ${\cal Z}_{1P}$ in \calZ\ with $|P|=4$:
\medskip
\eqnn\sevenZ
\settabs \+ \hskip 4.4cm \hfill & \hskip 5.1 cm \hfill & \hskip 6.1cm & \cr
\+ $z_{15}z_{12}z_{13}z_{14} \rightarrow 2345$, &
   $z_{15}z_{12}z_{34}z_{14} \rightarrow (2\shuffle 3)45$ ,&
   $z_{15}z_{23}z_{24}z_{14} \rightarrow 3245$ , \hskip0.6cm\sevenZ\cr
\+ $z_{15}z_{13}z_{23}z_{14} \rightarrow 2345$ ,&
   $z_{15}z_{13}z_{24}z_{14} \rightarrow (2\shuffle 3)45$ ,&
   $z_{15}z_{24}z_{34}z_{14} \rightarrow 3245$, \cr
\+ $z_{15}z_{12}z_{13} z_{45} \rightarrow (23\shuffle 4)5$ ,&
   $z_{15}z_{14}z_{24} z_{35} \rightarrow (24\shuffle 3)5$ ,&
   $z_{15}z_{13}z_{14} z_{25} \rightarrow (34\shuffle 2)5$,\cr
\+ $z_{15}z_{13}z_{23} z_{45}  \rightarrow (23\shuffle 4)5$ ,&
   $z_{15}z_{12}z_{14} z_{35} \rightarrow (24\shuffle 3)5$ ,&
   $z_{15}z_{14}z_{34} z_{25} \rightarrow (34\shuffle 2)5$ ,\cr
\+ $z_{15}z_{12}z_{35} z_{45} \rightarrow (43\shuffle 2)5$ ,&
   $z_{15}z_{13}z_{24} z_{25}  \rightarrow (42\shuffle 3)5$ ,&
   $z_{15}z_{14}z_{25} z_{35} \rightarrow (32\shuffle 4)5$ ,\cr
\+ $z_{15}z_{12}z_{34} z_{35} \rightarrow (43\shuffle 2)5$ ,&
   $z_{15}z_{13}z_{25} z_{45} \rightarrow (42\shuffle 3)5$ ,&
   $z_{15}z_{14}z_{23} z_{25} \rightarrow (32\shuffle 2)5$ ,\cr
\+ $z_{15}z_{25}z_{35}z_{45} \rightarrow 4325$ ,&
   $z_{15}z_{25}z_{23}z_{45} \rightarrow (3\shuffle 4)25$ ,&
   $z_{15}z_{25}z_{24}z_{34} \rightarrow 3425$ ,\cr
\+ $z_{15}z_{25}z_{34}z_{35} \rightarrow 4325$ ,&
   $z_{15}z_{25}z_{24}z_{35} \rightarrow (3\shuffle 4)25$ ,&
   $z_{15}z_{25}z_{24}z_{23} \rightarrow 3425$ . \cr
\medskip \noindent
Another seven-point topology can be derived from products
${\cal Z}_{1P} {\cal Z}_{6Q}$ with $|P|=3, \ |Q|=1$:
\medskip
\eqnn\seventopB
\settabs \+ \hskip 4.4cm \hfill & \hskip 5.1 cm \hfill & \hskip 6.1cm & \cr
\+ $z_{12}z_{13}z_{14}z_{56} \rightarrow 234 \shuffle 5$ ,&
   $z_{12}z_{34}z_{14}z_{56} \rightarrow ((2\shuffle 3)4)\shuffle5$ ,&
   $z_{23}z_{24}z_{14}z_{56} \rightarrow 324\shuffle5$ , \hskip0.3cm\seventopB\cr
\+ $z_{13}z_{23}z_{14}z_{56} \rightarrow 234\shuffle5$ ,&
   $z_{13}z_{24}z_{14}z_{56} \rightarrow ((2\shuffle 3)4)\shuffle5$ ,&
   $z_{24}z_{34}z_{14}z_{56} \rightarrow 324\shuffle5$ , \cr
\+ $z_{12}z_{13}z_{15}z_{46} \rightarrow 235 \shuffle 4$ ,&
   $z_{12}z_{35}z_{15}z_{46} \rightarrow ((2\shuffle 3)5)\shuffle4$ ,&
   $z_{23}z_{25}z_{15}z_{46} \rightarrow 325\shuffle4$ , \cr
\+ $z_{13}z_{23}z_{15}z_{46} \rightarrow 235\shuffle4$ ,&
   $z_{13}z_{25}z_{15}z_{46} \rightarrow ((2\shuffle 3)5)\shuffle4$ ,&
   $z_{25}z_{35}z_{15}z_{46} \rightarrow 325\shuffle4$ , \cr
\+ $z_{12}z_{14}z_{15}z_{36} \rightarrow 245 \shuffle 3$ ,&
   $z_{12}z_{45}z_{15}z_{36} \rightarrow ((2\shuffle 4)5)\shuffle3$ ,&
   $z_{24}z_{25}z_{15}z_{36} \rightarrow 425\shuffle3$ , \cr
\+ $z_{14}z_{24}z_{15}z_{36} \rightarrow 245\shuffle3$ ,&
   $z_{14}z_{25}z_{15}z_{36} \rightarrow ((2\shuffle 4)5)\shuffle3$ ,&
   $z_{25}z_{45}z_{15}z_{36} \rightarrow 425\shuffle3$ , \cr
\+ $z_{13}z_{14}z_{15}z_{26} \rightarrow 345 \shuffle 2$ ,&
   $z_{13}z_{45}z_{15}z_{26} \rightarrow ((3\shuffle 4)5)\shuffle2$ ,&
   $z_{34}z_{35}z_{15}z_{26} \rightarrow 435\shuffle2$ , \cr
\+ $z_{14}z_{34}z_{15}z_{26} \rightarrow 345\shuffle2$ ,&
   $z_{14}z_{35}z_{15}z_{26} \rightarrow ((3\shuffle 4)5)\shuffle2$ ,&
   $z_{35}z_{45}z_{15}z_{26} \rightarrow 435\shuffle2$ . \cr
\medskip \noindent
The seven-point topology of ${\cal Z}_{1P} {\cal Z}_{6Q}$ with $|P|=|Q|=2$ for both factors
gives rise to the following integration orders,
\medskip
\eqnn\sevenE
\settabs \+ \hskip 4.4cm \hfill & \hskip 5.1cm \hfill & \hskip 6.1cm & \hfill \cr
\+ $z_{12}z_{13} z_{46} z_{56} \rightarrow 23\shuffle 54$ ,&
   $z_{13}z_{23} z_{46} z_{56} \rightarrow 23\shuffle 54$ ,&
   $z_{12}z_{13} z_{45} z_{46} \rightarrow 23\shuffle 54$ ,\hskip0.4cm\sevenE\cr
\+ $z_{13}z_{23} z_{45} z_{46} \rightarrow 23\shuffle 54$ ,&
   $z_{12}z_{14} z_{36} z_{56} \rightarrow 24\shuffle 53$ ,&
   $z_{14}z_{24} z_{36} z_{56} \rightarrow 24\shuffle 53$ ,\cr
\+ $z_{12}z_{14} z_{35} z_{36} \rightarrow 24\shuffle 53$ ,&
   $z_{14}z_{24} z_{35} z_{36} \rightarrow 24\shuffle 53$ ,&
   $z_{13}z_{14} z_{26} z_{56} \rightarrow 34\shuffle 52$ ,\cr
\+ $z_{14}z_{34} z_{26} z_{56} \rightarrow 34\shuffle 52$ ,&
   $z_{13}z_{14} z_{25} z_{26} \rightarrow 34\shuffle 52$ ,&
   $z_{14}z_{34} z_{25} z_{26} \rightarrow 34\shuffle 52$ ,\cr
\+ $z_{12}z_{15} z_{36} z_{46} \rightarrow 25\shuffle 43$ ,&
   $z_{15}z_{25} z_{36} z_{46} \rightarrow 25\shuffle 43$ ,&
   $z_{12}z_{15} z_{34} z_{36} \rightarrow 25\shuffle 43$ ,\cr
\+ $z_{15}z_{25} z_{34} z_{36} \rightarrow 25\shuffle 43$ ,&
   $z_{13}z_{15} z_{26} z_{46} \rightarrow 35\shuffle 42$ ,&
   $z_{15}z_{35} z_{26} z_{46} \rightarrow 35\shuffle 42$ ,\cr
\+ $z_{13}z_{15} z_{24} z_{26} \rightarrow 35\shuffle 42$ ,&
   $z_{15}z_{35} z_{24} z_{26} \rightarrow 35\shuffle 42$ ,&
   $z_{14}z_{15} z_{26} z_{36} \rightarrow 45\shuffle 32$ ,\cr
\+ $z_{15}z_{45} z_{26} z_{36} \rightarrow 45\shuffle 32$ ,&
   $z_{14}z_{15} z_{23} z_{26} \rightarrow 45\shuffle 32$ ,&
   $z_{15}z_{45} z_{23} z_{26} \rightarrow 45\shuffle 32$ ,\cr
\medskip \noindent
and the remaining topologies of the simpset basis at seven points follow from
\sevenZ\ and \seventopB\ via parity $z_{j} \rightarrow z_{7-j}$. The seven-point
$\int^\eom$-integrals are sufficient to determine the $\Phi^6$-order of the
$Z$-theory equation of motion \fullBG\ to any order in $\ap$ and the $\ap^4$-order
of disk integrals at any multiplicity.

\ninerm
\footskip12.5pt
\listrefs
\bye